\documentclass[aps,prb,twocolumn,showpacs]{revtex4}
\usepackage{ifthen}
\newcommand{\usepdfs}{false}
\ifthenelse{\equal{\usepdfs}{true}}{
  \usepackage[pdftex]{graphicx}
}
{
  \usepackage[dvips]{graphicx}
}
\ifthenelse{\equal{\usepdfs}{true}}{
  
}
{
  
}

\usepackage{amssymb,amsmath,amsfonts,hyperref}
\usepackage{wasysym}
\usepackage{latexsym}
\usepackage{subfigure}

\newcommand{\ket}[1]{| #1 \rangle}
\newcommand{\bra}[1]{\langle #1 |}
\newcommand{\dirac}[2]{\langle #1 | #2 \rangle}

\renewcommand{\u}{\uparrow}
\renewcommand{\d}{\downarrow}

\newcommand{\be}{\begin{equation}}
\newcommand{\ee}{\end{equation}}

\newcommand{\Jx}{J_{\perp}}

\newcommand{\cdaf}{CdCr$_2$O$_4$}
\newcommand{\hgaf}{HgCr$_2$O$_4$}

\begin{document}

\title{Degenerate perturbation theory of quantum fluctuations in a
  pyrochlore antiferromagnet} \date{\today}

\author{Doron L. Bergman$^1$, Ryuichi Shindou$^1$, Gregory
  A. Fiete$^2$, and Leon Balents$^1$}
\affiliation{${}^1$Department of Physics, University of California,
  Santa Barbara, CA 
93106-9530\\$^2$Kavli Institute for Theoretical Physics, University of 
California, Santa Barbara, CA 93106-4030}

\begin{abstract}
  We study the effect of quantum fluctuations on the half-polarized
  magnetization plateau of a pyrochlore antiferromagnet.  We argue
  that an expansion around the easy axis limit is appropriate for
  discussing the ground state selection amongst the classically
  degenerate manifold of collinear states with a 3:1 ratio of spins
  parallel/anti-parallel to the magnetization axis.  A general
  approach to the necessary degenerate perturbation theory is
  presented, and an effective quantum dimer model within this
  degenerate manifold is derived for arbitrary spin $s$.  We also
  generalize the existing semiclassical analysis of Hizi and
  Henley [Phys. Rev. B {\bf 73}, 054403 (2006)] to the easy axis limit, and show that both
  approaches agree at large $s$.  We show that under rather general
  conditions, the first non-constant terms in the effective
  Hamiltonian for $s\geq 1$ occur only at {\sl sixth} order in the
  transverse exchange coupling.  For $s\geq 3/2$, the effective Hamiltonian
  predicts a magnetically ordered state. For $s\leq 1$ more exotic
  possibilities may be realized, though an analytical solution of the
  resulting quantum dimer model is not possible.
  \end{abstract}
\date{\today}



\maketitle

\section{Introduction}\label{intro}

Magnetism is an inherently quantum mechanical effect that ultimately
arises from the exchange processes in an interacting many-particle
quantum system.  Thus, magnetic systems can reveal much about the
richness of quantum mechanics itself, especially when cooperative
effects are operational and emergent effective low-energy theories
describe the relevant physics.  A few notable theoretical examples are the spin
liquid phases obtained in frustrated
magnets\cite{Fazekas:pm74,Hermele:prb04,Alicea:prl05,Moessner:prb01,Sachdev:prb92}.
Quantum effects may play a central role in other types of order as well.
Recent experiments on a number of insulating chromate compounds,
namely \cdaf\, and \hgaf,\, have shown peculiar features in the
low-temperature magnetization as a function of applied magnetic field.
At low temperatures the magnetization grows linearly with magnetic
field up to some critical value at which point there is a sharp jump
in magnetization onto a rather wide plateau with half the full
saturation magnetization.\cite{Ueda:prb06,Ueda:prl05,Matsuda:06} With
sufficiently large fields it is possible to observe a smooth
transition off the half-magnetization plateau and a gradual increase
in magnetization up to what may be a fully polarized plateau
state.\cite{Ueda:prb06} As described in
Ref.~\onlinecite{Bergman:prl05}, it is expected that the magnetism in
these compounds is well described by the Heisenberg antiferromagnet
(AFM) of spin $s = \frac{3}{2}$ on the pyrochlore lattice.

Motivated by these experimental examples of a half polarization
plateau in the pyrochlore Heisenberg AFM, we conduct a theoretical
study of the quantum pyrochlore Heisenberg AFM for any spin value $s$,
in a strong magnetic field, focusing on a half polarization plateau.
While the physics of \hgaf\; is probably determined to a large degree by
the classical physics of spin-lattice interactions,\cite{Bergman:prb06,Penc:prl04,Penc:05,Shannon:05,Fennie:06} it
may be that in other similar compounds where coupling to phonons is
weak, quantum effects could play a significant role.  

In any case, the general problem of determining the spin state on
plateaus of non-zero magnetization in frustrated magnets occurs in a
large number of materials\cite{Chandra:prb04,Narumi:el04,Matsumoto:prb03,Tanaka:02,Miyahara:03,Misguich:prl01,Uchida:jpsj01,Momoi:prb00}.  
At fields large enough to induce
substantial magnetization, the ground state is expected to be very
different from the zero field state and one would ideally pursue a
theoretical approach that takes advantage of the large external field.
The methods developed indeed use this explicitly, and the particular
application to the half-polarized pyrochlore magnetization plateau
provides a rather non-trivial test bed.  We make use of the large
field to justify an easy-axis approximation to a nearest-neighbor XXZ
antiferromagnetic in an external field. Physically, at large fields
the spin is oriented on average more along the field axis than
transverse to it.  Furthermore, specifically on a magnetization
plateau, general arguments imply that the static transverse moment
vanishes {\sl on every site}, $\langle S_i^\pm\rangle=0$.  Thus we
expect that Degenerate Perturbation Theory (DPT) about the easy-axis
(Ising) limit should be justified on the plateau, and one may thereby
derive an effective Hamiltonian.  This effective Hamiltonian acts in
the {\sl constrained} ``3:1'' space of states with $3$ majority spins with
$S_i^z=+s$ and $1$ minority spin with $S_i^z=-s$ on each tetrahedron.
This space is macroscopically degenerate, and all its members have
half the saturation magnetization.

The reader may well wonder whether there is any need for an approach
of this sort, given the successes of the large $s$ semiclassical
spin-wave method in many other contexts.  Indeed, for
unfrustrated antiferromagnets, it is known that the $1/s$ expansion
gives reasonably convergent results even down to $s=1/2$.  However,
this convergence is strongly dependent upon the lattice -- large
corrections to the spin-wave dispersion have recently been obtained
even for the rather weakly-frustrated triangular
lattice.\cite{Starykh:06,Zheng:06} In highly frustrated magnets such as the
pyrochlore, another approach is warranted.  Particularly worrisome in
the $1/s$ expansion is the difficulty of treating {\sl tunneling},
which is non-perturbative in this method\cite{Henley:prb93,Hizi:prb06,Henley:prl06}.  By contrast, in the
easy-axis expansion, tunneling and virtual exchange are treated on the
same footing.  Of course, for large $s$ both approaches must agree,
and we will indeed check that this is the case in our specific
application.   

The effective Hamiltonian describing splittings within the degenerate
manifold of Ising ground states will generally take the form of a
constrained quantum Ising model.  As explained in a previous
publication,\cite{Bergman:prl05} for the case of the pyrochlore
half-magnetization plateau in the easy-axis limit, it can be cast in
the form of a ``Quantum Dimer Model'' (QDM) on the bipartite diamond
lattice.  Such QDMs are known to display both ordered and disordered
(spin-liquid) ground states in different regions of their phase
diagrams.\cite{RK:prl88,Moessner:prb01,Hermele:prb04,Moessner:prl01,Huse:prl03,Syljuasen:05}  
We derive the parameters of this QDM for general $s$, and
discuss several limits and the expectations for the plateau ground
state.  For simplicity of presentation we perform this calculation
here for the simplest XXZ spin model with no additional anisotropy or
other spin interaction terms.  However, the method is
straightforwardly generalized to include other on-site (e.g. uniaxial
anisotropy\cite{Damle:06}) or nearest-neighbor (e.g. biquadratic) interactions
without substantial increase in computational complexity.  
More generally, the flexibility to include such effects allows one to
consider the quantum effects upon the ground state selection within a
magnetization plateau even when the dominant mechanism of plateau {\sl
  stabilization} is a classical one.

A remarkable feature of the DPT is that all diagonal terms describing
splitting of the low energy manifold {\sl vanish below sixth order}!
For $s\geq 1$, off-diagonal tunneling terms also vanish up to this
order, so that the entire effective Hamiltonian is determined by terms
of sixth order and higher.  This behavior is similar to a result of
Henley\cite{Hizi:prb06} that in the large-$s$ limit, the effective
Hamiltonian is expressed entirely in terms of a ``spin flux'' variable
involving a product of $6$ spins around an elementary loop of the
lattice.  We show here that our result is rather general, and
originates from two basic features: the absence of non-trivial loops
of length less than $6$ links on the pyrochlore lattice, and the fact
that all low-energy spin states on a single tetrahedron are
permutations of one another.  From our proof of this result, it can
readily be seen that similar behavior holds for any lattice of corner
sharing simplexes with only on-site and nearest-neighbor interactions
and permutation-related ground states on a single simplex.  We will
apply the methods of this paper to other such problems of interest in
future work. 

For the pyrochlore magnetization plateau and QDM studied here, the
conclusions are as follows.  For $s>3/2$, we find that {\sl diagonal}
terms in the QDM are much larger than off-diagonal ones.  In this
case, the latter are negligible, and because the diagonal QDM is
effectively classical, it is soluble and the ground state is {\sl
  ordered}.  We discuss the preferred spin ordered states as a
function of $s$.  For $s\leq 3/2$, the off-diagonal terms are
non-negligible, and a simple solution is no longer availed.  For
$s=3/2$, various arguments lead us to still expect an ordered state.
After correction of an error in Ref.\onlinecite{Bergman:prl05}, we
find two candidate states for this case.  One of these is the {\bf R}
state discussed previously in Ref.\onlinecite{Bergman:prl05}, which is
the state containing the maximal number of hexagonal loops with
alternating spins (flippable plaquettes in the QDM language).  Another
candidate is a $\sqrt{3}\times\sqrt{3}$ state with a planar structure.
In fact, the diagonal terms in the effective Hamiltonian do not
entirely fix the relation between adjacent planes in the latter state,
so there is additional degeneracy whose breaking we cannot resolve at
the present time.  Because the off-diagonal and diagonal terms are
comparable in this case, however, some other states may also be
possible, and a definite conclusion must await more serious
computational (e.g. quantum Monte Carlo) analysis.  For $s\leq 1$, the
off-diagonal term in the QDM is dominant.  In this case, either the
${\bf R}$ state or $U(1)$ spin liquid is the most likely candidate
ground state.  Indeed, as argued in Ref.\onlinecite{Bergman:prb05}, it
is quite possible that the simplest QDM displays a direct quantum
phase transition between these two states.

The ground state of the QDM just discussed is determined only by the
{\sl dimensionless} ratios of coupling constants.  However, the DPT
calculation also gives the overall scale of the effective interaction
in terms of the microscopic exchange $J$.  For $s=3/2$, the largest
interaction energy (extrapolated from the easy axis perturbation
theory to the Heisenberg limit) generated by quantum fluctuations is
only $\approx 0.02J$.  Were this the true scale for ground state
selection in the degenerate 3:1 manifold in \hgaf, the magnetic
ordering would occur at a temperature of this order, i.e. $\approx
0.2K$.  Experimentally, however, magnetic ordering is observed at a
substantial fraction of the temperature of onset of the plateau
formation, which is around $6K$.  The closeness of the ordering and
plateau scales in experiment suggests that both are determined by
the same physical mechanism, and argues against the importance of
quantum fluctuations in the ground state selection in \hgaf.  Indeed,
we have recently shown\cite{Bergman:prb06} that the same spin-lattice
coupling which leads to plateau formation can also account for the
state selection.  Curiously, the {\bf R} state is also stabilized by
the lattice mechanism.  This is symptomatic of the very strong
constraints defining the 3:1 QDM states, which lead rather different
microscopic interactions to favor the same ground state.  For $s=1$
and $s=1/2$, the DPT gives much larger characteristic scales for the
QDM, the off-diagonal term being of order $0.16J$ and $1.5J$ in the
two cases.  Thus such $s\leq 1$ antiferromagnets, if realized
experimentally, would be promising systems to observe quantum
fluctuation effects.

This manuscript is organized as follows. In Section~\ref{easy_axis},
we describe our theoretical model, the nearest-neighbor quantum
Heisenberg antiferromagnet on a pyrochlore lattice in an external
field.  An easy-axis limit is taken under the assumption of the
suppression of transverse spin fluctuations in large magnetic fields.
After applying degenerate perturbation theory (DPT) in the transverse
spin fluctuations, an effective dimer model emerges in Section~\ref{sec:DPT_Leon} that can be used
to obtain an approximate ground state of the original model. In
Section~\ref{large_s} we carry out a large $s$ analysis of the XXZ
model, deriving a different effective Hamiltonian splitting the 3:1
manifold of degenerate states.  This new effective Hamiltonian turns out
to coincide with the $s \rightarrow \infty$ limit of the effective
Hamiltonian from the DPT analysis. In Section~\ref{diagonal_gs} we
explore the ground state of the diagonal part of the
effective Hamiltonian from DPT. In Section~\ref{Effective_QDM} we
explore in more generality the appropriate Quantum Dimer Model (QDM)
of which all our effective Hamiltonians are special cases. 
We conclude the main text of this paper with a discussion of our results in
Section~\ref{sec:discussion}.
In appendix~\ref{app:PlateauWidth} we analyze how the half polarization plateau 
is modified by quantum fluctuations. An alternative method of performing DPT
is presented in Appendix~\ref{app:Other_DPT}, and shows perfect agreement with 
the result of Section~\ref{sec:DPT_Leon}. Finally, in Appendix~\ref{app:root3_degeneracy}
we explore the states degenerate with the $\sqrt{3}\times\sqrt{3}$ states, found for $s = \frac{3}{2}$.

\section{Models}
\label{easy_axis}

\subsection{Hamiltonians and Limits}

We begin with the simple spin-$s$ Heisenberg antiferromagnet (AFM) residing 
on the sites of the pyrochlore lattice in the presence of a magnetic field ${\bf H}$,
\begin{equation}
{\mathcal H} = J \sum_{\langle i j \rangle} {\bf S}_i \cdot {\bf S}_j - {\bf H} \cdot \sum_j {\bf S}_j
\; .
\end{equation}
On the pyrochlore lattice one may recast the nearest-neighbor exchange
in terms of the total spin on tetrahedra using the identity
\begin{equation}
2 \sum_{\langle i j \rangle} {\bf S}_i \cdot {\bf S}_j = 
\sum_t ({\bf S}_t)^2 - \sum_t \sum_{j \in t} ({\bf S}_j)^2
\; ,
\end{equation}
where ${\bf S}_t = \sum_{j \in t} {\bf S}_j$ is the sum of spins on a
tetrahedron labeled by $t$, and $({\bf S}_j)^2|j\rangle =
S(S+1)|j\rangle$.  This gives the more convenient form
\begin{equation}\label{Heisenberg}
{\mathcal H} =  \frac{J}{2} \sum_t \left[ ({\bf S}_t - {\bf h})^2  - {\bf h}^2 \right]
\; ,
\end{equation}
where we have introduced the dimensionless magnetic field ${\bf h} =
{\bf H}/2J = h {\hat z}$, 
and ignored a trivial constant term in the Hamiltonian. 

\subsubsection{Classical limit}

The form in Eq.\eqref{Heisenberg} makes the behavior in the large $s$
limit apparent.  In this limit the spins behave classically, and one
may replace ${\bf S}_i\rightarrow s\hat{n}_i$, where $\hat{n}_i$ is a
unit vector. The ground states then consist simply of all states for
which ${\bf S}_t = s\sum_{i\in t} \hat{n}_i = {\bf h}$ on every
tetrahedron.  This set has a large continuous degeneracy.  
Furthermore, since the magnetization is simply half the sum of the ${\bf S}_t$
(because each spin is contained in two tetrahedra), this implies a
continuous linear behavior of the magnetization with field.  Thus, 
in this model magnetization plateaus can emerge only from quantum corrections to the
classical limit.  

\subsubsection{Easy axis limit}

An alternative approach exploits the fact that, with the application of
the magnetic field, the global $SU(2)$ symmetry of the bare Heisenberg
model is broken down to a $U(1)$ symmetry (rotations about the magnetic
field direction).   Moreover, when the magnetization per spin is
substantial, on average the transverse components $S_i^\pm$ are smaller
in magnitude than the longitudinal ones.  It is therefore natural to
treat transverse and longitudinal exchange couplings on a different
footing, with the latter taking the dominant role.  Formally, this is
accomplished by replacing the isotropic Heisenberg Hamiltonian by an XXZ
model:
\begin{equation}\label{XXZ}
{\mathcal H} = {\mathcal H}_0 + {\mathcal H}_1\;,
\end{equation}
where
\begin{equation}\label{H_0}
{\mathcal H}_0 = \frac{J_z}{2} \sum_t \left[ (S_t^z - h)^2  - h^2 \right] - J_z \sum_i \left( S_i^z \right)^2
\; ,
\end{equation}
and 
\begin{equation}\label{H_1}
{\mathcal H}_1 = \frac{\Jx}{2} \sum_{\langle i j \rangle} \left( S_i^+ S_j^- + h.c. \right)
\; .
\end{equation}
We use the notation
$
S_t^z = \sum_{i \in t} S_i^z
$ ,
and we have made use of the identity
\begin{equation}
\sum_{\langle i j \rangle} S_i^z S_j^z = 
\frac{1}{2} \sum_t (S_t^z)^2 - \sum_i \left( S_i^z \right)^2
\; .
\end{equation}
In the equations above, and elsewhere in this manuscript, ${\langle i
  j \rangle}$ denotes a sum over nearest neighbor sites on the
pyrochlore lattice, and $S_i^{\pm}$ are the spin ladder operators.
Note that in the Heisenberg model $J_\perp=J_z=J$, but the more
general XXZ model has all the same symmetries as the former even when
this condition is not obeyed.  From the above reasoning, we expect
that the transverse terms involving $J_\perp$ may be treated as
``small'' perturbations in the strong-field regime of interest.  We
note that this is expected to be a particularly good approximation
when the system exhibits a magnetization plateau.  This is because, as
described in the introduction, $\langle S_i^\pm\rangle$ {\sl must}
vanish in such a state.  Formally, this ``easy axis'' limit consists
of taking $\Jx \ll J_z$ and doing degenerate perturbation theory in
$\alpha = \frac{\Jx}{J_z}$.  We will assess the validity of this
approximation later by considering the {\sl magnitude} the
perturbative corrections extrapolated to $\alpha=1$.  Finally, we note
that several other effects can stabilize a collinear state.  One is the
addition of easy axis anisotropy,
\begin{equation}
  \label{eq:6}
  {\mathcal H}_0^\prime = - K \sum_i \left( S_i^z\right)^2,
\end{equation}
with $K\gg J_\perp>0$.  A second mechanism is biquadratic exchange,
\begin{equation}
  \label{eq:biquad}
  {\mathcal H}_0^{\prime\prime} = - b J \sum_{\langle ij\rangle}
  \left( {\bf S}_i\cdot  {\bf S}_j\right)^2,
\end{equation}
with $b>0$.  A term of this form can be generated dynamically from
spin-lattice interactions, known to be strong in \hgaf.  The DPT
treatment discussed below can readily be generalized to include either
or both of the terms in Eqs.(\ref{eq:6},\ref{eq:biquad}).  For
simplicity of presentation we do not do so here.  While easy-axis terms
similar to Eq.\eqref{eq:6} are allowed for $s>1/2$, this particular
simple form, with the same spatial direction for the local easy axis
of all spins, is not physically appropriate for the cubic pyrochlore
spinels, and the proper anisotropy terms allowed by symmetry in these
materials are likely to be very small in any case.

\subsection{Magnetization process in the Ising model} 
\label{plateau_structure}

The evolution of the ground state with field in the extreme easy-axis
limit $\alpha=0$ is less trivial than in the classical limit.  The
system is then described by the Ising Hamiltonian, Eq.\eqref{H_0}.  We
shall focus on the $h \geq 0$ case, as the case $h \leq 0$ is
equivalent. The expression Eq.\eqref{H_0} can be written as a sum over
tetrahedra ${\mathcal H}_0 = \sum_t {\mathcal H}_t$ with
\begin{equation}\label{H_t}
{\mathcal H}_t = \frac{J_z}{2} \left[ (S_t^z - h)^2  - h^2 - \sum_{j \in t} \left( S_j^z \right)^2 \right]
\; ,
\end{equation}
and therefore if one can minimize this energy on each single tetrahedron, one will have attained the minimum energy of the many-body system.

The magnetization $S_t^z$ of any individual tetrahedron is \emph{quantized} to the values 
$ S_t^z = 0, \pm 1, \pm 2, \ldots \pm 4 s$.
The 1st term in \eqref{H_t} favors the magnetization of the tetrahedron to take on a value close to the integer part of the dimensionless magnetic field. The 2nd term in \eqref{H_t} favors larger $z$-components of the spin values $S_j^z = \pm s$. 
This 2nd term is trivial only in the spin $\frac{1}{2}$ case where the spin-1/2 Pauli matrices square to the identity.

A state with the magnetization $ S_t^z = m = [h] $ (the integer part of $h$) clearly minimizes the energy of the first 
term in \eqref{H_t}.
However, given the magnetization, there is some freedom for the values of the spins on each tetrahedron. The 2nd term in \eqref{H_t} reduces this freedom by adding an energy cost for small $S^z$ components. 

Consider 4 spins on a tetrahedron with total magnetization $m$, and individual values of $S_1^z,S_2^z,S_3^z,S_4^z$.
Now compare the energy of this state with that of $S_1^z -1,S_2^z +1,S_3^z,S_4^z$. The only energy difference comes from the second term in \eqref{H_t}
\begin{equation}
\begin{split} 
\Delta E = & - \left( S_1^z + 1 \right)^2 - \left( S_2^z - 1 \right)^2 + \left( S_1^z \right)^2 + \left( S_2^z \right)^2
\\
= & -2 \left( 1 - S_1^z + S_2^z \right)\,.
\end{split}
\end{equation}
$ $From this expression one deduces that if one begins with $S_2^z > S_1^z$, then it is energetically favorable to increase $S_2^z$
even more at the expense of $S_1^z$. This increase in $S_2^z$ can only be halted by one of $S_{1,2}^z$ reaching an \emph{extreme}
spin value of $\pm s$. From this reasoning we conclude that the lowest energy state on a single tetrahedron with 
a fixed total magnetization $m$ has a spin configuration with the largest possible number of extreme valued spins.
This also makes intuitive sense, since ideally \emph{all} the spins should take on extreme values if possible.
In three particular choices of $m$, all the spins take on extreme values.
Zero magnetization $m=0$ with $s,s,-s,-s$, half polarization of $m=2s$ with $s,-s,-s,-s$ and full polarization
$m= 4s$ with $s,s,s,s$. For $m < 2s$ we find the lowest energy configuration for the spins $s,s,-s,m - s$,
and for $m > 2s$, we find $s,s,s, m - 3s$.
Finally, we can find the minimal energy for given magnetization $m$ by using the spin configurations described above for every value of $m$, and plugging them into \eqref{H_t}. For $m < 2s$ we find
\begin{equation}
E_m = \frac{J_z}{2} \left[ 2 m (s - h) - 4 s^2 \right],
\end{equation}
and for $m > 2s$ we find
\begin{equation}
E_m = \frac{J_z}{2} \left[ 2 (m - 2s) (3s - h) - 4 h s \right].
\end{equation}
$ $From the above expressions it is easy to see that at $h=s$ all $m<2s$ yield the \emph{same} energy - all these states are \emph{degenerate} at this field value. Similarly, for $h = 3s$ all $m>2s$ yield the \emph{same} energy. For other 
values of the magnetic field, we find for $h<s$ the lowest energy state is the $m=0$ zero magnetization state,
for $h < s < 3s$ the lowest energy state is the $m=2s$ half polarized state, and for $h>3s$ the lowest energy state 
is the $m=4s$ fully polarized state.

The three lowest energy states in the various magnetic field ranges have all spins at extreme values $\pm s$,
and can be realized on every tetrahedron in the pyrochlore lattice. The $m=0$ state induces a degenerate
manifold of states with every tetrahedron having $s,s,-s,-s$ on it. This 2:2 proportionality is well known 
as the ``ice rules'' encountered in a particular phase of water ice,\cite{Pauling:35} 
as well as spin ice compounds.\cite{Bramwell:01} The half 
polarization states are also massively degenerate, with every tetrahedron in 
a 3:1 proportionality of $S_j^z = +s$ to $S_j^z = -s$ spins (or 3 up one down). This particular degenerate manifold will be the focus of the remainder of our discussion. To summarize, the magnetization curve for \eqref{H_0} exhibits 3 plateaus at zero, half and full polarization, for all values of $s$.

\section{Easy axis degenerate perturbation theory}
\label{sec:DPT_Leon}

\subsection{Structure of perturbation theory}

\subsubsection{Basic formulation}
\label{sec:dpt_formulation}

In the previous section, we observed that the extreme easy axis limit of
the Heisenberg model exhibits a broad magnetization plateau at half
polarization.  However, the ground states on this plateau are
macroscopically degenerate, consisting of all states with a 3:1 ratio of
majority and minority spins on each tetrahedron.  In this section, we
study the splitting of this degeneracy by perturbation theory in
$J_\perp$.  We employ the following formulation of Degenerate
Perturbation Theory (DPT).  Define the projection operator, ${\mathcal
  P}$, onto any degenerate manifold of states $M$.  Consider any exact
eigenstate $|\Psi\rangle$.  Its projection $|\Psi_0\rangle={\mathcal
  P}|\Psi\rangle$ satisfies the ``effective Schr\"odinger equation''
\begin{equation}
  \label{eq:effshrod}
   \left[ E_0 + 
{\mathcal P} {\mathcal H}_1 \sum_{n=0}^{\infty} {\mathcal G}^n 
{\mathcal P} \right] |\Psi_0\rangle = E |\Psi_0\rangle = {\mathcal H}_{\textrm{eff}} |\Psi_0\rangle,
\end{equation}
where the operator ${\mathcal G} = \frac{1}{E - {\mathcal H}_0} \left( 1 -
  {\mathcal P} \right) {\mathcal H}_1 $.  Because the
resolvent contains the exact energy $E$, Eq.~(\ref{eq:effshrod}) is
actually a non-linear eigenvalue problem.  However, to any given order
of DPT, $E$ may be expanded in a series in $\Jx$ to obtain an equation
with a true Hamiltonian form within the degenerate manifold.  Each
factor of ${\mathcal G}$ is at least of $O(J_\perp)$ due to the
explicit factor in ${\mathcal H}_1$, with higher order corrections
coming from the expansion of $E$.  Once
$|\Psi_0\rangle$ and $E$ are known, the full wavefunction can be reconstructed
as $\ket{\Psi} = (1-{\mathcal G})^{-1} \ket{\Psi_0} =
\sum_{n=0}^{\infty} {\mathcal G}^n \ket{\Psi_0}$. 

Considering the lowest order term in DPT that breaks the degeneracy, the
precise energy $E = E_0 + O(\alpha)$ in the resolvent can be replaced by
$E_0$, where $O(\alpha)$ represents possible energy shifts from lower
order terms that do \emph{not} break the degeneracy, and $E_0$ is the
0-th order energy of the degenerate manifold of states.



\subsubsection{Order of off-diagonal terms}
\label{sec:order-diagonal-terms}

Every order in DPT can in principle have {\sl diagonal} (in the $S_i^z$
basis) as well as {\sl off-diagonal} terms in which the degeneracy is
removed.  Any {\sl off-diagonal} term in the effective Hamiltonian must
flip spins in such a way as to preserve the 3:1 constraint on each
tetrahedron.  This can only be accomplished by flipping spins around a
non-trivial closed loop on the pyrochlore lattice (see, e.g.
Ref.\onlinecite{Hermele:prb04}).  The smallest such loop involves
flipping spins on 3 different bonds, and flipping a spin from $S^z = \pm
s$ to $S^z = \mp s$ requires ${\mathcal H}_1$ to act $2s$ times, so
off-diagonal processes occur first at order $O(\Jx^{6s})$.  Therefore,
below this order of DPT, one need consider only diagonal terms.  In 
subsection~\ref{symbolic-red}
we will demonstrate that the lowest order diagonal energy
splitting term {\sl for any $s$} can occur only at 6th order.  For spin
$s=\frac{1}{2}$ an off-diagonal term appears at 3rd order in DPT, and no
diagonal energy splitting occurs at this order, resulting in a purely
off-diagonal effective Hamiltonian. For spin $s=1$ the lowest order
diagonal and off-diagonal terms simultaneously appear at 6th order. For
any higher value of $s$, the diagonal energy splitting appears at a
lower order than any off-diagonal term can occur, and therefore the
leading order effective Hamiltonian is purely diagonal in the 3:1
states.  We will nevertheless compute the first non-vanishing
off-diagonal term for various values of $s$ in Section~\ref{Effective_QDM}, to use its magnitude for an assessment of
the validity of the truncation of the DPT expansion.

\subsubsection{Unitarily transformed formalism for diagonal terms}

We next develop a scheme to compute the diagonal terms, by unitarily
transforming the expression in Eq.\eqref{eq:effshrod} to obtain a
formula for the diagonal effective Hamiltonian with all dependence upon
the spin state explicit.  The 3:1 manifold can be described using Ising
variables to indicate which spins are minority sites.  That is, in the
3:1 states, we denote $S_j^z = \sigma_j s$ with $\sigma_j = \pm 1$ the
Ising variable.  At $n^{\rm th}$ order, presuming that all lower order
terms are constants, the diagonal terms in the effective Hamiltonian
constitute the function of the set $\{\sigma_i\}$ given by
\begin{equation}
  \label{eq:7}
 {\mathcal H}_n[\{\sigma_i\}] = (-1)^{n+1} \langle
 \psi[\{\sigma_i\}]|\left({\mathcal H}_1 {\mathcal R}{\mathcal Q}\right)^{n-1} {\mathcal
   H}_1  |\psi[\{\sigma_i\}]\rangle, 
\end{equation} 
where the resolvent ${\mathcal R} =({\mathcal H}_0 - E_0)^{-1}$,
${\mathcal Q}= 1 - {\mathcal P}$, and
\begin{equation}
  \label{eq:8}
    |\psi[\{\sigma_i\}]\rangle=\otimes_i | S_i^z=\sigma_i
      s\rangle.
\end{equation}
The assumption that all lower order terms are
constant allows us to replace $E$ by $E_0$ in the denominators in
Eq.\eqref{eq:effshrod}, since the constant corrections to $E$ lead to
higher order terms in the effective Hamiltonian.  

The dependence upon the $\sigma_i$ in Eq.\eqref{eq:7}, is not explicit,
but, following Hizi {\rm et al.}~\cite{Hizi:prb06}, it can be made so by a
unitary transformation.  The operator
\begin{equation}\label{unitary}
{\hat U} = e^{+ i \pi \sum_j \frac{\left( 1-\sigma_j \right)}{2} {\hat S}_j^x}
\; 
\end{equation}
effects a rotation about the $x$-axis in spin space only for the minority
spins.  This interchanges raising and lowering operators, and reverses
the orientation of $S_i^z$ for these sites.  We may therefore write
\begin{equation}
  \label{eq:9}
  |\psi[\{\sigma_i\}]\rangle= U |\psi_0\rangle,
\end{equation}
where
\begin{equation}
  \label{eq:12}
  |\psi_0\rangle = \otimes_i |S_i^z=s\rangle.
\end{equation}
is the fully polarized state, which is now independent of $\sigma_i$.
Then we have
\begin{equation}
  \label{eq:7a}
 {\mathcal H}_n[\{\sigma_i\}] = (-1)^{n+1} \langle
 \psi_0|\left(\tilde{\mathcal H}_1 \tilde{\mathcal R}\tilde{\mathcal
     Q}\right)^{n-1} \tilde{\mathcal 
   H}_1  |\psi_0\rangle,
\end{equation} 
where 
\begin{equation}
  \label{eq:13}
  \tilde{\mathcal O} = U^\dagger {\mathcal O} U,
\end{equation}
for any operator ${\mathcal O}$.  In what follows, all the operators
appearing in Eq.\eqref{eq:7a} above, will be simplified so that their dependence upon
$\sigma_i$ becomes explicit.  

First consider $\tilde{\mathcal H}_1$.  It consists, from
Eq.\eqref{H_1}, of a sum of operators
transferring spin 1 between two nearest neighbor sites, i.e. a bond of the
pyrochlore lattice.  We define the nearest-neighbor connectivity matrix
of the lattice $\Gamma_{i j} = \Gamma_{j i}=1$ when $i$ and $j$ are
nearest neighbors, and $\Gamma_{i j}=0$ otherwise.  With this terminology 
we write \eqref{H_1} as
\begin{equation}
{\mathcal H}_1 = J_z \frac{\alpha}{4} \sum_{i j} \Gamma_{i j} \left(
  S_i^{+} S_j^{-} +h.c. \right)  
\; . 
\end{equation}
After the unitary transformation, one obtains
\begin{equation}
\label{rotated_H_1}
\begin{split}
{\tilde {\mathcal H}}_1 = U^{\dagger} {\mathcal H}_1  U & = J_z \frac{\alpha}{4} \sum_{i j} \Gamma_{i j} \left( S_i^{+\sigma_i} S_j^{-\sigma_j} +h.c. \right) 
\\
& =
J_z \frac{\alpha}{4} \sum_{i j} \Gamma_{i j} 
\Bigg[
\frac{\left( 1 + \sigma_i \sigma_j \right)}{2}
\left( S_i^+ S_j^- + h.c. \right)
\\ &
+
\frac{\left( 1 - \sigma_i \sigma_j \right)}{2}
\left( S_i^+ S_j^+ + h.c. \right)
\Bigg].
\end{split}
\end{equation}
Here the expressions $\frac{\left( 1 \pm \sigma_i \sigma_j \right)}{2}$
denote ``Ising delta functions'' that select the cases in which the two
$\sigma_{i,j}$ have the same or opposite sign.

Assuming the lowest order term in DPT that splits the 3:1 configurations
is a diagonal term of order $n_0$, the only 3:1 configuration which can 
be reached as an intermediate state in
Eq.\eqref{eq:7} for any $n \leq n_0$ is the 
starting state $|\psi[\{\sigma_i\}]\rangle$.
Under the unitary transformation, this state maps to $|\psi_0\rangle$,
and therefore the projection operator $\tilde{\mathcal Q}$
may be replaced by
\begin{equation}
  \label{eq:16}
  \tilde{\mathcal Q} \rightarrow 1- |\psi_0\rangle \langle \psi_0|
\end{equation}
in Eq.\eqref{eq:7a}.  

Finally, we consider the resolvent.  Using $U^\dagger S_i^z
U = \sigma_i S_i^z$, one finds
\begin{equation}
  \label{eq:19}
  \tilde{\mathcal R}^{-1} = \frac{J_z}{2}\sum_{ij} \Gamma_{ij} \sigma_i
  \sigma_j S_i^z S_j^z - 2 J_z h \sum_j \sigma_j S_j^z - E_0.
\end{equation}
First we note that because both ${\mathcal H}_0$ and ${\mathcal H}_1$
conserve the total magnetization of the lattice (this is just the conserved quantity
arising from the global $U(1)$ symmetry), the term $\sum_j \sigma_j S_j^z$
remains unchanged at every stage in a DPT process, and 
we can therefore absorb this term into the constant energy $E_0$.
Clearly, the inverse resolvent should vanish when acting upon the fully
polarized state $|\psi_0\rangle$.  Hence we may absorb the constant
energy $E_0$ into the sum as
\begin{equation}
  \label{eq:20}
   \tilde{\mathcal R}^{-1} = \frac{J_z}{2}\sum_{ij} \Gamma_{ij} \sigma_i
  \sigma_j \left(S_i^z S_j^z - s^2\right).
\end{equation}
We can simplify the resolvent in the
restricted space of virtual states which will be accessed in evaluating
Eq.\eqref{eq:7a}.  In particular, the $\sigma_i$ configurations are 
restricted to the 3:1 manifold.
Furthermore we note that all intermediate states
will have only some small finite set of spins whose $S_i^z$ quantum
numbers are different from $s$, due to the action of $\tilde{\mathcal
  H}_1$.  Let us consider then the action of the resolvent on a state
for which this set of sites is denoted by ${\sf F}$.  In this case, only
terms in Eq.\eqref{eq:20} for which at least one of $i$ or $j$ is in
${\sf F}$ can contribute.  Thus
\begin{eqnarray}
  \label{eq:22}
   \tilde{\mathcal R}^{-1} & = & \frac{J_z}{2}\sum_{ij\in {\sf F}}
   \Gamma_{ij} \sigma_i \sigma_j \left(S_i^z S_j^z - s^2\right)
   \nonumber \\
   && + J_z s \sum_{i\in {\sf F}}\sum_{j \not\in {\sf F}} \Gamma_{ij}
   \sigma_i \sigma_j \left(S_i^z - s\right). 
\end{eqnarray}
One may replace the sum over $j$ by $\sum_{j \not\in {\sf F}} =\sum_j -
\sum_{j \in {\sf F}}$ to obtain
\begin{eqnarray}
  \label{eq:23}
     \tilde{\mathcal R}^{-1} & = & \frac{J_z}{2}\sum_{ij\in {\sf F}}
   \Gamma_{ij} \sigma_i \sigma_j \left(S_i^z -s\right)\left(S_j^z - s\right)
   \nonumber \\
   && + J_z s \sum_{i\in {\sf F}}\sigma_i \left(\sum_{j}
     \Gamma_{ij}\sigma_j \right) \left(S_i^z - s\right). 
\end{eqnarray}
The crucial observation is that the 3:1 constraint implies
\begin{equation}
  \label{eq:24}
  \sum_j \Gamma_{ij}\sigma_j = 4-2\sigma_i.
\end{equation}
This is because once $\sigma_i$ is specified, the {\sl set} of its
neighbors is also specified (see also Fig.~\ref{fig:configs}).  Eq.\eqref{eq:24} allows one to eliminate
the latter sum and obtain
\begin{eqnarray}
       \tilde{\mathcal R}^{-1} & = & \frac{J_z}{2}\sum_{ij\in {\sf F}}
   \Gamma_{ij} \sigma_i \sigma_j \left(S_i^z -s\right)\left(S_j^z - s\right)
   \nonumber \\
   && +2 J_z s \sum_{i\in {\sf F}}(2 \sigma_i - 1)\left(S_i^z - s\right). 
\end{eqnarray}

Using again the observation that $\sum_j \sigma_j S_j^z$ remains unchanged throughout the stages of any
DPT process, it is equal to the constant $\sum_j \sigma_j S_j^z = \sum_j \sigma_j s$. Using this fact,
we finally obtain
\begin{eqnarray}
  \label{eq:25}
       \tilde{\mathcal R}^{-1} & = & \frac{J_z}{2}\sum_{ij\in {\sf F}}
   \Gamma_{ij} \sigma_i \sigma_j \left(S_i^z -s\right)\left(S_j^z - s\right)
   \nonumber \\
   && - 2 J_z s \sum_{i\in {\sf F}} \left(S_i^z - s\right). 
\end{eqnarray}

By successive action of $\tilde{\mathcal H}_1$, $\tilde{\mathcal Q}$,
and $\tilde{\mathcal R}$ using
Eqs.~(\ref{rotated_H_1},\ref{eq:16},\ref{eq:25}), one can obtain
explicit expressions for any intermediate state in the DPT expression of
Eq.\eqref{eq:7a}, with $n \leq n_0$.  For example, one action of each of these operators gives
\begin{eqnarray}
  \label{eq:oneact}
  &&  \tilde{\mathcal R}\tilde{\mathcal Q}\tilde{\mathcal H}_1
  |\psi_0\rangle =  \\
  && \frac{\alpha s}{4(4s-1)}\sum_{a_1 a_2} \Gamma_{a_1 a_2}
  (1-\sigma_{a_1}\sigma_{a_2}) \left| 1_{a_1}
    1_{a_2}\right\rangle, \nonumber  
\end{eqnarray}
where we have introduced the compact notation
\begin{eqnarray}
  \label{eq:notation}
  \!\!\!\left| (m_1)_{a_1}\cdots (m_n)_{a_n}\right\rangle & = &
  |S_{a_1}^z=s-m_1\rangle \cdots |S_{a_n}^z=s-m_n\rangle\nonumber \\
&&  \otimes_{i\neq
    a_1\cdots a_n}|S_i^z=s\rangle. 
\end{eqnarray}
Acting twice with the same sequence of operators gives
\begin{widetext}
\begin{eqnarray}
  \label{eq:twoacts}
    \left(\tilde{\mathcal R}\tilde{\mathcal Q}\tilde{\mathcal
        H}_1\right)^2 &&
  |\psi_0\rangle  =  \frac{\alpha^2 s}{16(4s-1)} \sum_{a_1 a_2} \Gamma_{a_1 a_2}
  (1-\sigma_{a_1}\sigma_{a_2}) \left| 2_{a_1}
    2_{a_2}\right\rangle \\
&& + \frac{\alpha^2 s^2}{4(4s-1)} \sum_{a_1 a_2 a_3}\frac{\Gamma_{a_1
  a_3}\Gamma_{a_2 a_3} \eta_{a_1 a_2}}{4s-\Gamma_{a_1 a_2}}[(\sigma_{a_1}+\sigma_{a_3})\sigma_{a_2}
-\sigma_{a_1}\sigma_{a_3}-1]\left| 1_{a_1} 1_{a_2}\right\rangle
\nonumber \\
&& + \frac{\alpha^2 s^{3/2} \sqrt{2s-1}}{4(4s-1)} \sum_{a_1 a_2 a_3}
\frac{\Gamma_{a_1 a_2}\Gamma_{a_1 a_3} \eta_{a_2
    a_3}}{8s-4+\Gamma_{a_2 a_3}}
\left[1+\sigma_{a_2}\sigma_{a_3}-\sigma_{a_1}(\sigma_{a_2}+\sigma_{a_3})\right]
\left|2_{a_1} 1_{a_2} 1_{a_3}\right\rangle \nonumber \\
&& + \frac{\alpha^2 s^2}{16(4s-1)} \sum_{a_1\cdots a_4}
\frac{\Gamma_{a_1 a_2} \Gamma_{a_3 a_4} \eta_{a_1 a_4} \eta_{a_2 a_3}
  \eta_{a_2 a_4}}{8s-2+\sigma_{a_1}\sigma_{a_3}\left(\Gamma_{a_1
      a_3}-\Gamma_{a_1 a_4}-\Gamma_{a_2 a_3}+\Gamma_{a_2 a_4}\right)}
(1-\sigma_{a_1} \sigma_{a_2})(1-\sigma_{a_3}\sigma_{a_4})
\left|1_{a_1}1_{a_2}1_{a_3}1_{a_4}\right\rangle, \nonumber
\end{eqnarray}
\end{widetext}
where we have introduced the ``non-coincident'' symbol
\begin{equation}
  \label{eq:nco}
  \eta_{ab} = 1-\delta_{ab}.
\end{equation}
The corresponding expressions for more successive actions of these
operators upon $|\psi_0\rangle$ can also be obtained, but are too
unwieldy to present here.  

Using such expressions, one may readily evaluate the terms, ${\mathcal
  H}_n[\{\sigma_i\}] $ in the diagonal effective Hamiltonian,
Eq.\eqref{eq:7a}. For $n_0$ an even number, a convenient way to calculate 
the $n_0$-th order term is to consider the state
\begin{equation}
\ket{\Psi} = 
{\tilde {\mathcal R}}^{1/2} \tilde{\mathcal Q} {\tilde {\mathcal H}}_1 
\left({\tilde {\mathcal R}} \tilde{\mathcal Q}{\tilde {\mathcal
      H}}_1\right)^{\frac{n_0}{2}-1}
\ket{\psi_0}\label{eq:26}
\end{equation}
and then find the magnitude of this wavefunction:
\begin{equation}
  \label{eq:mag6}
  {\mathcal
  H}_{n_0}[\{\sigma_i\}] =  -{\dirac {\Psi}{\Psi}}.
\end{equation}
Note that the square-root of $\tilde{\mathcal R}$ in Eq.\eqref{eq:26}
is easily evaluated by just taking the square-root of
Eq.\eqref{eq:25}, since it is diagonal in the basis of 3:1 configurations.  Other terms can be obtained
similarly.  

\subsection{Restricting the Hilbert space to the 3:1 manifold}
\label{symbolic-red} 

Calculating each such magnitude as defined in the previous subsection
leads to an explicit expression for the corresponding term in DPT.
These expressions appear to be extremely complex and formidable
functions of the Ising spin variables $\{ \sigma_i \}$. In this
subsection, we show that the projection of these functions to the 3:1
manifold of allowed $\{ \sigma_i\}$ configurations affords a
tremendous simplification.  In fact, we will demonstrate that all
terms in DPT below 6th order can give only constant functions --
i.e. no splitting -- within the 3:1 states.  At 6th order, the full
functional dependence can be characterized by only 3 independent
numbers which may be defined on plaquettes of the pyrochlore lattice.
We show how these numbers can be extracted from the expressions
obtained by the analysis of the previous subsection.

\subsubsection{Functional form of diagonal DPT terms}

$ $From the analysis of the previous subsection, the $n^{\rm th}$ order
effective diagonal Hamiltonian in DPT must take the form of a multiple 
sum of $n$ site indices $a_1\cdots a_n$, where each site index 
is summed over all lattice sites.
The summand is a function only of
$s$ and of the set of $\sigma_i, \Gamma_{ij}, \eta_{ij},\delta_{ij}$
where $i,j$ {\sl must belong to the set of the site indices}.  
The general form can be somewhat simplified by noting first that the
dependence upon the $\eta_{ij}$ can be eliminating by rewriting them
in terms of $\delta_{ij}$ using Eq.\eqref{eq:nco}, and then
eliminating all $\delta_{ij}$ by collapsing the sums
containing these factors.  Finally, we note that any factors of
$\Gamma_{ij}$ in the denominators in these expressions can be moved to
the numerator using the identity
\begin{equation}
g(\Gamma_{i j}) = g(0) + \Gamma_{i j} \left[ g(1) - g(0) \right]
\; ,
\end{equation}
for any function $g$ (which may also depend upon any other set of
variables), since $\Gamma_{i j} = 0,1$.

\begin{figure}
	\centering
	\subfigure{
	\label{fig:disc1}
\includegraphics[width=1.0in]{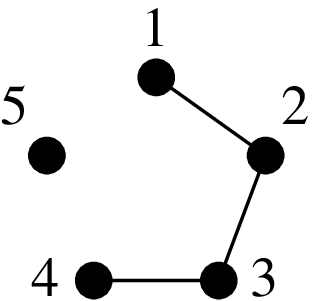}}
	\subfigure{
	\label{fig:disc2}		
		\includegraphics[width=1.0in]{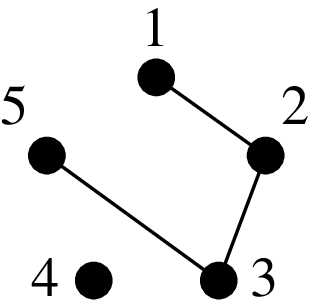}}
  	\caption{Examples of contractible diagrams.}
	\label{fig:disc_g}
\end{figure}

By these manipulations, one may write the effective Hamiltonian
${\mathcal H}_{\rm eff}[\{\sigma_i\}] = \sum_n  {\mathcal H}_n[\{\sigma_i\}]$ as
\begin{eqnarray}
  \label{eq:1}
&& {\mathcal H}_{\rm eff}[\{\sigma_i\}]  =  \\
&&  \sum_n  \sum_{G_n} \sum_{a_1 \ldots
  a_n}\!\!\! \left(\prod_{(ij)\in G_n} \Gamma_{a_i a_j}\right)
  f_{G_n}(\sigma_{a_1},\ldots,\sigma_{a_n}) .\nonumber
\end{eqnarray}
Here we have divided the effective Hamiltonian into terms involving
$n$ independent sites variables, $a_1\ldots a_n$ that are summed over
the lattice sites.
A given order $N$ in DPT contributes terms with $n\leq N$.  For a
given $n$, all possible products of $\Gamma_{a_i a_j}$ can appear.
The different such products are specified by $G_n$, which may be
considered as a ``diagram'' in the following fashion.  Each $G_n$ can be
represented by drawing $n$ points, corresponding to $i=1\ldots n$, and
connecting some arbitrary set of pairs of these points by lines.  For
each (unordered) pair of points $(ij)$ which is connected in $G_n$, we
include one factor of $\Gamma_{a_i a_j}$.  Since there are $n(n-1)/2$
pairs of points, and each pair may or may not be connected, there are
$2^{n(n-1)/2}$ distinct diagrams $G_n$.  
For example,  
in our conventions, $\Gamma_{a_1a_2}\Gamma_{a_2a_3}\Gamma_{a_3a_4}$ and 
$\Gamma_{a_1a_2}\Gamma_{a_2a_3}\Gamma_{a_3a_5}$ are represented
by different diagrams (see Fig.~\ref{fig:disc_g}), which means that $f_{G_n}(\sigma_{a_1},\ldots,\sigma_{a_6})$ is 
not necessarily symmetric with respect to swapping   
$\sigma_{a_4}$ and $\sigma_{a_5}$.
We will refer to the number $n$ as the
{\sl order} of the given term, even though it can come from a term of
that order or higher in DPT.

\subsubsection{Contractible diagrams}
\label{sec:contraction-rules}

\begin{figure}
	\centering
		\includegraphics[width=3.0in]{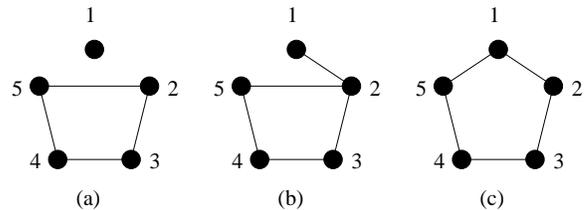}
	\caption{Examples of contractible ((a) and (b)) and non-contractible (c) diagrams.}
	\label{fig:contractible_examples}
\end{figure}

First we would like to show that any such term represented by a diagram
containing a point $i$ with less than two connections to other points
can be reduced to a term of one lower order. These diagrams are
``contractible'' (see Fig.~\ref{fig:contractible_examples} for
examples).  We prove this by showing that the sum over $a_i$ can be
carried out explicitly to obtain an expression of the same form of
Eq.\eqref{eq:1} in terms of the $n-1$ remaining sum variables.  There
are two cases.  Suppose in $G_n$ the point $i$ in question has no lines
connected to it.  Taking $i=n$ without loss of generality, we note that
the sum on $a_n$ is unconstrained, i.e. it runs over all lattice sites.
Thus we may write
\begin{eqnarray}
  \label{eq:3}
& &  \hspace{-0.1in}2 \sum_{a_n}
  f_{G_n}(\sigma_{a_1},\ldots,\sigma_{a_n}) =  \sum_t \sum_{a\in t}
  f_{G_n}(\sigma_{a_1},\ldots,\sigma_{a})   \\
  & = & N_t \left[3 f_{G_n}(\sigma_{a_1},\ldots,\sigma_{a_{n-1}},+)+
    f_{G_n}(\sigma_{a_1},\ldots,\sigma_{a_{n-1}},-) \right].\nonumber
\end{eqnarray}
The second line applies because on every tetrahedron there is the same
set of four single-spin states.  By inserting Eq.\eqref{eq:3} into
Eq.\eqref{eq:1}, one reduces the order of this term, as asserted above.

\begin{figure}[hbt]
	\centering
	\subfigure[Site adjoining the tetrahedra is the only minority site.]{
	\label{fig:conf1}
		\includegraphics[width=1.5in]{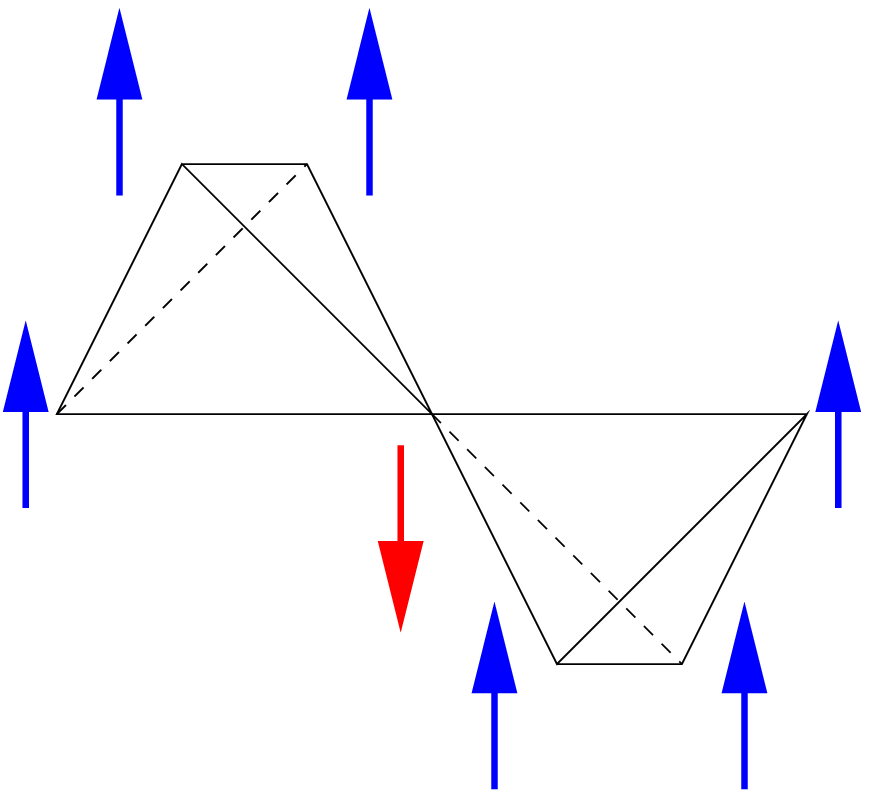}}
	\subfigure[Two minority sites connected by two parallel links.]{
	\label{fig:conf2}		
		\includegraphics[width=1.5in]{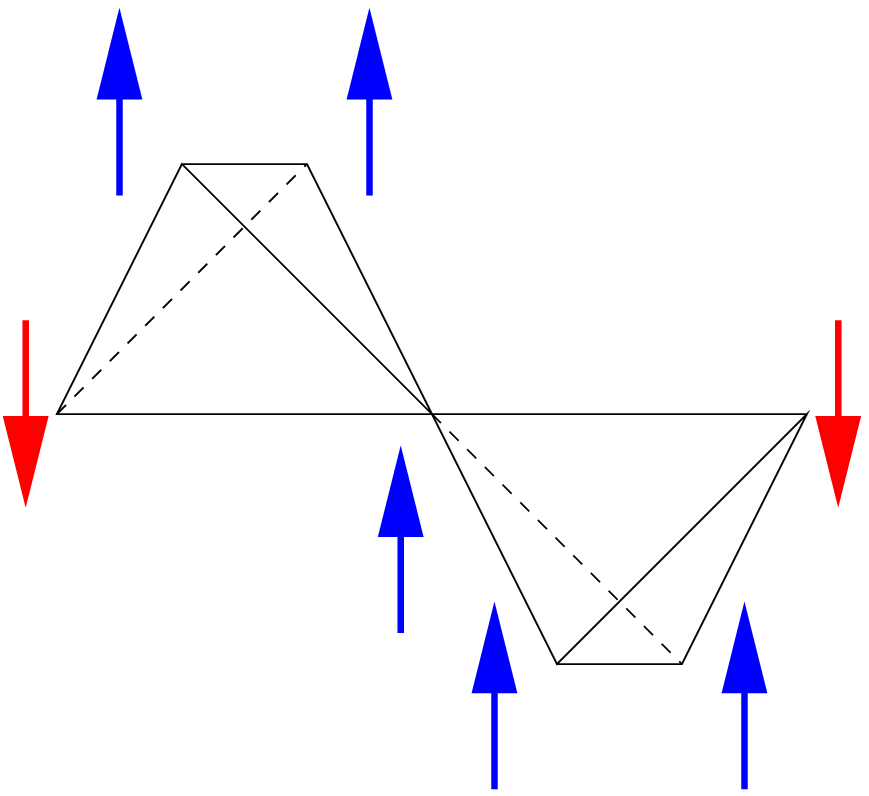}}
  \hspace{2.0in}	
  \subfigure[Two minority sites connected by two links bending.]{
	\label{fig:conf3}
		\includegraphics[width=1.5in]{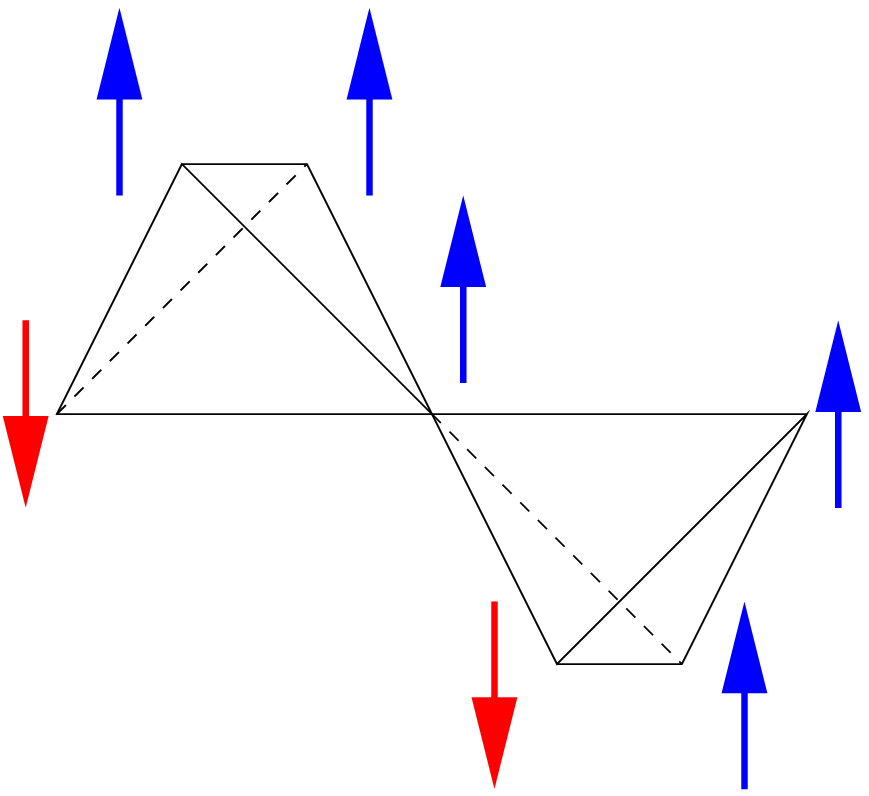}}
              \caption{(Color online) The three possible configurations
                of minority sites (red arrow) on two adjacent
                tetrahedra, in the 3:1 manifold of states.}
	\label{fig:configs}
\end{figure}

Consider the second case, in which there is one connection to the
point $i=n$.  We may suppose this connection is to the point $j<n$.
The sum over $a_n$ is then constrained {\sl only} by the requirement that
$a_n$ be a nearest-neighbor of $a_j$.  For fixed $a_j$, this includes
just 6 sites on the pyrochlore lattice.  Moreover, the {\sl set} of
spins on these six sites is entirely determined by the spin at site
$a_j$.  In particular, if $\sigma_{a_j}=+1$, the sum contains $4$
terms with $\sigma_{a_n}=+1$ and $2$ terms with $\sigma_{a_n}=-1$; if
$\sigma_{a_j}=-1$, the sum contains $6$ spins with $\sigma_{a_n}=+1$.
This can easily be understood from Fig.~\ref{fig:configs}.
Therefore the sum can again be carried out explicitly:
\begin{eqnarray}
  \label{eq:5}
&&    \sum_{a_n}
  \Gamma_{a_n a_j} f_{G_n}(\sigma_{a_1},\ldots,\sigma_{a_n}) = \nonumber \\
&& 
  \frac{1+\sigma_{a_j}}{2}
  \left[4f_{G_n}(\sigma_{a_1},\ldots,+)+2f_{G_n}(\sigma_{a_1},\ldots,-)\right]
  \nonumber \\ 
  && +  \frac{1-\sigma_{a_j}}{2} 6 f_{G_n}(\sigma_{a_1},\ldots,\sigma_{a_{n-1}},+) .
\end{eqnarray}
Once again, Eq.\eqref{eq:5} can be inserted into Eq.\eqref{eq:1} 
to reduce the order by one.

\subsubsection{Non-contractible diagrams}

\begin{figure}
	\centering
		\includegraphics[width=2.5in]{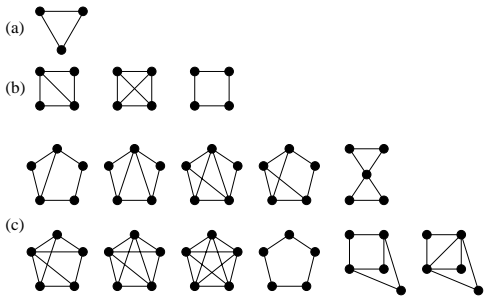}
                \caption{All $n\leq 5$ non-contractible diagrams.
                  (a)The triangle diagram is the only possible such
                  diagram for $n=3$. (b) The square framed diagrams are
                  all the possibilities for $n=4$. (c) The
                  pentagon-framed diagrams together with the three
                  right-most diagrams comprise all the possibilities for
                  $n=5$.}
	\label{fig:diagrams}
\end{figure}

Since all contractible diagrams can be reduced using the above rules
until they become either non-contractible or constant, we therefore
need to consider only non-contractible diagrams.  In these diagrams, 
each point in $G_n$ is connected to at least two other points.  Let us
first make a few general observations about these diagrams.
One can readily see that for these diagrams at order $n\leq 5$, all points
must be connected, i.e. it is possible to pass from one point to any
other by a sequence of links.  It is useful to consider the notion of
a {\sl loop}, or sequence of points, each connected to the next by a
link, which visits no point twice and returns to the first point of
the sequence.  For $n\leq 4$, there is always at least one loop which
includes all $n$ points.  For $n=5$, all but three non-contractible
diagrams contain a loop of length $5$.  The three remaining diagrams
at $n=5$ contain smaller loops
(see part (c) of Fig.~\ref{fig:diagrams}).
All the non-contractible single loop diagrams for $n \leq 5$ are shown
in Fig.~\ref{fig:diagrams}.
For $n=6$, there is one possible
\emph{disconnected} diagram, which
contains two disjoint loops of length $3$.  Apart from this last diagram,
all others are fully connected.

\begin{figure}
	\centering
		\includegraphics[width=2.5in]{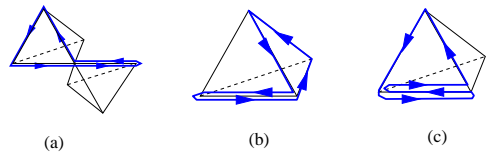}
	\caption{(Color online) 3 example topologies of closed paths of
          five steps on the pyrochlore lattice.  These are all the
          possible topologies of clusters corresponding to diagrams at
          order $n=5$ containing a loop of length 5.}
	\label{fig:path5}
\end{figure}

Let us consider the physical pyrochlore sites which are summed over in
a given term.  They comprise a set ${\sf S}(G_n) = \{
(a_1^{(1)},\ldots,a_n^{(1)}),(a_1^{(2)},\ldots,a_n^{(2)}),\ldots\}$ of
solutions, $(a_1^{(i)},\ldots,a_n^{(i)})$, to the conditions 
\begin{eqnarray}
  \label{eq:conds}
  \Gamma_{a_i a_j}=1 \qquad {\rm for}\,\, (ij)\in G_n .
\end{eqnarray}
We will call these solutions ``clusters''.  In an infinite system,
${\sf S}$ is of course infinite because of translational symmetry, but
this is immaterial.  A given term may then be written simply as
\begin{eqnarray}
  \label{eq:termset}
  \sum_{(a_1,\ldots,a_n)\in {\sf S}(G_n)} f(\sigma_{a_1},\ldots,\sigma_{a_n}).
\end{eqnarray}
We note that all the clusters for $n\leq 5$ are confined to one or two
adjacent tetrahedra.  This can be seen by considering the constraints
imposed on clusters by the non-contractibility of the diagram.  For
instance, all but three diagrams at order $n=5$ contain a loop of length
5, and this allows only three topologies of clusters, which are
illustrated in Fig.~\ref{fig:path5}. The remaining three diagrams only allow
clusters that are confined to two or less adjacent tetrahedra.
 We will show more generally that
any term containing only clusters confined to three or fewer adjacent
tetrahedra is a constant.

The set ${\sf S}$ can therefore be broken up into three components,
comprising clusters which contain $1$, $2$, or only  $3$
multiply-occupied tetrahedra,
\begin{eqnarray}
  \label{eq:Ssplit}
  {\sf S}(G_n) = {\sf S}_1(G_n) + {\sf S}_2(G_n) + {\sf S}_3(G_n).
\end{eqnarray}
The sum in Eq.\eqref{eq:termset} can be carried out separately over
these three sets.  Let us consider first the sum over ${\sf S}_1$.
The clusters in ${\sf S}_1$ can be divided into subsets of those
residing on a specific tetrahedron ${\sf S}_1^t$.  

An arbitrary permutation $P$ of the 4 sites on tetrahedron $t$
leaves the set ${\sf S}_1^t(G_n)$ invariant. This is because
each solution obeys Eq.\eqref{eq:conds}, and 
$\Gamma_{a_i a_j} = \Gamma_{P(a_i) P(a_j)}$ for $a_i,a_j \in t$
(this is a set of permutations that leaves 
nearest neighbor pairs invariant).

The contribution of all clusters on $t$ to the term in question 
can only be a function of the 4 Ising variables of the 4 sites $q = 1,2,3,4 \in t$
\begin{eqnarray}
  \label{eq:cl1}
\sum_{(a_1,\ldots,a_n)\in {\sf S}_1^t}
f_{G_n}(\sigma_{a_1},\ldots,\sigma_{a_n}) = F(\sigma_1,\sigma_2,\sigma_3,\sigma_4)
\; .
\end{eqnarray}
Now we can use the fact that the spin configurations on one
tetrahedron are always constrained to be of the 3:1 form, i.e. they
are a permutation $P$ of the specific configuration $+++-$:
\begin{eqnarray}
  \label{eq:permsig}
  \sigma_q = \sigma^0_{P(q)},
\end{eqnarray}
with $(\sigma^0_1,\sigma^0_2,\sigma^0_3,\sigma^0_4)=(+,+,+,-)$.  Here
$q\rightarrow P(q)$ is a permutation of the $4$ sites.
The specific (cyclic) permutation $P$ now encodes the spin state on
this tetrahedron
\be\label{eq:cl3}
\begin{split}
& F(\sigma_{P(1)}^0,\sigma_{P(2)}^0,\sigma_{P(3)}^0,\sigma_{P(4)}^0) 
\\ = & 
\sum_{(a_1,\ldots,a_n)\in {\sf S}_1^t}
f_{G_n}(\sigma_{P(a_1)}^0,\ldots,\sigma_{P(a_n)}^0)
\\ = & 
\sum_{(a_1,\ldots,a_n)\in P^{-1}({\sf S}_1^t)}
f_{G_n}(\sigma_{a_1}^0,\ldots,\sigma_{a_n}^0)
\; .
\end{split}
\ee
Since the set ${\sf S}_1^t(G_n)$ is invariant under
these permutations, we find from the last expression, that 
$F(\sigma_1,\sigma_2,\sigma_3,\sigma_4)$ is also invariant under the permutations.
Hence this contribution is identical for {\sl all} spin
configurations, and is a constant within the 3:1 manifold.

Let us next consider the clusters in ${\sf S}_2$.  For each cluster,
there are two neighboring tetrahedra $t,t'$ which each contain two or
more sites $a_i$.  These tetrahedra share one specific site, which we
call $A$.  The pair of tetrahedra in question are determined by $A$ (the
tetrahedra $t$ and $t'$ are determined by the choice of $A$).  For one
such cluster, the sites $a_i$ with $i=1\ldots n$ may be partitioned into
three groups: the site $A$, and those which are on $t$ or $t'$ but are
not $A$:
\begin{eqnarray}
  \label{eq:14}
  {\sf t} & = & \{ a_i | a_i \in t, a_i \neq A\}, \\
  {\sf t'} & = & \{ a_i | a_i \in t', a_i \neq A\}.
\end{eqnarray}
Similarly to ${\sf S}_1$, we can divide ${\sf S}_2$ into subsets ${\sf S}_2^A$
residing on tetrahedron pairs defined by the site $A$.
We can then rewrite the sum by summing $A$ over all lattice sites, 
and summing the set of sites $a_1 \ldots a_n$ over ${\sf S}_2^A$ 
\be
  \label{eq:15}
  \begin{split} &
\sum_{(a_1,\ldots,a_n)\in {\sf S}_2}
 f_{G_n}(\sigma_{a_1},\ldots,\sigma_{a_n}) \\ & =
\sum_A \sum_{(a_1,\ldots,a_n)\in {\sf S}_2^A}
 f_{G_n}(\sigma_{a_1},\ldots,\sigma_{a_n}).
 \end{split}
\ee

We now observe that the set of solutions ${\sf S}_2^A$ is invariant
under any permutation $P_t$ ($P_{t'}$) of the 3 sites in ${\sf t}$
(${\sf t'}$).  Exactly as for ${\sf S}_1^t$ this is because each
solution in ${\sf S}_2^A$ obeys Eq.\eqref{eq:conds}, and $\Gamma_{a_i
  a_j} = \Gamma_{P_t(a_i) P_t(a_j)}$ for $a_i,a_j \in A\cup {\sf t}\cup
{\sf t'}$ (and the same holds if $P_t$ is replaced by $P_{t'}$).

The sum
\be\label{SevenIsing}
\sum_{(a_1,\ldots,a_n)\in {\sf S}_2^A}
 f_{G_n}(\sigma_{a_1},\ldots,\sigma_{a_n})
\ee
can only be a function of the 7 Ising variables of the sites in $A\cup
{\sf t}\cup{\sf t'}$. Due to the 3:1 constraint, if 
$\sigma_A=+$, then the Ising variables $\sigma_q$ for $q \in {\sf t}$ must be a
permutation $P_t$ of $\sigma_q^{(1)}=(++-)$.  If $\sigma_A=-$, then all
the $\sigma_q=+$. 
Hence we may write 
\begin{eqnarray}
  \label{eq:17}
  \sigma_{q} = \left\{ \begin{array}{ll} \frac{(1+\sigma_A)}{2}
      \sigma^{(1)}_{P_t(q)} + \frac{(1-\sigma_A)}{2} (+1) & {\rm for}\;
      q\in {\sf t} \\
\frac{(1+\sigma_A)}{2}
      \sigma^{(1)}_{P_{t'}(q)} + \frac{(1-\sigma_A)}{2} (+1) & {\rm for}\;
      q\in {\sf t'}\end{array}\right. .
\end{eqnarray}
Using these expressions, and the fact that ${\sf S}_2^A$ is invariant
under these two permutations, the sum in Eq.~\eqref{SevenIsing} is found
to depend only on $\sigma_A$.

This leaves finally 
\begin{eqnarray}
  \label{eq:18}
  \sum_{(a_1,\ldots,a_n)\in {\sf S}_2} = \sum_{A} \tilde{f}(\sigma_A),
\end{eqnarray}
where $\tilde{f}(\sigma_A)$ is a complicated function obtained from the
above manipulations -- which however does not depend upon $A$ itself.
The sum is clearly then constant, as the number of + and - spins are
fixed for the lattice.  Thus all terms in ${\sf S}_2$ are also
constants.

Finally, consider ${\sf S}_3$.  In these clusters there are three
adjoining tetrahedra, and one may identify a ``central'' tetrahedron $t$
which shares a site with each of the other two tetrahedra $t',t''$.
Here one may divide the sum variables into five groups: two
corresponding to the site shared by $t,t'$ and the site shared by
$t,t''$, and three others corresponding to the sites on $t,t',t''$ but
not shared.  One can again sum over the unshared sites on $t'$ and
$t''$, and obtain an expression for the cluster sum which involves sites
only on $t$.  By manipulations of the type used to analyze ${\sf S}_1$,
one finds that this remaining single-tetrahedron sum must also be
constant.

\begin{figure}
	\centering
		\includegraphics[width=1.0in]{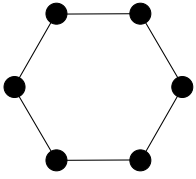}
	\caption{The only diagram at order $n=6$ giving a non-constant
          diagonal contribution in degenerate perturbation theory.}
	\label{fig:plaquette_loop}
\end{figure}

We conclude that any term for which the corresponding clusters are
confined to three or fewer adjacent tetrahedra must be constant.  Therefore 
all terms up to and including 5th order are constant.  At
sixth order, amongst the non-contractible diagrams there are a few
exceptions.  First, there is one disconnected diagram containing 
two loops of length three. In
this term, the sum over variables in the first and second groups is
independent, and therefore each can be carried out separately as for a
third order term.  This gives immediately a constant contribution.
The remaining diagrams are connected.  All but one of these diagrams
contains a loop of length $5$ or less (possibly in addition to other
larger loops).  Such terms are confined to three or fewer tetrahedra,
and are constant by the above arguments.  What remains is the single
diagram consisting of {\sl only} a single
loop of length six, shown in Fig.\ref{fig:plaquette_loop}.

This ``large loop'' diagram is thus the sole non-trivial contribution.  It
can be written in the form
\begin{eqnarray}
  \label{eq:form6}
&&  {\mathcal H}_6^{L}[\{\sigma_i\}] = \sum_{a_1\ldots a_6}
\left(\prod_{i=1}^6 \Gamma_{a_i a_{i+1}}\right) f_L(\sigma_{a_1},\ldots,\sigma_{a_6}),
\end{eqnarray}
where we identify $a_7=a_1$.  To analyze each term (given a particular set $a_1 \ldots a_6$), 
we employ a trick:
multiplying it by a carefully chosen representation of the
identity 
\begin{eqnarray}
  \label{eq:10}
  1=\prod_{\langle\langle ij\rangle\rangle} \left( \delta_{a_i a_j} +
    \eta_{a_i a_j}\right),
\end{eqnarray}
with $\eta_{ab}=1-\delta_{ab}$.  Here the product is over distinct pairs
$i,j$ which are {\sl not} connected in the loop diagram.  We multiply
the loop term by this expression and expand the product fully.  All but
one term involves at least one Kronecker $\delta$-function.  In each of
these summand terms, at least one sum can be collapsed, leading to a lower-order
term, which we have already shown is necessarily a constant.  The
remaining non-vanishing part is the original summand term multiplied by the
product,
\begin{eqnarray}
  \label{eq:11}
  \prod_{\langle\langle ij\rangle\rangle} \eta_{a_i a_j}.
\end{eqnarray}
This factor is non-zero if and only if all $n=6$ sites $a_i$ are
\emph{distinct}.  Thus the sites $a_i$ must comprise a closed walk on
the lattice in which each site on the walk is visited only once.  On
the pyrochlore lattice, this is exactly the set of hexagonal
plaquettes.  A specific plaquette on the lattice containing sites
$i_1\ldots i_6$ in sequence around the plaquette appears 12 times in
the sum in Eq.\eqref{eq:form6}, with $a_1\ldots a_6$ taking the six cyclic
permutations of $i_1\ldots i_6$ {\sl and} the six cyclic permutations
of these sites in reverse order.  Hence the non-constant contribution
to the diagonal energy at 6th order in DPT can be written:
\begin{eqnarray}
  \label{eq:Hnotconst}
 {\mathcal H}_6 & = & \sum_{\mathcal P} {\mathcal E}_{\mathcal
   P}(i_1,\ldots,i_6),
\end{eqnarray}
where $i_1,\ldots,i_6$ are the six sites moving clockwise around
plaquette ${\mathcal P}$, and 
\begin{eqnarray}
  \label{eq:Eform}
&&  {\mathcal E}_{\mathcal
   P}(\sigma_{i_1},\ldots,\sigma_{i_6})= \nonumber \\
&&  \sum_{k=1}^6 \left[ f_L(\sigma_{i_{k}},\ldots,\sigma_{i_{k+5}})
       +f_L(\sigma_{i_{k+5}},\ldots,\sigma_{i_{k}})  \right],
\end{eqnarray}
where $i_{k+6}\equiv i_k$.

\subsection{Results}
\label{DPT_result}

We have carried out the calculations detailed in the previous
subsections.  Specifically, by explicitly constructing $|\Psi\rangle$
in Eq.\eqref{eq:26}, we obtained ${\mathcal H}_6[\{\sigma_i\}]$ in
Eq.\eqref{eq:mag6}.  From this, we extracted the function $f_L$ in
Eq.\eqref{eq:form6} and thereby determined the plaquette energies
${\mathcal E}_{\mathcal P}$ using Eq.\eqref{eq:Eform}.  Using the 3:1
constraint, there are 5 configurations possible on any plaquette,
which we denote ``type 0'' to ``type 4''. These are enumerated in
Table~\ref{table1}.   The DPT calculation gives a specific energy
(proportional to $J_z \alpha^6$) for each type.  

\begin{table}[h]
\begin{tabular}{|c|c|c|}
\hline
Type  & Configuration & Fraction of    \\
      &               & minority spins \\
\hline 
$0$ & $\uparrow\uparrow\uparrow\uparrow\uparrow\uparrow$       &  $0$ \\
$1$ & $\downarrow\uparrow\downarrow\uparrow\downarrow\uparrow$ &  $\frac{1}{2}$\\
$2$ & $\downarrow\uparrow\uparrow\uparrow\uparrow\uparrow$     &  $\frac{1}{6}$\\
$3$ & $\downarrow\uparrow\downarrow\uparrow\uparrow\uparrow$   &  $\frac{1}{3}$\\
$4$ & $\downarrow\uparrow\uparrow\downarrow\uparrow\uparrow$   &  $\frac{1}{3}$\\
\hline
\end{tabular}
\caption{\label{table1} The different plaquette types, with the fraction of minority sites in each one.}
\end{table}

There is some freedom in the choice of these 5 energies.  That is,
certain changes of the plaquette energies leave the {\sl differences}
of total energy amongst distinct 3:1 states unchanged.  One obvious
such ``gauge'' change is a global shift of all 5 energies by the same
amount.  Another less obvious constraint comes directly from the 3:1
rule. If one denotes the fraction of plaquettes in the lattice in
configuration $a$ by $x_a$, the total fraction of minority sites must
always be $1/4$.  Each plaquette configuration has a fixed fraction of
minority sites $M_a$, given in Table~\ref{table1}.  Thus
\begin{equation}\label{magnetization}
\frac{1}{4} = \sum_{a=0}^4 M_a x_a .
\end{equation}
The energy per plaquette is then 
\begin{equation}\label{diagonal_energy}
{\mathcal H}_6 = \sum_{a=0}^4 {\mathcal E}_a x_a
\end{equation}
Using \eqref{magnetization}, one sees that a shift $\Delta{\mathcal
  E}_a = c M_a$, with arbitrary constant $c$ shifts the energy by a
constant.  The obvious global energy shift remarked on above derives
similarly from the normalization condition 
$\sum_a x_a = 1$.  Using these two constraints, we see there are only
3 independent plaquette fractions.
We (arbitrarily) choose to keep $x_{1,2,4}$ as our independent variables.
Substituting the solutions for the other fractions ($x_{0,3}$) into
\eqref{diagonal_energy}, we find
\begin{equation}
{\mathcal H}_6 =x_1 V_1 + x_2 V_2 + x_4 V_4,
\end{equation}
with the 3 ``gauge invariant'' physical energy parameters 
\begin{eqnarray}
  \label{eq:ginv}
  V_1 & = &  \frac{1}{2} \left({\mathcal E}_0+2 {\mathcal E}_1-3
    {\mathcal E}_3\right) , \nonumber\\
  V_2 & = & \frac{1}{2} \left(-{\mathcal E}_0+2 {\mathcal
      E}_2-{\mathcal E}_3\right), \nonumber\\
  V_4 & = & \left( {\mathcal E}_4-{\mathcal E}_3 \right).
\end{eqnarray}
\begin{widetext}
Our DPT results are:
\begin{equation}
\begin{split}
V_1 = & - J_z \alpha^6
\frac{3 s^4(98304 s^5-139648 s^4+79136 s^3-22040 s^2+3006s-165)}{32 (2 s-1) (4 s-1)^5 (8 s-3)^2 (12 s-5)} ,
\\
V_2 = & J_z \alpha^6
\frac{s^3 \left(256 s^3-51 s+9\right)}{32 (4 s-1)^3 (8 s-3)^2},
\\
V_4 = & J_z \alpha^6
\frac{s^4 \left(272 s^2-136 s+15\right)}{16 (4 s-1)^5 (8 s-3)^2}
\; .
\end{split}
\end{equation}
\end{widetext}
We have made several checks on the above calculation.  First, we have
carried out a more direct scheme which
sums the terms in DPT in a completely different manner from the
methods described in this section.  We leave the vast details of this
calculation to Appendix~\ref{app:Other_DPT}. The results of this
alternative method agree perfectly with those quoted above.  Second,
in the following section we will compare the $s \rightarrow \infty$
limit of the above result with the result of a large $s$ calculation
for the XXZ model. The large $s$ limit of the energies we find in DPT
becomes
\begin{equation}\label{Infty_S_lim_DPT}
\begin{split}
\mathop {\lim }\limits_{s \to \infty } \frac{V_1}{s} = & 0,
\\
\mathop {\lim }\limits_{s \to \infty } \frac{V_2}{s} = & \frac{J_z \alpha^6}{512} ,
\\
\mathop {\lim }\limits_{s \to \infty } \frac{V_4}{s} = & 0
\; .
\end{split}
\end{equation}
We shall see that this result indeed agrees exactly with the
corresponding limit of the large $s$ expansion.

\subsection{Off diagonal term}

In this section we describe how the lowest order off diagonal term in the DPT effective Hamiltonian
is calculated. As explained in Section~\ref{sec:dpt_formulation}, this term appears at order $O(\alpha^{6 s})$.

The lowest order off diagonal term acts only on a hexagonal plaquette in the flippable configuration 
(type 1 plaquette, as in Table~\eqref {table1}). It changes the plaquette configuration from one 
flippable configuration to the other flippable configuration. Therefore the off diagonal term has 
the following general form 
\be
\begin{split} &
{\mathcal H}_{\textrm{off diagonal}} 
= (-1)^{6s+1} \alpha^{6 s} J_z K 
\sum_P 
\left({\centering \includegraphics[width=0.4in]{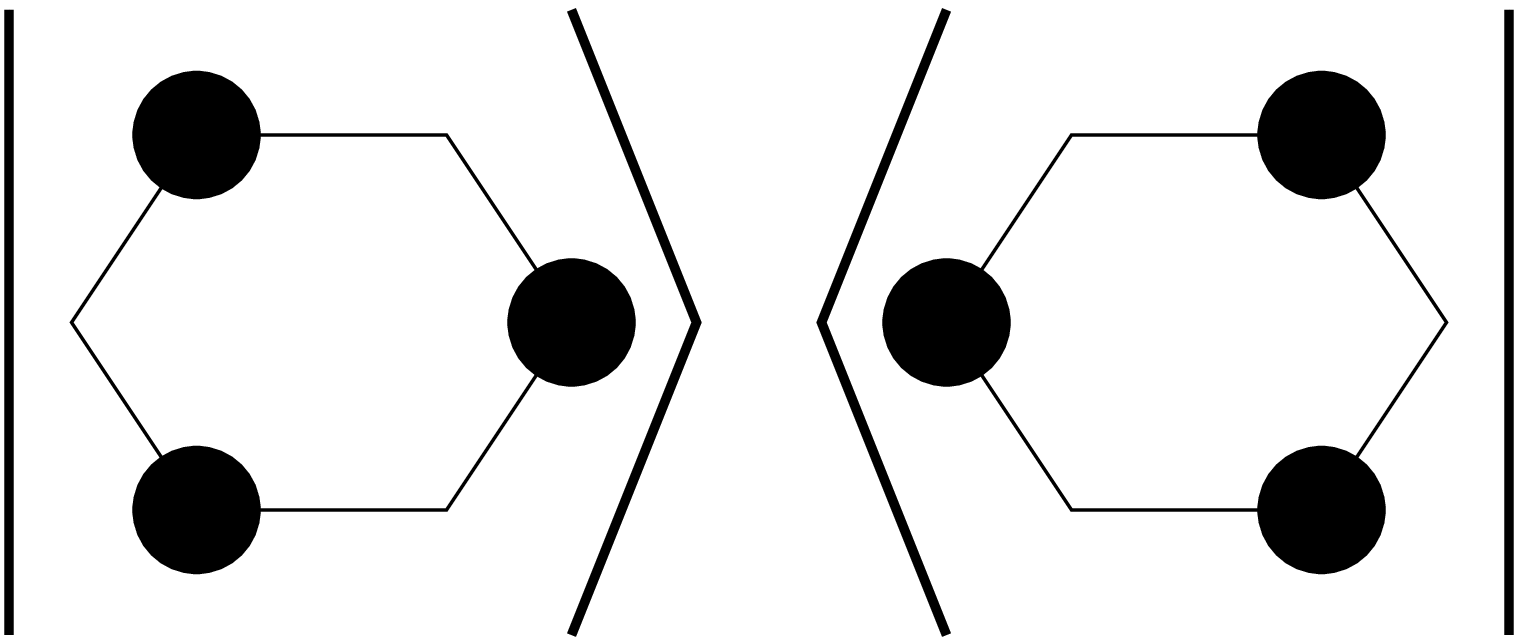}} + {\rm h.c.} \right)
\\
= & (-1)^{6s+1} \alpha^{6 s} J_z K 
\sum_P \left( \ket{\d\u\d\u\d\u} \bra{\u\d\u\d\u\d}  + {\rm h.c.} \right)
\; ,
\end{split}
\ee
where we denote the two flippable configurations of the plaquette by $\includegraphics[width=0.2in]{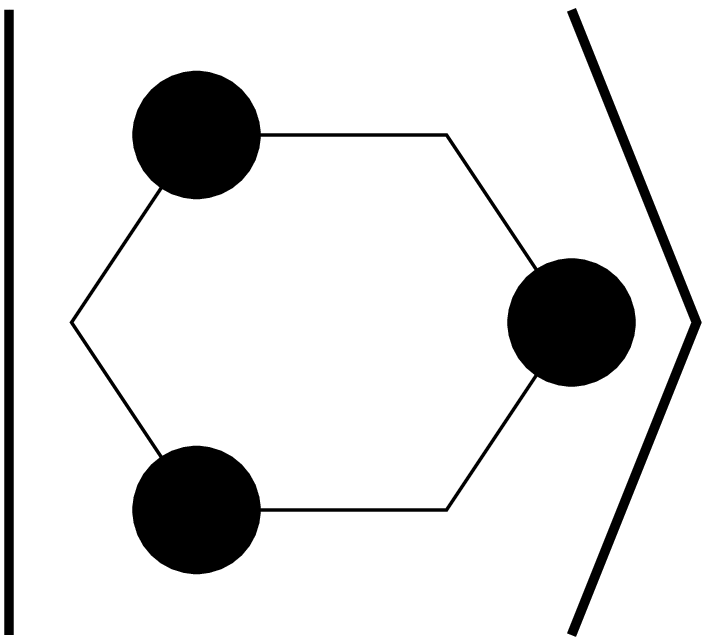}$,
$\includegraphics[width=0.2in]{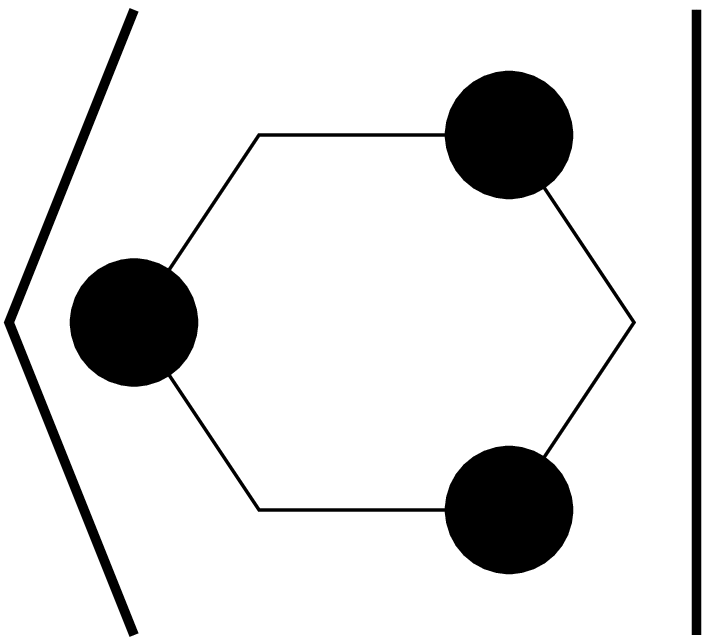}$ 
for the sake of brevity. Note that we can change the $(-1)^{6s+1}$ factor into a $(-1)$ by
a unitary transformation similar to that employed in Ref.~\onlinecite{Hermele:prb04}.
We shall now 
describe how the coefficient $K$ is calculated. 

\begin{figure}
	\centering
		\includegraphics[width=2.0in]{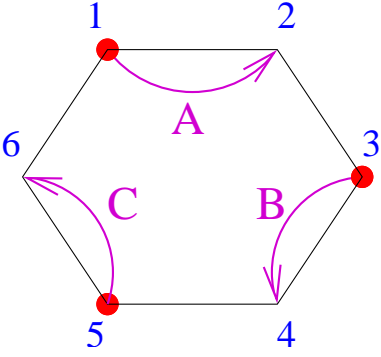}
	\caption{(Color online) Off diagonal process on a single plaquette. The (red) circles denote minority sites.}
	\label{fig:off_diagonal}
\end{figure}

Each one of the DPT processes contributing to the off-diagonal term
consists of $2s$ spin transfer operations along each one of 3 links of a
hexagonal plaquette of the pyrochlore lattice (see
Fig.~\ref{fig:off_diagonal}), acting in some particular order.
 
We can calculate $K$ by adding the contributions from all the DPT
processes occurring on a single plaquette, starting in the state
$\ket{\d\u\d\u\d\u}$ and ending in the state $\ket{\u\d\u\d\u\d}$.

In every such process 3 spins go from an initial state of $+s$ to $-s$, and 3 start with $-s$ and end up as $+s$.
The spins change via ladder operators $S_j^{\pm}$, and therefore we get ``angular momentum factors'' from the
action of these operators. The same set of operators $S_j^{\pm}$ act in every process, and so these
factors are always the same. For the $S^+$ operators taking a single site from $-s$ to $+s$ we find
\be
\prod_{m = -s}^{s-1} \sqrt{s(s+1)-m(m+1)} = (2s)!
\; ,
\ee
and for the $S^-$ operators taking a single site from $+s$ to $-s$ we find
\be
\prod_{m = -s+1}^{s} \sqrt{s(s+1)-m(m-1)} = (2s)!
\; .
\ee
In total from all the ladder operators, we find a common factor
$ ((2s)!)^6 $. From the 6 spin transfer operators we have another common factor of $1/2^{6s}$.

All that remains to be calculated for a single DPT process is the product of resolvents 
of each stage in the spin transfer process.
First let us classify the different processes on a single plaquette. We can choose one
of two sets of three links on which spin transfer will occur (one such choice is shown in Fig.~\ref{fig:off_diagonal}). 
The contribution from each one of these 
two cases is identical, so we shall calculate the contributions for one set of three links, and 
multiply the final result by 2. The processes we will sum over only differ by the order in which the 
spin transfer operators act on the 3 predetermined links. We call the three links $A$,$B$, and $C$, and 
then each process is described by a string of $6s$ letters $q_1 \ldots q_{6s}$ which contain
$2s$ instances of each one of the three letters $A$,$B$,$C$. For example, a possible string for $s = 1$
is $AABBCC$.
$ $From this classification, it is evident that in total there are $\frac{(6s)!}{ (2s)! (2s)! (2s)!}$ different processes.
At this point we can write a formal expression for the coefficient $K$
\be
K = 2 \frac{((2s)!)^6}{2^{6s} }  \sum_{ \{ q_n\} } \prod_{\ell = 1}^{6s-1} \tilde{\mathcal R}_{\ell}(\{ q_n\})
\; ,
\ee
where $\tilde{\mathcal R}_{\ell}(\{ q_n\})$ denotes the resolvent at step $\ell$ of the DPT process described by the 
string $\{ q_n\}$.

Now we turn to formulating the resolvent in a convenient manner that will facilitate the summation
over all processes.
Starting from Eq.\eqref{eq:25}, in this case the set ${\sf F}$ consists only of the 6 sites surrounding  
the hexagonal plaquette. We shall denote these 6 sites $1$ through $6$, as in Fig.~\ref{fig:off_diagonal}
so that $A$ denotes the link $(1,2)$, $B$ denotes the link $(3,4)$, and $C$ denotes the link $(5,6)$. 
Since the 6 sites have alternating initial states $\pm s$,
any pair of nearest neighbor sites has $\sigma_i \sigma_j = -1$.
We can therefore rewrite the inverse resolvent operator as
\be
\tilde{\mathcal R}^{-1}  = -J_z \sum_{j = 1}^6
\left(S_j^z -s\right)\left(S_{j+1}^z - s\right) 
- 2 J_z s \sum_{j = 1}^6 \left(S_j^z - s\right)
\; ,
\ee
where the indices are defined modulo 6, so that $S_{6+1}^z = S_1^z$. From this point on, all 
index arithmetic is defined modulo 6 as well, for ease of presentation.

To further simplify the resolvent, we introduce $n_A(\ell,\{ q_n\})$ as the number of times the 
link $A$ has had spin transfer occur on it up to stage $\ell$ in the process 
desribed by the string $\{ q_n\}$.  
The same numbers can be also introduced for $B$ and $C$.  
Then, by definition, the total number of spin transfer operations is
$n_A(\ell,\{ q_n\}) + n_B(\ell,\{ q_n\}) + n_C(\ell,\{ q_n\}) = \ell $. 
In what follows we will show that the resolvent can be described only by these 3 numbers. 
To see this, notice first that, regardless
of the order of spin transfer operations, a spin transfer operator on the link 
$(j,j+1)$  changes $\left(S_j^z - s\right) \rightarrow \left(S_j^z - s - 1\right)$
and $\left(S_{j+1}^z - s\right) \rightarrow \left(S_{j+1}^z - s - 1\right)$. 
Note also that, in
the initial state, all $\left(S_{j+1}^z - s\right) = 0$.
Thus, at every stage of any process,
$\left(S_1^z - s\right) = \left(S_2^z - s\right) = - n_A(\ell,\{ q_n\})$
Similarly $\left(S_3^z - s\right) = \left(S_4^z - s\right)= - n_B(\ell,\{ q_n\})$,
and $\left(S_5^z - s\right) = \left(S_6^z - s\right) = - n_C(\ell,\{ q_n\})$.
Using these variables, one can then rewrite the resolvent as 
\be
\begin{split} &
\tilde{\mathcal R}_{\ell}(\{ q_n\}) = 4 J_z s \ell \\ &
- \frac{J_z}{2} \left[ \left( n_A + n_B \right)^2 + \left( n_B + n_C \right)^2 + \left( n_C + n_A \right)^2 \right]
\; ,
\end{split}
\ee
where we have suppressed the explicit dependence of the $n_{A,B,C}$ numbers on $\ell,\{ q_n\}$
for clarity. It is more convenient to derive a recursion relation for the resolvent at stage $\ell$
\be\label{recursion}
\tilde{\mathcal R}_{\ell + 1}(\{ q_n\}) =
\tilde{\mathcal R}_{\ell}(\{ q_n\}) +  
J_z \left( 4s - 1 - \ell - n_{q_{\ell +1}}(\ell,\{ q_n\}) \right)
\; .
\ee
The initial condition for this recursive series is $\tilde{\mathcal R}^{-1}_0 = 0$.
Using Eq.\eqref{recursion}, we can calculate the product $\prod_{\ell = 1}^{6s-1} \tilde{\mathcal R}_{\ell}(\{ q_n\})$ 
for a given process. For every process, 
we need to keep track of only the 3 numbers $n_{A,B,C}$
in the various steps of the the process.

We have calculated the coefficient $K$ explicitly for a number
of interesting values of $s$. The results are summarized in Table~\ref {table3}.
\begin{table}[h]
\begin{tabular}{|c|c|}
\hline
$s$  & $K$  \\
\hline
$\frac{1}{2}$ & $\frac{3}{2}$ \\
$1$ & $0.166$ \\
$\frac{3}{2}$ & $0.00839536$ \\
$2$ & $0.000304464 $ \\
$\frac{5}{2}$ & $9.1747 \times 10^{-6}$ \\
\hline
\end{tabular}
\caption{\label{table3} Values $K$ of the coefficient for the lowest  order off-diagonal term, for various values of $s$.}
\end{table}

\section{Large $s$ expansion}
\label{large_s}

A large $s$ analysis has recently been employed in
Ref.~\onlinecite{Hizi:prb06} to explore the magnetic order for the
general spin $s$ Heisenberg AFM on the pyrochlore lattice.  Restricting
the Hilbert space to collinear spin configurations, the authors of
Ref.~\onlinecite{Hizi:prb06} derived an effective Hamiltonian out of the
harmonic spin wave energy contribution, to order ${\cal O}(s)$.  The
effective Hamiltonian prefers spin products around hexagonal plaquettes
to be $-s^6$ in the zero magnetic field, and $+s^6$ in the
half-polarized plateau region. Following a terminology inspired by Ising
gauge theory, these are denoted ``$\pi$ flux'' configurations and ``zero
flux'' configurations, respectively.  In order to compare this approach,
which is justified in the large $s$ limit, with the DPT analysis of
Section~\ref{sec:DPT_Leon}, we have repeated the same type of effective
Hamiltonian calculation for the the XXZ model.  Our derivation follows
closely that of Ref.~\onlinecite{Hizi:prb06}.

The large $s$ expansion consists of expressing the spin degrees of
freedom in terms of Holstein-Primakoff bosons and expanding in
decreasing powers of $s$. The lowest order term in the large $s$
expansion is of order $s^2$, and corresponds to the classical spin
version of the quantum XXZ Hamiltonian 
\be\label{H_cl_aniso}
\begin{split}
{\mathcal H}_{\textrm{cl}} = &
J_z \sum_{\langle i j \rangle}
\left[ \alpha \left( {\bf S}_i \cdot {\bf S}_j \right) +
\left( 1 - \alpha \right) \left( {\bf S}_i \cdot {\hat z}\right)
\left( {\bf S}_j \cdot {\hat z}\right)
\right]
\\ &
- 2 J_z h \sum_j S^z_j
\; ,
\end{split}
\ee where as before $\alpha = \Jx/J_z$.  

\begin{figure}
	\centering
		\includegraphics[width=3.6in]{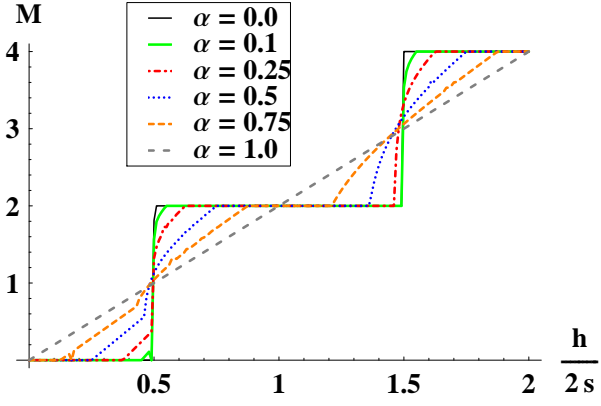}
	\caption{(Color online) Magnetization (in units of $s$) of a single tetrahedron of classical spins with an anisotropic XXZ interaction, parametrized by $\alpha$. For any $\alpha <1$ a half polarization plateau exists.}
	\label{fig:Classical_Anisotropic_Magnetization2}
\end{figure}

In order to analyze the ground state of this anisotropic classical model \eqref{H_cl_aniso}, we first
calculate the minimum energy
configuration for a \emph{single} tetrahedron. For the single tetrahedron we obtain the magnetization 
curve shown in Fig.~\ref{fig:Classical_Anisotropic_Magnetization2}. We find that for
$\alpha < 1$ a half polarization plateau opens up, and the plateau
becomes wider as $\alpha$ decreases from $1$. 
In this plateau, the classical spins on the single tetrahedron being analyzed are in a collinear configuration,
with three ${\bf S}_j = s{\hat z}$ and one ${\bf S}_j = -s{\hat z}$ spins.  
This is just the classical analog of the 3:1 configuration on a single tetrahedron
found in Section~\ref{easy_axis}.
A 3:1 spin configuration can be realized on each and every tetrahedron of the lattice simultaneously.
We therefore conclude that in the range of magnetic fields where 
the single tetrahedron is in a half polarized state,
the ground state of the many body system~\eqref{H_cl_aniso}
is the manifold of 3:1 configurations.  This means that the
plateaus in the classical XXZ model on the complete pyrochlore lattice
are at least as wide as in
Fig.\ref{fig:Classical_Anisotropic_Magnetization2}.

In the following we will discuss only this half magnetization plateau. We then
assume the collinear 3:1 states,
which allows us to
describe the magnetic configuration in terms of
the same Ising variables $\sigma_j = \pm 1$ as in Section~\ref{easy_axis}.

As in Ref.~\onlinecite{Hizi:prb06} we use the unitary transformation~\eqref{unitary} so that we can define the
Holstein-Primakoff bosons, which amounts to replacing the rotated spin operators as follows
\be
\begin{split}
S_j^z = & s - {\hat m}_j
\\
S_j^+ = & \sqrt{2 s - {\hat m}_j} \, {\hat b}_j \approx \sqrt{2 s} \, {\hat b}_j
\; ,
\end{split}
\ee
where ${\hat b}_j$ are canonical bosonic operators, and
${\hat m}_j = {\hat b}_j^{\dagger} {\hat b}_j$ is the boson number operator.
We plug these into the Hamiltonian \eqref{XXZ}, and keep only the quadratic terms in the bosonic operators.

Since the spin configurations are now restricted to the 3:1 manifold,
the magnetic field term is the \emph{same} for every 3:1 configuration as the magnetization
is constant on the plateau.
In terms of the \emph{unrotated} spin variables {\bf $S_j^z$}, this amounts to
$\sum_j S_j^z = \frac{s}{2} N$ where $N$ is the
number of sites in the pyrochlore lattice. Varying the magnetic field in the plateau region
causes an overall shift in the spin wave energies of all the 3:1 states,
and thus will not alter the energy differences between different 3:1 states.
Similarly the Ising variables have a sum of $\sum_j \sigma_j = \frac{1}{2} N$,
and we can use these two identities to derive $\sum_j \sigma_j {\hat m}_j = 0$, which is useful in simplifying other terms.
Therefore, we can ignore the magnetic field term, since we are searching for an effective Hamiltonian splitting the energies of
different 3:1 states. The effect of the magnetic field is to determine the energy gap for spin wave excitations.
The vanishing of the spin wave gap signifies an instability of the 3:1 manifold, corresponding to the
edges of the half polarization plateau.

$ $From Eqs.\eqref{virtualE},\eqref{rotated_H_1},  the resulting
harmonic spin wave term reads
\begin{eqnarray}
  \label{eq:harmxxx}
&& {\mathcal H}^{3:1}_{\textrm{harm}} = 
J_z \frac{\alpha}{2} s \sum_{i,j}
\Gamma_{i j} \Big[ 
\left( \frac{1 + \sigma_i \sigma_j}{2} \right)
\left( {\hat b}_j^{\dagger} {\hat b}_i + h.c. \right)
\nonumber \\ && + 
\left( \frac{1 - \sigma_i \sigma_j}{2} \right)
\left( {\hat b}_j {\hat b}_i + h.c. \right)
\Big]+ J_z 2 s \sum_j {\hat m}_j
\; .
\end{eqnarray}
Following the derivation Ref.~\onlinecite{Hizi:prb06},
the zero point energy of this
harmonic term for a given 3:1 spin configuration (described by $\{ \sigma_{j}\}_{j=1}^N$) is
\be
E_{\textrm{harm}} = J_z s \sum_{k=1}^{N} \frac{|\lambda_k|}{2},
\ee
where $\lambda_k$ are the solutions of the eigenvalue equation
\be
\left( \frac{\lambda}{2} \right)^2 {\bf v} =
\left[
\mathbf{1}  +
\frac{\alpha}{2} \left( {\hat \sigma} {\hat \Gamma} {\hat \sigma} + {\hat \Gamma} \right)
+ \left( \frac{\alpha}{2} {\hat \sigma} {\hat \Gamma} \right)^2
\right] \cdot {\bf v}
\; .
\ee
In the right hand side $\hat \Gamma$ denotes the same $N \times N$ connectivity matrix
introduced in Section~\ref{easy_axis}, and $\hat \sigma$ is
a diagonal $N \times N$ matrix with $\sigma_j$ as its diagonal elements.
Without specifying the 3:1 configuration, we can
write an expression for the harmonic energy in terms of $\sigma_j$
\be\label{Eharm}
E_{\textrm{harm}} = J_z s {\textrm{Tr}}\left[
\sqrt{\mathbf{1} + \frac{\alpha}{2} \left({\hat \sigma}{\hat \Gamma}{\hat \sigma} + {\hat \Gamma}\right)
+ \frac{\alpha^2}{4} \left({\hat \sigma}{\hat \Gamma}\right)^2
}
\right]
\;.
\ee

One can calculate the spin wave energies by assuming a particular spin
configuration and computing the trace exactly.  However, as in
Ref.~\onlinecite{Hizi:prb06}, if one does not know which candidate spin
configurations to consider, one can derive an effective Hamiltonian to
determine which spin configuration gives the lowest harmonic energy, and
find a favorable spin configuration.

The square root in \eqref{Eharm} can be expanded in powers of matrix
operators.  We first observe that $\alpha$ only appears as a multiplier
of the matrix $\Gamma$.  Therefore, an expansion in powers of matrix
operators is \emph{equivalent} to expansion in the parameter $\alpha$.
In the present context, this expansion is justified due to the easy axis
anisotropy $\alpha <1 $.

The terms in the expansion can be organized as a sum of traces over
products of $\Gamma$ matrices and Ising variables $\sigma_j$. The order
of $\alpha$ for each term also specifies the number of connectivity
matrices $\Gamma$ appearing in that term.

Due to the trace operation, the product of $\Gamma$ matrices represents
closed loops on the lattice. The Ising variables appearing in each such
term can only involve the sites on the loops defined by the product of
$\Gamma$ matrices.  Using the results of
Section~\ref{sec:contraction-rules}, which discuss functions of Ising
variables and $\Gamma$ matrices precisely of the form appearing in this
expansion, it is evident that all terms involving less than six $\Gamma$
matrices will result in constants, which will not split energies of the
3:1 states. As in Section~\ref{easy_axis}, the lowest order term in the
expansion in $\alpha$ causing energy splitting in the 3:1 manifold
involves loops around hexagonal plaquettes of the pyrochlore lattice.
For simplicity, we consider only these terms, and ignore any higher
order term in the expansion in $\alpha$. After extensive simplification,
the $6$-th order term reads \be\label{eff_large_s}
\begin{split}
{\mathcal H}_{\textrm{harm}} = &
J_z s \left( \frac{\alpha}{2} \right)^6 \frac{1}{512} \Big[
14 \text{Tr}\left(\sigma .\Gamma.\sigma .\Gamma ^5\right)
+ 14 \text{Tr}\left(\sigma.\Gamma ^2.\sigma .\Gamma ^4\right)
\\ &
+ 7 \text{Tr}\left(\sigma .\Gamma ^3.\sigma .\Gamma^3\right)
-6 \text{Tr}\left(\sigma .\Gamma .\sigma.\Gamma .\sigma .\Gamma .\sigma .\Gamma ^3\right)
\\ &
- 3 \text{Tr}\left(\sigma .\Gamma ^2.\sigma .\Gamma .\sigma.\Gamma ^2.\sigma .\Gamma \right)
\\ &
- 6 \text{Tr}\left(\sigma .\Gamma ^2.\sigma .\Gamma^2.\sigma .\Gamma .\sigma .\Gamma \right)
\\ &
+ \text{Tr}(\sigma .\Gamma .\sigma .\Gamma .\sigma.\Gamma .\sigma .\Gamma .\sigma .\Gamma .\sigma .\Gamma)\Big]
+ O(\alpha^8)
\; .
\end{split}
\ee
$ $From the expression one extracts only those terms corresponding to loops around hexagonal plaquettes.
Eq.\eqref{eff_large_s} takes the form of the function in
Eq.\eqref{eq:1}, with $n = 6$ and the ``loop'' diagram
$G_n=\{(12),(23),(34),(45),(56),(61)\}$. 
The corresponding function $f(\sigma_{a_1},\ldots,\sigma_{a_6})$ reads
\begin{eqnarray}
&& f(\sigma_{a_1},\ldots,\sigma_{a_6}) = 14\ \sigma_{a_1}\sigma_{a_2}+ 14\ \sigma_{a_1}\sigma_{a_3}
+ 7\ \sigma_{a_1}\sigma_{a_4} \nonumber \\
&&\ -\ 6\ \sigma_{a_1}\sigma_{a_2}\sigma_{a_3}\sigma_{a_4}
- 3\ \sigma_{a_1}\sigma_{a_3}\sigma_{a_4}\sigma_{a_6}
- 6\ \sigma_{a_1}\sigma_{a_3}\sigma_{a_5}\sigma_{a_6} \nonumber \\
&& \ +\  \sigma_{a_1}\sigma_{a_2}\sigma_{a_3}\sigma_{a_4}\sigma_{a_5}\sigma_{a_6}. \label{eq:f}
\end{eqnarray}
The effective Hamiltonian therefore describes all possible
spin interactions on the hexagonal plaquette of the pyrochlore lattice
-- 2,4 and 6 spin interactions. It is far more convenient to express this complicated Hamiltonian in terms
of energies of plaquette configurations, in the same way we formulated the results of the DPT
in Section~\ref{easy_axis}  as ${\mathcal H}_{\textrm{harm}} = \sum_P
{\mathcal E}_P$ (using the same 5 plaquettes in Table~\ref{table1}).

As in Section~\ref{DPT_result}, there are only $3$ independent plaquette
configuration energies, $V_{1,2,4}$ which to $O(\alpha^6)$ are
\be\label{large_S_ergs}
\begin{split}
V_1 = &0,
\\
V_2 = &
\frac{J_z \alpha^6 s}{512},
\\
V_4 = & 0
\; .
\end{split}
\ee

\begin{figure}
\centering
\includegraphics[width=2.5in]{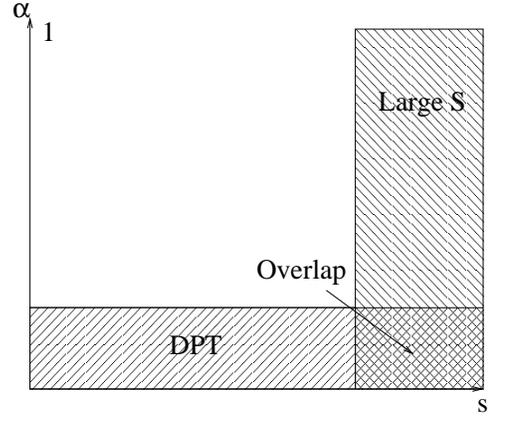}
\caption{This figure shows the regions of parameter space where the DPT
  and large S expansions are justified, and their region of overlapping
  validity.}
\label{fig:overlap}
\end{figure}

Comparing \eqref{large_S_ergs} with \eqref{Infty_S_lim_DPT}, we find
complete agreement between the DPT of Section~\ref{easy_axis} and the
large $s$ expansion of this section, in the limit of both $\alpha
\rightarrow 0$ and $s \rightarrow \infty$, where both approaches are
justified (see Fig.~\ref{fig:overlap}). This serves an excellent check
on the correctness as well as validity of our calculations, in the
parameter regime where the approximations overlap.

\section{Low energy states of the effective Hamiltonian} 
\label{diagonal_gs}

\subsection{Strict easy axis limit for $s\geq 3/2$}
\label{sec:strict-easy-axis}

In this subsection, we consider the $\alpha=J_\perp/J_z \ll 1$ limit,
for which the lowest-order non-vanishing terms in the effective
Hamiltonian are dominant.  For any $s\geq 3/2$, this is just the sixth
order diagonal contribution.

\subsubsection{large-$s$ case}
We first consider the large $s$
limit.  
As is clear from Eq.~\eqref{large_S_ergs}, at order $s$ only the type 2
plaquette suffers from a \emph{positive} energy correction, while
$V_{1}$ and $V_4$ are only nonzero at order $s^{0}$ or lower.  Hence 
the type 2 plaquette is strongly disfavored for large $s$.  
This in combination with the 3:1 constraint allows us to restrict ourselves 
to the ``$0$-flux manifold'' in the large $s$ region (See Section~\ref{large_s} for 
the definition of a 0-flux manifold.)
. 

\begin{figure}
	\centering
		\includegraphics[width=2.4in]{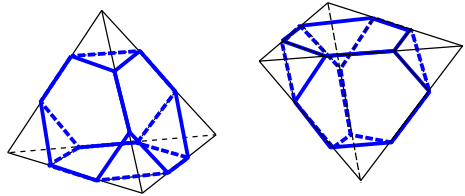}
	\caption{(Color online) A cell is composed of four link-sharing hexagonal plaquettes (The polyhedron bounded by thick (blue) lines). The cell on the left is up-headed, and the one on the right is a down-headed cell.}
	\label{fig:cell}
\end{figure}

To see this, let us first introduce a ``cell'' comprised by 4
link-sharing plaquettes. Choose 4 hexagonal plaquettes such that any
pair of two plaquettes out of these 4 plaquettes always share a link.
Then these four plaquettes form a single polyhedron (a truncated
tetrahedron), with 4 hexagonal faces, and 4 triangular faces (see
Fig.~\ref{fig:cell}).  We will refer to this polyhedron as a cell. In
the pyrochlore lattice, one may distinguish two kinds of cells -- when
one completes the tetrahedra enclosing a cell, we can identify {\it
  up-headed} and {\it down-headed} cells, according to the direction at
which the tetrahedra are pointing (see Fig.~\ref{fig:cell} for examples
of both kinds ).  Each up/down-headed cell shares its faces (hexagonal
plaquettes) with 4 nearest neighboring down/up-headed cells. Thus,
centers of cells constitute a diamond lattice, where those of up-headed
cells take part of one FCC lattice and those of down-headed cells form
the other FCC lattice. It suffices to determine the spin configuration
on only the up-headed (down-headed) cells in order to specify the spin
configuration on all sites of the lattice.
 
Observing the local constraint, one can readily enumerate the various
minority spin configurations of a cell. In Table.~\ref{cell_table}, all
possible cell configurations allowed in the 3:1 manifold are listed.
Each cell type is described by the configurations of its 4 hexagonal
plaquettes.

To see that the ground state manifold in the large-$s$ region is
composed only of type 0, 3 and 4 plaquettes (i.e. $0$-flux states),
notice from Table.~\ref{cell_table} that any cell type which contains a
type 1 plaquette always contains at least one type 2 hexagonal
plaquette. This implies \be 0 \le x_{1} \le x_2 \ee where $x_a$ are the
plaquette type fractions, as introduced in Section~\ref{DPT_result}.
Disallowing the type 2 plaquette inevitably leads to excluding the type
1 plaquette, and therefore, the positive $V_2$ in leading order of
$s^{-1}$ expansion allows us to conclude that the classical ground state
spin configurations in the large $s$ limit consist of \emph{only} the
$0$-flux states.



This ground state, however, is massively degenerate.  Higher order
quantum corrections in $s^{-1}$ can select a particular classical state
out of this $0$-flux manifold. To see this, let us expand the plaquette
energies in ${s^{-1}}$ \be\label{growing_s}
\begin{split}
\frac{V_1}{J_z \alpha^6} = & -\frac{3}{512} 
+ O \left( s^{-1} \right),
\\
\frac{V_2}{J_z \alpha^6} = & \frac{s}{512}+\frac{3}{1024}
+O \left( s^{-1} \right)
\,,
\\
\frac{V_4}{J_z \alpha^6} = & \frac{17}{65536 s} 
+O \left( s^{-2} \right)
\; .
\end{split}
\ee Notice first that the ${\cal O}(1)$ negative energy correction to
$V_1$ plays no role in lifting the degeneracy of the $0$-flux manifold,
since this manifold does not contain any type 1 hexagonal plaquettes.
Thus, provided that $V_2$ dominates the other two, the most relevant
correction in the large $s$ limit is $V_4$, which
always disfavors the type 4 hexagonal plaquette, since it is
positive.

\par
\begin{table}[htbp]
\begin{center}
\begin{tabular}{|c||c|c|c|c|c|}
\hline
{\it cell}  $\setminus$  {\sf plaquette}   & {\sf type 1} & {\sf type 2} 
& {\sf type 3} & {\sf type 4} & {\sf type 0} \\
\hline \hline 
{\it type 1} & 1 & 1 & 1 & 1 & 0 \\
{\it type 2} & 1 & 3 & 0 & 0 & 0 \\
{\it type 3} & 0 & 4 & 0 & 0 & 0 \\
{\it type 4} & 0 & 2 & 2 & 0 & 0 \\
{\it type 5} & 0 & 2 & 1 & 1 & 0 \\
{\it type 6} & 0 & 2 & 1 & 0 & 1 \\
{\it type 7} & 0 & 2 & 0 & 1 & 1 \\ 
{\it type 8} & 0 & 2 & 0 & 0 & 2 \\ \hline 
{\it type 9} & 0 & 0 & 4 & 0 & 0 \\
{\it type 10} & 0 & 0 & 2 & 1 & 1 \\
{\it type 11} & 0 & 0 & 0 & 3 & 1 \\
{\it type 12} & 0 & 0 & 0 & 0 & 4 \\
\hline
\end{tabular}
\caption{\label{cell_table} The various cell configurations
are described by the number of each plaquette type included 
in the plaquettes comprising a cell.
Cell types are indicated by {\it italics} and plaquette type 
by {\sf sans serif}. In the zero-flux manifold, only the type 9, 10, 11 and 12 cells are  
allowed, since they do not contain type 1 and type 2 hexagonal plaquettes. 
Furthermore, a  cell must contain a type 2 plaquette 
whenever it contains a type 1 plaquette. This is quantified by $0\le x_1 \le x_2 $. 
 }
\end{center}
\end{table}
\par

\begin{figure}
	\centering
		\includegraphics[width=2.8in]{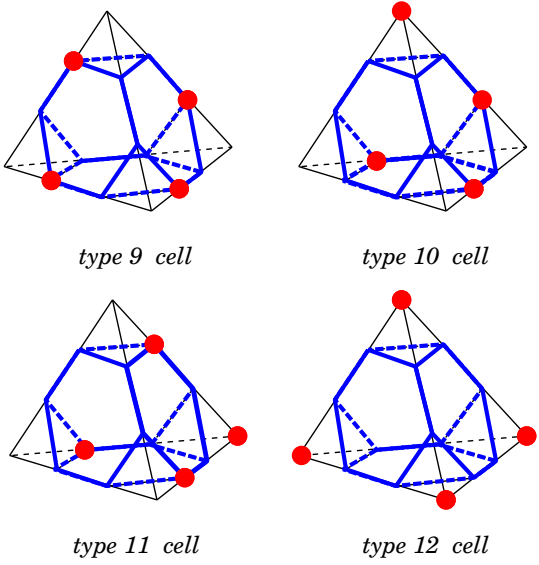}
	\caption{(Color online) The four cell types allowed in the zero-flux manifold. Minority sites are specified by 
	  the (red) circles.}
	\label{fig:celltype}
\end{figure}

Since the type 4 plaquette is disfavored, observing the $0$-flux condition 
on all plaquettes, we have only to minimize $x_4$
to obtain the ground state in the large $s$ region.  

However, in the 3:1 manifold, the $0$-flux condition becomes so strong
that $x_4$ is in fact bounded by $\frac{3}{28}$ from below ($x_4 \geq
\frac{3}{28} $). To see this, notice first that only the type 9, 10, 11
and type 12 cells drawn in Fig.\ref{fig:celltype} are allowed in the
$0$-flux manifold. Next, we denote by $y_{9,10,11,12}$ the fraction of
cell types $9 \ldots 12$ in the entire pyrochlore lattice (we use these
instead of plaquette type fractions $(x_3,x_4,x_0)$ for later
convenience).  In the $0$-flux manifold, only these cell types may
occur, and therefore $\sum_{j=9}^{12}y_{j}=1$.  Together with the
``global'' 3:1 constraint, i.e. Eq.~\eqref{magnetization} one finds
\begin{eqnarray}
  3y_{12} = y_{9},  \label{globalcondition}
\end{eqnarray}
or alternatively, 
\begin{eqnarray}
y_{12}=\frac{1}{4}(1-y_{10}-y_{11}), \label{y-magnetization}
\; .
\end{eqnarray}

An important step to identify the lower bound on $x_4$ is to note that
packing these four cell types into a pyrochlore lattice is highly
constrained by the \emph{local} 3:1 rule imposed on each tetrahedron.
For example, a type 12 cell can only have cell types 10,11 and 12 as
neighboring cells.  Each type 10 and 11 cell can neighbor at most one
type 12 cell, as they both have only one type 0 plaquette, and the type
12 cell consists only of type 0 plaquettes.  One can also show that a
type 12 cell can have at most one neighboring type 12 cell. the
remaining neighboring cells must be of type 10 or 11. These observations
are already sufficient to conclude that 
\be 
3y_{12} \leq y_{10}+y_{11}.
\ee 
Now using Eq.~\eqref{y-magnetization}, we obtain the lower bound on
the fraction of type 10 and type 11 cells: $y_{10} + y_{11}\ge
\frac{3}{7}$.  Since 
these two types of cell are the only cells allowed in the $0$-flux
manifold which have type 4 hexagonal plaquettes, this lower bound
immediately gives us that for the fraction of type 4 hexagonal
plaquette:
\begin{eqnarray}
x_{4}=\frac{1}{4}y_{10}+\frac{3}{4}y_{11}\geq \frac{1}{4}y_{10}+\frac{1}{4}y_{11} \geq \frac{3}{28}. \label{lowerbound}
\end{eqnarray}

$ $From the derivation above, one can easily see that the equal sign is
realized {\it if and only if} $y_{11} = 0$. Excluding type 11 cell
configurations, one can show that a type 12 cell \emph{always} neighbors
three type 10 cells, and one type 12 cell.  As a consequence, the
condition $3y_{12} = y_{10}$ is satisfied only when any type 10 cell has
a type 12 cell as its nearest neighboring cell, through its single type
0 plaquette.  Since the fraction of type 9 cells is uniquely determined
by the fraction of type 12 cells \eqref{globalcondition}, and we have
already excluded any type 11 cells from a state saturating the bound on
$x_4$, if a state saturating this bound exists, since it minimizes the
fraction $x_4$ it must be configuration with a maximum number of type 12
cells on the lattice.  Without any type 12 cells, we cannot have any
type 9 cells either, and are limited to type 10 and 11 cells.  A state
comprised only of type 10 and type 11 cell, has fraction of type 4
plaquettes that is always greater than $\frac{1}{4}$.

\begin{figure}
	\centering
		\includegraphics[width=3.6in]{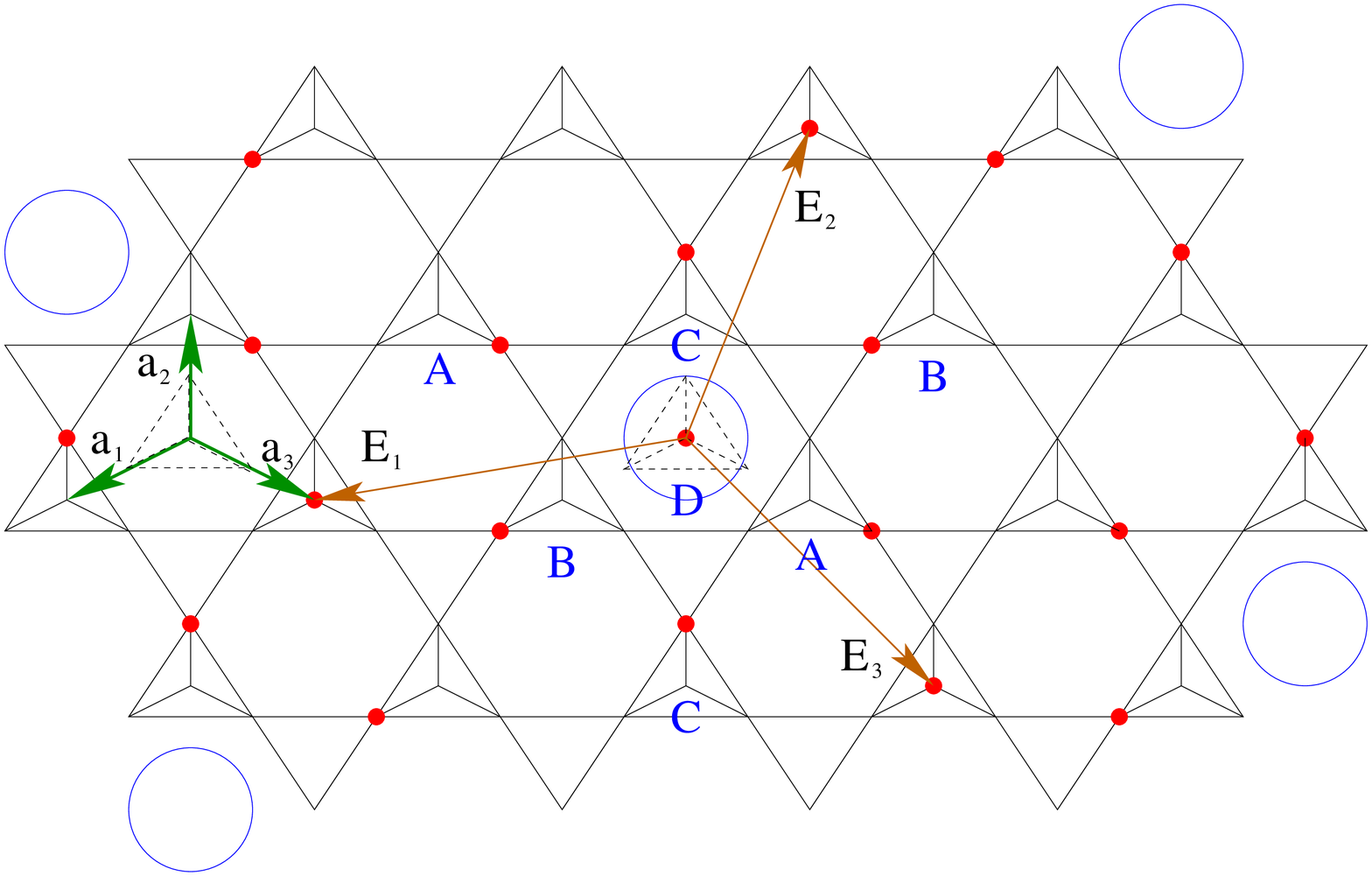}
	\caption{(Color online) The trigonal$_{7}$ state. The spin configuration of a planar layer
	of tetrahedra	is shown. Triangles with lines connected at their centers represent up pointing 
	tetrahedra, while the other triangles represent down pointing tetrahedra.
	Minority sites are denoted by (red) circles. 
	Two dashed triangles denote up pointing tetrahedra in the planar layer of tetrahedra immediately above the
	one depicted in this figure. These two tetrahedra are used to show the primitive vectors
	for the pyrochlore lattice ${\bf a}_{1,2,3}$ and for the magnetic unit cell of the trigonal$_7$ state
	${\bf E}_{1,2,3}$. The seven up pointing tetrahedra included in one valid choice of a magnetic unit cell for the 
	trigonal$_7$ state are marked by (blue) letters indicating one of 4 3:1 configurations for a single tetrahedron.
	The type 0 plaquettes residing between pairs of adjacent type 12 cells are marked by large (blue) circles. }
	\label{fig:trig7}
\end{figure}

In what follows, we will show that this lower bound for $x_4$ is {\it uniquely} 
(up to a finite degeneracy) realized by the {\it periodic} minority spin 
configuration depicted in Fig.~\ref{fig:trig7}. This collinear magnetic ordered state, 
which we shall refer to as the ``trigonal$_{7}$'' state,  
contains $7$ pyrochlore unit cells. It has a magnetic unit cell with primitive vectors  
${\bf E}_1 = 2{\bf a}_1 - {\bf a}_3,
{\bf E}_2 = 2{\bf a}_2 - {\bf a}_1$, and 
${\bf E}_3 = 2{\bf a}_3 - {\bf a}_2$, where 
${\bf a}_{1,2,3}$ are the primitive unit vectors of 
the pyrochlore lattice (FCC lattice vectors ${\bf a}_1 = \frac{a}{2} (0,1,1)$ and cyclic permutations). 
$ $From the unit cell vectors, we can find the volume of the magnetic unit cell 
\be
\left( {\bf E}_1 \times {\bf E}_2 \right) \cdot {\bf E}_3 =
7 \left( {\bf a}_1 \times {\bf a}_2 \right) \cdot {\bf a}_3
\ee 
These 3 primitive vectors are of equal length, and are not mutually
perpendicular. Therefore, the magnetic Bravais lattice is in the {\sl
  trigonal} crystal system -- whence the name trigonal$_7$ state.

\begin{figure}
	\centering
		\includegraphics[width=3.0in]{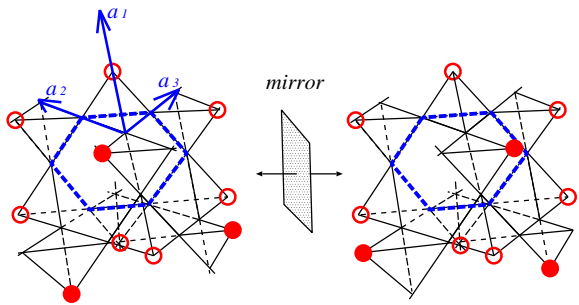}
	\caption{(Color online) The broken reflection symmetry. 
	Starting from a type 12 (up-headed) cell, and drawing the 10 tetrahedra surrounding it,
	we first choose the place of the nearest neighboring type 12 cell (down-headed) - the 
	two type 12 cells share the hexagonal plaquette marked by thick dashed (blue) lines. With this choice, 
  the minority sites on 7 tetrahedra are automatically determined (marked by open
	(red) circles). However, minority sites for the other 3 tetrahedra (solid (green) circles) are not fully determined 
	and we still have a ``mirror'' degree of freedom. For convenience, we also draw the 
	primitive vectors of pyrochlore lattice, in accordance with those defined in Fig.~\ref{fig:trig7}.}
	\label{mirror}
\end{figure}

$ $From the planar view in Fig.~\ref{fig:trig7} it is clear that this magnetic state has a three-fold 
rotation symmetry about the ${\bf a}_1 + {\bf a}_2 + {\bf a}_3 = a (1,1,1)$ axis perpendicular to the page. 
In the direction of the pyrochlore lattice directions, there is a periodicity of 7,	giving rise to a seven 
fold degeneracy due to FCC lattice translations alone. 
The trigonal$_7$ state breaks a reflection symmetry about a plane perpendicular to the Kagome plane, parallel to $a_2$
and passing through the point where the three vectors ${\bf E}_{1,2,3}$ originate in the figure
(see Fig.~\ref{mirror} for a another view of this symmetry operation).
Together with the 4-fold choice of the set of Kagome planes, 
it is evident that the degeneracy of this magnetic state
is $4 \times 7 \times 2 = 56$

As is clear from Fig.~\ref{fig:trig7}, the spin configuration satisfies both 
the local zero flux condition and the local 3:1 constraint. 
To see that this trigonal state saturates the lower bound for $x_4$, 
one has only to count the fraction of type 9, 10, 11 and 12 cells: 
$y_{9}:y_{10}:y_{11}:y_{12}=3:3:0:1$.  From Table~\ref{cell_table}, we find the trigonal$_7$ state  
realizes the lower bound $x_{4}=\frac{3}{28}$. We conclude the trigonal$_{7}$ state is {\it at least} 
one of the ground states in large-$s$ region. 

As argued above, any state saturating the bound must have every type 0 plaquette connecting between a type 12 cell
and a type 10 cell. Starting with a type 12 cell, and using this rule together with the 3:1 constraint 
and the zero flux condition, suffice to uniquely construct the trigonal$_7$ state, up to the finite degeneracy
described in above. Starting from the initial type 12 cell, the plaquette connecting this cell to another type 12
cell defines the Kagome plane in Fig.~\ref{fig:trig7}. Next, pick one of the 2 mirror image choices in Fig.~\ref{mirror}
of the type 10 cell configurations neighboring the first type 12 cell. From this point on, the three rules mentioned
above uniquely determine the rest of the magnetic configuration in the entire lattice.

Finally, the energy per plaquette of the trigonal$_7$ state is
\be
\frac{1}{N} E_{\textrm{trigonal}_{7}} = \frac{3}{28} V_4
\; . 
\ee

\subsubsection{Spin $s \geq 2$}

We expect that the trigonal$_7$ state described above is the
ground state for sufficiently large $s$. In the following, we shall
argue that this is indeed the case for $s \geq 2$.  For $s=5/2,2,3/2,1$, the
energy parameters in the effective Hamiltonian are given in Table~\ref{vals}.  

For all the cases in Table~\ref{vals}, $V_1$ is the largest and most negative energy.
This would suggest that the lowest energy 3:1 state is one with a maximum number of 
plaquettes of type 1.  However, the geometry of the lattice as well
as the 3:1 constraint pose stringent restrictions.
By enumerating all possible types of cells in Table~\ref{cell_table}, 
one finds that every type 1 plaquette must be accompanied by at least one type 2 plaquette
on the same cell. The configuration of the entire lattice can be determined by considering only up-headed cells,
and therefore, the existence of $M$ type 1 plaquettes demands that at
least $M$ type 2 plaquettes are present as well.
We deduce the following inequality $x_1 \leq x_2$, in any 3:1 configuration.
$ $From this we see that the energy of a type 1 plaquette is offset by the energy cost of a type 2
plaquette which is the \emph{highest} energy cost for all the $s$ values in Table~\ref{vals}.
Therefore, the number of type 1 plaquettes is not necessarily maximized
in the ground state even with small $s$.  

One observes that the magnitude of the energy $V_1$ is comparable
to $V_2$ already at $s=5/2$, and this trend continues to higher $s$ - $V_2$ becomes more dominant. 
Given the restriction $x_1 \leq x_2$, and the large energy cost of type 2 plaquettes, 
the analytic arguments in the above subsection suggest the trigonal$_7$ state may be
the lowest energy state for $s\geq 5/2$. The case $s=2$ is close to 
the boundary for a change in behavior.

In order to search for other candidate ground states, we have enumerated
all 3:1 states on a variety of periodic finite clusters, and determined
the exact lowest energy state for each one, for
$s=1,3/2,2,5/2,\cdots,6$. For $s \geq 2$ we find no states with lower
energy than that of the trigonal$_7$ state.  This strongly suggests that
the trigonal$_7$ state is the ground state for all $s \geq 2$, though of
course this limited numerical investigation does not constitute a proof
that this is the case.  Moreover, states with large numbers of type
1 plaquettes, are among the \emph{highest} energy states we have found,
which does give credence to our assessment that when the $V_2$ and $V_1$
are comparable energy scales (with opposite sign), because of the
condition $x_1 \leq x_2$ the energy $V_2$ is still dominant.  One can
conclude that {\sl if} there is a state with lower energy for $s \geq
2$, it must have a large unit cell which is incompatible with all the
clusters considered in Table~\ref{table2}.

\begin{table}
\begin{tabular}{|c|r|c|c|c|}
\hline \hline
energy& $s=\frac{5}{2}$& $s=2$  & $s = \frac{3}{2}$ & $s = 1$   \\
\hline 
$\frac{V_1}{J_z \alpha^6}$ & $-0.0113$ & $-0.0135$ &$-0.0188$ & $-0.0410$ \\
$\frac{V_2}{J_z \alpha^6}$ & $ 0.0090$ & $ 0.0084$ &$ 0.0083$ & $ 0.0099$ \\
$\frac{V_4}{J_z \alpha^6}$ & $ 0.0002$ & $ 0.0003$ &$ 0.0005$ & $ 0.0015$  \\
\hline
\end{tabular}	
	\caption{Energies $V_{1,2,4}$ of the plaquette configurations type $1,2,4$ for $s=2,\frac{3}{2},1$}
	\label{vals}
\end{table}

\begin{table}
\begin{tabular}{|c|c|c|c|c|c|c|c|}
\hline \hline
Number of  & Number of & \multicolumn{2}{c|}{$s=\frac{3}{2}$}\\
    \cline{3-4} unit cells & 3:1 states & E & gs\\
\hline
$2 \times 2 \times 1 = 4$   &  $36$        &  $1.3\cdot 10^{-4}$  & $4$    \\
$2 \times 2 \times 2 = 8$   &  $272$       &  $1.3\cdot 10^{-4}$  & $12$   \\
$4 \times 2 \times 1 = 8$   &  $708$       &  $1.3\cdot 10^{-4}$  & $4$    \\
$3 \times 3 \times 1 = 9$   &  $1,120$     &  $-2.9\cdot 10^{-4}$ & $24$   \\
$5 \times 2 \times 1 = 10$  &  $3,370$     &  $4.\cdot 10^{-4}$   & $4$    \\
$3 \times 2 \times 2 = 12$  &  $2,436$     &  $1.3\cdot 10^{-4}$  & $4$    \\
$4 \times 2 \times 2 = 16$  &  $23,696$    &  $1.3\cdot 10^{-4}$  & $12$   \\
$6 \times 3 \times 1 = 18$  &  $649,480$   &  $-2.9\cdot 10^{-4}$ & $192$  \\
$3 \times 3 \times 2 = 18$  &  $61,192$    &  $-2.9\cdot 10^{-4}$ & $30$   \\
$5 \times 2 \times 2 = 20$  &  $237,156$   &  $1.3\cdot 10^{-4}$  & $4$    \\
$4 \times 3 \times 2 = 24$  &  $1,685,508$ &  $1.3\cdot 10^{-4}$  & $4$    \\
$3 \times 3 \times 3 = 27$  &  $7,515,136$ &  $-2.9\cdot 10^{-4}$ & $216$  \\
\hline
\end{tabular}	
	\caption{3:1 configurations on periodic clusters. Energy is given in units of $J_z \alpha^6$}
	\label{table2}
\end{table}

\subsubsection{Spin $s=3/2$}
\label{sec:spin-s=32}

Spin $s=3/2$ is the smallest spin value for which in the extreme easy-axis limit $\alpha \ll 1$
the off-diagonal term in the effective Hamiltonian may be ignored.  The
corresponding plaquette energies are given in column 4 of
Table~\ref{vals}.  The energy for type $1$ plaquettes is approximately
50\% larger (more negative) than for $s=2$.  In the extreme limit of
very large and negative $V_1$, the ground state has been determined
previously in Ref.\onlinecite{Bergman:prl05}.  
The state, referred to as the {\bf R} state in Ref.\onlinecite{Bergman:prl05}
as well as in the remainder of this manuscript, 
maximizes the fraction of type 1 plaquettes, and is unique (up to lattice symmetries).  

\begin{figure}
	\centering
		\includegraphics[width=3.0in]{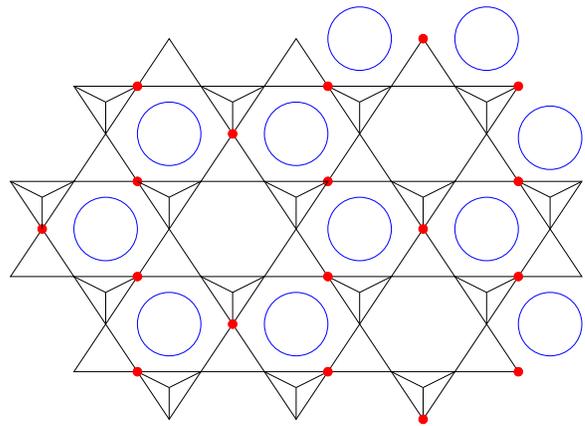}
	\caption{(Color online) 3:1 spin configuration of a single layer of tetrahedra in the ${\sqrt 3} \times {\sqrt 3}$
	state. Only minority spin sites are 
	marked by (red) solid circles.
	Flippable plaquettes (type 1) are denoted by a (blue) circle drawn at their center. The same 
	conventions as in Fig.~\ref{fig:trig7} are used here.}
	\label{fig:root3_state3}
\end{figure}

The numerical investigation mentioned in the previous subsection, shows that the 
{\bf R} state is {\sl not} the lowest energy state for the diagonal effective Hamiltonian at
$s=3/2$. Instead, we find a {\sl massively degenerate set of classical
ground states}. One example of these states has all the minority sites contained in a set of 
parallel Kagome layers of the pyrochlore lattice. Every Kagome plane will be have the same spin configuration
shown in Fig.~\ref{fig:root3_state3}. This example, and many other states in this degenerate manifold
all have a $\sqrt{3} \times \sqrt{3}$ structure in the Kagome planes, and therefore we shall refer to a large subset
of this manifold of states as the $\sqrt{3} \times \sqrt{3}$ states.

The analysis of this degenerate manifold of states is somewhat involved. We therefore leave the 
details to Appendix~\ref{app:root3_degeneracy}, and only mention a number of facts here.
All the states we have found numerically have 
plaquette type fractions of
$x_0 = \frac{1}{6}, x_1 = \frac{1}{6}, x_2 = \frac{1}{3}, x_3 = \frac{1}{6}$ 
and $x_4 = \frac{1}{6}$. As a consequence, the energy
per plaquette of these states is
\be
\frac{1}{N} E_{\sqrt{3} \times \sqrt{3}} =
\frac{1}{6} \left( V_1 + 2 V_2 + V_4 \right).  
\label{energy33}
\ee
In appendix~\ref{app:root3_degeneracy}, we show by explicit construction
that the degeneracy is at least
\be
18 \times 2^{\frac{N}{12L}} + 4 \times 3^L - 36,
\; .
\ee
which grows exponentially with system size.  We have not shown that the
above states exhaust the possibilities with energy given by
Eq.~\eqref{energy33}, so the above formula is only a lower bound for the
degeneracy.

\subsection{Effect of off-diagonal term}
\label{Effective_QDM}

In this subsection we add to the effective Hamiltonian the off-diagonal
term where it is likely to be important (low values of $s$).  For
$s=1/2$, the off-diagonal term is parametrically larger than the
diagonal terms in the $\alpha \ll 1$ limit.
For $s=1$, it is of the same order 
as the diagonal terms. However, our
explicit calculations demonstrate 
that even in this case, the off-diagonal term
is numerically more than four times larger than the largest 
diagonal plaquette energy.

For $s=3/2$, the off-diagonal term
is negligible in the $\alpha \ll 1$ limit, but extrapolating the DPT results to
the isotropic case $\alpha=1$ indicates that while it is not larger than the diagonal terms, it is likely
not negligible either.

To gauge the importance of the off-diagonal plaquette term, 
it is instructive to compare the diagonal energy of the various 
candidate ground states studied above.
Their energies per plaquette are shown for small values of $s$
in Table~\ref{tab:gse}.  We see that the energy {\sl differences} amongst these 
competing states are rather small on the scale of the
off-diagonal amplitude $K$.  For instance, for $s=3/2$, the energy
difference between the ``worst'' of these three states (the {\bf R}
state) and the ``best'' (the $\sqrt{3}\times\sqrt{3}$ state) is only
$0.0018$ per plaquette, approximately {\sl four times smaller} than the
diagonal coupling $K=0.008$.  Thus we can expect that adding the $K$
term can introduce sufficient quantum fluctuations to alter the balance
between these states, either destabilizing one in favor of the other, or
perhaps stabilizing a superposition of these orders in some form.

\begin{table}
  \centering
\begin{tabular}{|c|c|c|c|}
\hline
Spin  & R state & Trigonal$_7$ state & $\sqrt{3} \times \sqrt{3}$ state \\
\hline
$s=1$ &           $-2.8 \times 10^{-3}$   & $1.7 \times 10^{-4}$  & $-3.3 \times 10^{-3}$   \\
$s=\frac{3}{2}$ & $ 1.5 \times 10^{-3}$   & $5.7 \times 10^{-5}$    & $-2.9 \times 10^{-4}$  \\
$s=2$ &           $ 2.9 \times 10^{-3}$   & $3.1  \times 10^{-5}$    & $6.0 \times 10^{-4}$   \\
$s = \infty$ &    $ 1.5 \times 10^{-3} s$ & $0.0$                 & $6.5 \times 10^{-4} s$ \\
\hline
\end{tabular} 
  \caption{Diagonal energy per plaquette in various classical ground states. Energies are given in units of $J_z \alpha^6$}
  \label{tab:gse}
\end{table}

We cannot hope to establish the result of such subtle energetics here,
particularly given the non-trivial nature of the effective QDM
Hamiltonian including the off-diagonal term.  However, we will discuss
several {\it natural candidate ground states} from the perspective of
order-by-disorder and the general theoretical framework of QDM-type
models.  

\subsubsection{Purely off-diagonal QDM  {\bf --- $s=1/2$ case ---}}
\label{sec:purely-diagonal-qdm}

Let us consider first the simple case of $s=1/2$, for which the
Hamiltonian is well-approximated by including the off-diagonal {\sl
  only}.  Clearly low-energy ground states of this Hamiltonian must
have significant amplitude for type 1 plaquettes, as other plaquettes
are annihilated by the off-diagonal term.  We note that the trigonal 
state has {\sl no} type 1 plaquettes.  
This implies that it is an exact zero energy eigenstate of the 
purely kinetic Hamiltonian.  Since 
it is straightforward to construct states with significantly negative
energy per plaquette, the classical trigonal$_7$ state is clearly an 
excited state in this case.  It seems difficult to imagine any
way that the ground state could be adiabatically connected to the
trigonal state (or any other zero energy state with no type 1
plaquettes).  

Let us instead consider what sorts of states might naturally minimize
the energy of the kinetic term.  This sort of pure QDM problem has
been considered in numerous places in the literature.  Specifically
for the QDM on the diamond lattice, the question has been discussed in
Ref.\cite{Bergman:prb05} (see references therein for a guide to QDMs).
Roughly speaking, the energy is minimized by delocalizing the
wavefunction as much as possible amongst different dimer
configurations.  However, the non-trivial connectivity in the
constrained space of dimer coverings makes the nature of this
delocalization subtle.

One possibility in such a 3d QDM is that
the ground state is a $U(1)$ {\sl spin liquid}, in which the
delocalization is sufficiently complete as to prevent any symmetry
breaking (the meaning of the $U(1)$ is discussed in-depth in e.g.
Ref.~\onlinecite{Hermele:prb04}).  Roughly 
speaking, the wavefunction has support for all possible dimer 
coverings, with 
equal amplitude for all topologically equivalent configurations.  The
existence and stability of such a state can be established in a QDM
with a particular form of diagonal interaction, in the neighborhood of
the so-called ``Rokhsar-Kivelson'' (RK) point.  While this point
(corresponding to $V_1=K>0$, $V_2=V_4=0$) is not physically relevant
to the pyrochlore antiferromagnets, it is possible that such a $U(1)$
spin liquid state remains the ground state for the purely off-diagonal
QDM.

A second possibility is that the delocalization is incomplete, due to
``order-by-disorder'' physics.  In particular, it may be favorable to
delocalize only over a limited set of classical states, amongst which
the connections are greater than those amongst generic classical
configurations.  In this case there is generally some
symmetry-breaking induced by the selection of the states involved.
Two sorts of such ordering have been proposed and observed in other
similar QDM models.  The first type of order-by-disorder state is 
one in which the set of classical states for which the ground state
wavefunction has the largest amplitude are 
``centered'' about a single 
classical state having
the maximal number of type 1 plaquettes. 
Such a wavefunction may be ``selected'' by the kinetic energy, 
since under the action of the kinetic term of the QDM, this is the 
classical state is connected to the \emph{largest number} of other classical
states. In our problem, this classical state is just 
the {\bf R} state mentioned above and discussed at length in
Refs.\onlinecite{Bergman:prl05,Bergman:prb06}. 
A simple form  for such a wavefunction is 
\begin{eqnarray}
  \label{eq:varwfR}
  |{\bf R},\{\gamma_P\}\rangle &=& \exp\left[ \sum_{P} 
    \gamma_P \left({\centering \includegraphics[width=0.4in]{fig34.eps}} + {\rm h.c.} \right)    \right] |{\bf R}\rangle, 
\end{eqnarray}
where $|{\bf R}\rangle$ is the classical {\bf R} state (with definite
$S_i^z=S\sigma_i$), and $\gamma_P$ are variational parameters which can
be used to optimize the quantum state $ |{\bf R},\{\gamma_P\}\rangle$.

The second type of order-by-disorder state is one in which there are a
maximal number of {\sl independently resonating plaquettes}.  
This is based on the observation that the exact ground state for the kinetic
term on a single plaquette is simply an equal amplitude superposition
of the two type 1 states. 
However, neighboring plaquettes share sites, and therefore it is not possible to
form a direct product of such resonances on {\it all} plaquettes.  
Instead, the best one can na\"ively do along these lines is 
to find the classical state with the largest number 
of type 1 plaquettes which can be {\it independently flipped}, and 
on these type 1 plaquettes form an equal amplitude superposition of these
two states.  

\begin{figure}
	\centering
		\includegraphics[width=3.0in]{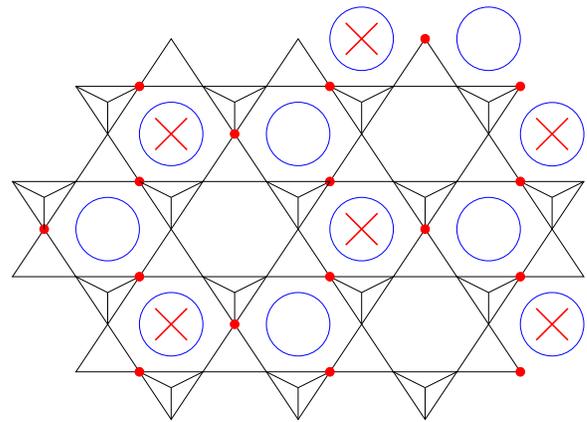}
	\caption{(Color online) Choice of non-overlapping flippable plaquettes to resonate in a plane of the 
	$\sqrt{3} \times \sqrt{3}$ state. The chosen plaquettes are marked with (red) crosses in the middle.}
	\label{fig:RPS}
\end{figure}

A state with the maximal number of independently flippable plaquettes can be 
described, starting from the $\sqrt{3} \times \sqrt{3}$ states introduced in
Sec.\ref{sec:spin-s=32}. The largest set of independently flippable plaquettes is 
a subset of all the type 1 plaquettes. An
appropriate choice in a single plane is demonstrated in
Fig.~\ref{fig:RPS}, which includes half of the flippable plaquettes in
the plane. It is interesting to point out that in each plane there are
2 possible choices of the plaquettes to be resonated (one half or the
other), so out of each $\sqrt{3} \times \sqrt{3}$ state we can
construct $2^L$ different choices of the plaquettes that will be
resonating.  The degeneracy of these state therefore is $2^L \times
3^L \times 4 = 6^L \times 4$. Other states realizing this maximum 
number of independently resonating plaquettes may be possible, but 
we have not pursued this further. We refer to these states as
``Resonating Plaquette States'' (RPS). A precise wavefunction
describing the RPSs we have derived from the 
$\sqrt{3} \times \sqrt{3}$ states is
\begin{equation}
\label{RPS} \ket{RPS} = \prod_{P \in
  G} \frac{1}{\sqrt 2} \left( 1 + \left( \ket{\hexagon_A}
    \bra{\hexagon_B} + {\rm h.c.} \right) \right) \ket{\Psi} \; , 
\end{equation}
where $G$ denotes the set of non-overlapping resonating plaquettes,
and $\ket{\Psi}$ denotes one of the $\sqrt{3} \times \sqrt{3}$ states.
There are $4 \times 3^L$ choices for $\ket{\Psi}$, and $2^L$ choices
for $G$ given $\ket{\Psi}$.  We note that the {\sl symmetry} of the
RPS is distinct and lower than that of the
$\sqrt{3}\times\sqrt{3}$ state -- even 
in a single layer.
Thus there is a precise distinction between these
two states independent of the detailed form of their wavefunctions,
for which the above explicit forms are of course only crude 
approximations. 

While potentially there might be some other state we have not
anticipated, we think that most likely one of the three above states
obtains in the purely kinetic QDM valid for $s=1/2$.  We will, however,
refrain from making any definite statement as to which of these is the
true ground state.  One may imagine comparing the energies of the
wavefunctions in Eqs.(\ref{eq:varwfR},\ref{RPS}) to gauge the relative
favorability of the ${\bf R}$ and RPS states.  Unfortunately, even
evaluating the variational energy of the $|{\bf R}\rangle$ state in
Eq.\eqref{eq:varwfR} is rather challenging.  Another difficulty is the
considerable freedom in choosing the RPS wavefunctions.  Furthermore, a
good variational wavefunction for the spin liquid is also needed for a
more complete comparison.  As always, there is much arbitrariness in
defining each variational wavefunction, making the predictive power of
such an approach unclear.  We believe this issue is more likely to
reliably resolved in the future thorough numerically exact methods such
as quantum Monte Carlo or exact diagonalization.

\subsubsection{$s>1/2$ QDMs}
\label{sec:s12-qdm}

For $s=1$ and $s=3/2$, significant diagonal terms enter the QDM
Hamiltonian.  These act to alter the balance between the three
candidate states discussed above, and also potentially to introduce
the possibility of other states disfavored in the purely kinetic
Hamiltonian.  For both $s=1$ and $s=3/2$, the ground state of the
classical diagonal term alone is actually a massively degenerate set of states
discussed briefly in Section~\ref{sec:spin-s=32}, 
and in more detail in Appendix.~\ref{app:root3_degeneracy}.
$ $From all the $\sqrt{3}\times\sqrt{3}$ states, 
we can construct RPSs. For $s=3/2$, as we have
shown, however, the ${\bf R}$ state is also quite low in diagonal
energy and indeed only slightly worse
than the $\sqrt{3}\times\sqrt{3}$ states, as far as the diagonal term is concerned. 
Thus we expect that introducing the 
diagonal terms tends to favor both the RPS and the ``renormalized'' {\bf R} state 
over the $U(1)$ spin liquid.   
If their effects are strong enough, they could 
also stabilize the ``non-resonating'' $\sqrt{3}\times\sqrt{3}$ states (or any one 
of the other states with the same energy).
We speculate that a spin liquid is unlikely to be realized in these
cases, but that the RPS, ${\bf R}$, and $\sqrt{3}\times\sqrt{3}$
states (or more precisely all the states degenerate with the $\sqrt{3}\times\sqrt{3}$
states) remain very reasonable candidate ground states for these values
of $s$ in the isotropic limit.

\section{Discussion}
\label{sec:discussion}

Since the development of the DPT and its analysis in this paper is
rather involved, we begin in the first subsection by recapitulating the
central points.  In the second subsection we will then turn to a brief
discussion of the implications on experiments and future directions of
this work.

\subsection{Summary}
\label{sec:summary}

As a prototypical model of a magnetization plateau in a strongly
frustrated quantum antiferromagnet, we considered in this paper a
nearest-neighbor spin-$s$ model on the pyrochlore lattice at half the
saturation magnetization.  Such plateaus have been observed in the
spinel materials \hgaf\;, and \cdaf.  We argued that a useful starting
model is the easy-axis XXZ Heisenberg model in an external field,
Eq.~\eqref{XXZ}.  This model possesses all same symmetries as the
isotropic Heisenberg model in an external field, and indeed we were able
to extrapolate our results to this limit. This model 
has the advantage that the transverse spin fluctuations can be 
treated systematically as a perturbation
to the underlying Ising model.  The resulting Ising model can be
rewritten as a sum over the elementary tetrahedra of the pyrochlore
lattice.  In this Ising limit on the plateau, the spins on each
tetrahedron satisfy a 3:1 constraint, comprising a set of 3 majority
spins fully polarized parallel to the field, and 1 minority spin
antiparallel to the field.  The half-polarized state has a macroscopic
degeneracy corresponding to the number of possible positions for all the
down pointing spins in the lattice.  It is expected that the transverse
spin fluctuations will play
a role in selecting a ground state or set
of ground states from the massively degenerate 3:1 manifold.  
In this way, we are lead to a theoretical model involving a ``constrained''
degenerate perturbation theory in the 3:1 manifold.  Our paper is
devoted to a detailed analysis of such a theory and many parts are
couched in sufficiently general
language to be applicable to a broad class of
systems.

We began our discussion of the constrained easy-axis degenerate
perturbation theory by deriving the general structure of the effective
Hamiltonian that occurs at each order of dimensionless coupling
$\alpha=J_\perp/J_z$, Eq.~\eqref{eq:7}.  We found that the effective
Hamiltonian could be cast into a convenient form by performing a unitary
transformation that rotates all down pointing (minority) spins to up 
pointing spins and also by introducing a connectivity matrix (whose elements are one
for nearest neighbor spins and zero otherwise). 
The latter makes it possible to convert the sums over
nearest-neighbor lattice sites to sums over the entire lattice
\eqref{rotated_H_1}.  These transformations 
cast the terms of the
effective Hamiltonian coming from each order of perturbation theory into
a form rather convenient for analysis. The 
resulting terms are expressed
explicitly in terms of the Ising variables on the lattice sites, the
spin $s$, and the connectivity matrix.  These terms were studied
order-by-order in perturbation theory.  We found that diagrams 
representing these terms naturally fell into two categories:
contractible and non-contractible.  Contractible diagrams are those 
whose dependence on some of the Ising variables 
is eliminated by summing with respect to their site index
over all lattice points.  
Thus, a function of $N$ Ising variables can be reduced to a
function of less than $N$ 
Ising variables after this 
``contraction'' process.  The allowed contractions depend on 
the lattice geometry, the 3:1 constraint, and the Ising nature of the spin variables.  
Diagrams for which it is not possible to perform a contraction (equivalently, a reduction in the
number of relevant Ising variables) we termed non-contractible.

The central result of the analysis of contractible and non-contractible
diagrams is that all terms in the constrained degenerate perturbation
theory up to and including 5th order are constant {\it within the 3:1 manifold}.
Individual terms are shown to be constant by first contracting the
diagrams as much as possible and then noting that the value of the
diagram is unchanged under permutation among site indices associated with 
the Ising variables. The latter statement implies that the value of the diagram 
is independent of spin configurations allowed in the 3:1 manifold
and hence a constant.  
In a similar manner, most terms at 6th order are also shown to be
constant. However, we also observe that, at 6th order, there appears a ``single large loop'' diagram 
which cannot be contracted, and also defies the permutation arguments mentioned above. 
In fact, this loop diagram brings about
non-constant contributions to the effective Hamiltonian in the 3:1 manifold. 
Therefore, this is the lowest order term which 
lifts the degeneracy of the 3:1 manifold (at least for $s > 1$). The 
non-constant 6th order term includes effective interactions among spins on each
hexagonal plaquette of the pyrochlore lattice.
Depending on the arrangement of minority sites, there
are five distinct kinds of plaquettes that may appear 
and we label them 0,1...4 (See Table~\ref{table1}). Using the results of our degenerate
perturbation theory, we evaluate the energy of each of these plaquettes
as a function of $\alpha$ and $s$, and correct a mistake in
Ref.~\onlinecite{Bergman:prl05}.  The 3:1 condition constrains the
allowed ratios of the various plaquettes in the lattice and allows us to
express the total energy of the system (up to an overall constant),
Eq.~\eqref{diagonal_energy}, in terms of only 3 energies \eqref{eq:ginv}.

As a check on the results immediately above and as a further test of the
robustness of those results, we also performed a large-$s$ expansion in
the easy-axis limit.  As with the fully quantum theory, we {expanded 
the harmonic spin wave energy
in powers of $\alpha$ up to the 6th order, applied the diagrammatic
analysis above involving contractible and non-contractible diagrams, and
studied the resulting energy of the non-constant 6th order terms.  
The result \eqref{large_S_ergs} agrees exactly with the 
${\cal O}(s)$ term obtained from the quantum degenerate perturbation theory, 
\eqref{Infty_S_lim_DPT}. This satisfying consistency tells us that 
the large-$s$ limit and small-$\alpha$ limit commute, and thus 
our analysis is likely well controlled.

In the final section of the paper we used the results of the degenerate
perturbation theory to determine the low energy states on the plateau as
a function of $s$.  Our result that the first non-constant diagonal term
in perturbation theory comes at 6th order is independent of the spin
value $s$.  However, terms that allow plaquettes (such as type 1) to
resonate occur at order $6s$, which can be either larger or smaller than
6 depending on $s$.  In the strict easy-axis limit, therefore, for $s\geq 3/2$, 
the low energy states are therefore determined only by a diagonal
effective Hamiltonian, which can be analyzed classically.  In the large 
but finite $s$ limit, we are able
to resolve the degeneracy 
of the ``zero flux'' manifold 
found in the large-$s$ analysis (extended from that of Hizi and Henley\cite{Hizi:prb06} to the XXZ model).  
We predict a ``trigonal$_{7}$'' state (see Fig.~\ref{fig:trig7}) 
to be the exact ground state in this easy-axis limit and for large $s$, 
and numerical analysis suggests this obtains for $s\geq 2$.  
For $s=3/2$, the lowest  energy configuration we have found in the Ising 
limit is a massively degenerate set of states (for example the $\sqrt{3}\times\sqrt{3}$ states, see Fig.\ref{fig:root3_state3}). For $s\leq 1$, and for $s=3/2$ extrapolated to the isotropic limit, we find
that the off-diagonal term in the effective Hamiltonian becomes
significant, and we suggest several likely candidates for the ground states in these 
cases. This includes a possible $U(1)$ spin liquid state, which would be quite
remarkable if realized.

\subsection{Implications and future directions}
\label{sec:impl-future-direct}

First let us comment briefly upon the relevance to the spinel chromites.
For \hgaf, it is known that the temperature at which the plateau forms
($\approx 7^\circ K$) is comparable to the highest temperature at which
magnetic order is observed.  The theoretical estimate of the magnitude
of the couplings in the effective Hamiltonian due to quantum
fluctuations for $s=3/2$ is however small, e.g. $V_1 \approx 0.02J$ from
Eq.\eqref{eq:ginv}.  Thus the temperature at which quantum fluctuations
are expected to induce magnetic ordering would be very low.  A crude
estimate based on the measured Curie-Weiss temperature in
\hgaf\cite{Ueda:prb06} would 
predict an ordering temperature $\lesssim 0.2K$.  This strong quantitative
disagreement with experiment indicates that a stronger classical
mechanism -- i.e. physics outside the Heisenberg model -- must be behind
the plateau formation.  Indeed, a recent study of a simple model of
spin-lattice coupling gives a reasonable explanation of the plateau and
its order, predicting stabilization of the ${\bf R}$
state.\cite{Bergman:prb06}   It would be
quite interesting to see whether quantum fluctuations might however play
a role in the other chromite spinels, e.g. \cdaf.

We now move away from the experiments on \hgaf, where the pure
nearest-neighbor Heisenberg antiferromagnet neglecting spin-lattice
interactions is clearly inadequate.  Instead, we would like to address a
basic question that may be in the mind of the reader.  In the pure
spin-$s$ isotropic Heisenberg model (i.e. $J_\perp=J_z=J$), is there a
plateau at half-magnetization?  At $s=\infty$, i.e. the strict classical
limit, the answer is {\sl no}, and indeed the magnetization is a simple
linear function of field in this case.  In principle this question can
be addressed by the $1/s$ expansion.  However, to the order studied, the
situation remains unclear: the leading-order spin-wave spectrum remains
gapless even in a field.  Higher-order calculations in $1/s$ are
required to resolve this question via that approach.  Within the XXZ
model, for any amount of anisotropy ($\alpha<1$), a plateau is expected
even in the classical limit, so by continuity it is likely to persist at
smaller $s$.  However, the extrapolation to $\alpha=1$ is not clear.  In
Appendix~\ref{app:PlateauWidth}, we present some simple calculations
aimed at addressing the plateau width.  In particular, we show that the
plateau narrows both from above and below upon perturbing away from the
Ising limit, where it is maximal.  The plateau edges are determined by
the points at which the gap to excitations with non-zero $S^z$ vanishes.
Unfortunately, unlike the calculation of the splitting {\sl within} the
plateau states (the main focus of this paper), the energy difference
{\sl between} the plateau ground state and excited states with
higher/lower $S^z$ is non-vanishing already at quadratic/linear order in
$\alpha$.  Hence, a high-order calculation of this gap becomes much more
involved than those in the bulk of this paper, and an extrapolation to
the isotropic limit is probably not reliable.  The existence of a
plateau in the isotropic limit is a subject worthy of study by other
methods.

Next we turn to future applications of the formalism developed here to
other problems.  From our exposition, it should be evident that our
methods generalize rather straightforwardly to other models of quantum
antiferromagnets with Ising anisotropy, provided a few 
conditions hold.
First, the lattices should be composed of site-sharing simplexes.  A
simplex is a collection of sites in which every pair of sites is
connected by a bond; examples include triangle, crossed square,
and tetrahedron.  
Second, the ground states of the Ising part of the Hamiltonian on a single
simplex should all be permutations of one another.  This allows Ising
exchange, single-site anisotropies, biquadratic and other interactions.
Third, the interactions should be the same on each bond, but could
include quite arbitrary combinations of exchange, biquadratic couplings
etc.  There are quite a number of
interesting models of frustrated magnetism which share these features.
For instance, the XXZ models on the Kagome and checkerboard lattices can
be studied this way at several values of the magnetization.  The XXZ
model on the pyrochlore lattice at zero field is also such a system.
It will be interesting to explore the behavior of these models at
various values of $s$ using the methods of this paper.  

More generally, the methods of this paper are possible because of a key
simplification: in a strong magnetic field, the symmetry of the spin
Hamiltonian is 
$U(1)$ rather than $SU(2)$.  Many
more theoretical methods are available to treat systems with {\sl
  abelian} 
conserved charges than for $SU(2)$-invariant spin
models.  Furthermore, in the interesting search for spin-liquid states
of quantum antiferromagnets, much theoretical success has been achieved
in recent years in realizing such states in $U(1)$-symmetric models,
while examples of $SU(2)$-invariant spin liquids, even in models, are
much more limited.  Therefore it seems likely that quantized
magnetization plateaux may be an excellent hunting ground for such
exotic states of matter, and moreover there is hope for theory and
experiment to meet on this plain.

\acknowledgments

This work was supported by NSF Grant DMR04-57440, PHY99-07949, and the
Packard Foundation. R.S. is supported by JSPS as a Postdoctoral Fellow.

\appendix

\section{Edges of the magnetization plateau}
\label{app:PlateauWidth}

In this appendix we analyze the gap to spin excitations above the 3:1
plateau.  To lowest order in the transverse spin fluctuations, in the
field range corresponding to the plateau in the Ising model, excitations
carrying $\Delta S^z=\pm 1$ are are separated by an energy gap of order
$J_z$ from the manifold of half-polarized states.  As the magnetic field
is varied this gap decreases.  At zero temperature, the magnetization
plateau ends when the energy gap to one of these spinful excitations
vanishes.  We shall consider the spin $\Delta S = \pm 1$ ``single
particle'' excitations above the 3:1 manifold only, and find a rough
perturbative estimate for the limits of the half magnetization plateau
region.  Clearly, the plateau region cannot extend beyond the magnetic
field values at which the gap to these excitations vanishes, and so can
only be more narrow than the extent we shall find here.

Consider $\Delta S = +1$ excitations above the ground state.  These
excitations are more favorable in higher magnetic fields.  The simplest
way to insert such an excitation is by raising one minority spin by one
unit, i.e. obtained by acting with $S_j^+$ on a minority site to change
$S_j^z = -s$ to $S_j^z = 1-s$.  This creates two ``defective''
tetrahedra (shared by this site), which no longer have the optimal 3:1
spin configuration.  We will call the space of such states ``manifold
A''.  For $s>1/2$, manifold A comprises all the {\sl lowest energy}
$\Delta S=+1$ excitations in the Ising Hamiltonian.  Other $\Delta S=+1$
states which involve more spin raising/lowering operators, e.g. those
created by $S_i^+ S_j^+ S_k^-$, have higher energy because they either
create more defect tetrahedra or make the two defect tetrahedra more
energetically costly.  The exception is $s=1/2$, for which manifold A is
actually incomplete, and there are other $\Delta S=+1$ excitations
outside it with the same $0^{\rm th}$ order energy.  We will henceforth
assume $s>1/2$ in the remainder of this appendix to avoid this
complication.

The $0^{\rm th}$ order energy difference from the ground state manifold
is \be E_A^0 - E_{3:1}^0 = 2 J_z (3 s - h) \; .  \ee The gap shrinks as
$h$ increases, and vanishes at a magnetic field $h_A = 3 s$.  This
corresponds to the high field edge of the half polarized plateau in the
$\alpha = 0$ limit.  The single site excitation is however highly
degenerate.  The excitation can reside on any one of $\frac{N}{4}$
minority sites ($N$ is the number of pyrochlore sites).  We expect
breaking this massive degeneracy (i.e.  having this magnon ``flat band''
acquire some dispersion) will lower the energy of this excitation
(closing the gap at lower magnetic field). One therefore needs to take
into account corrections from higher orders in DPT.

To first order in $\alpha$, only spin $\Delta S_i^z=\pm 1$ can be
transferred from one site to a nearest neighbor. In any 3:1
configuration, the nearest minority spins reside on next nearest
neighboring sites \emph{in the lattice sense}, which means these sites
are a distance of two \emph{links} apart.  Therefore this process always
leaves manifold A, and the first order term has no contribution.

To 2nd order in $\alpha$ there are various viable processes. Apart from
hopping spin 1 between neighboring sites, and then back, there is also a
non-trivial process in which the defect site (the $-s+1$ spin) swaps with one
of the (next-) nearest neighbor minority spins. This is also a
legitimate member of the A manifold. Processes not involving the defect
directly are affected by the presence of the defect only when involving
the nearest neighbors of the defect site. All other sites have the same
contribution to the 2nd order correction as in the ground state
manifold.

We immediately conclude the form of the effective ``magnon Hamiltonian''
to 2nd order is
\be
{\mathcal H}_{A}^2 - {\mathcal H}_{3:1}^2 = - \left[
c_1 + c_2 \sum_{\langle \langle i j \rangle \rangle} \left(
  e^{\dagger}_i e_j + h.c. \right) 
\right]
\; ,
\ee
where $\langle \langle i j \rangle \rangle$ denotes next nearest
neighbor sites on the pyrochlore lattice  
that are both minority sites, $c_{1,2} > 0$
are coefficients depending on the physical couplings, and the operator
$e^{\dagger}_j$ creates a local 
excitation -- it replaces a spin $S_j^z = -s$ with a spin $S_j^z = 1-s$
on site $j$:
\be
e^{\dagger}_j = \ket{-s}_j \bra{1-s}_j\;.
\ee

An arduous yet straightforward calculation for the A manifold results in
\be\label{H_A_2}
\begin{split} &
{\mathcal H}_A^2 - {\mathcal H}_{3:1}^2 = 
\\ &
- \frac{J_z \alpha^2}{2s-1} \Bigg\{
\frac{3 s \left(4 s^2-3 s+1\right)}{4 s-1} + 
\frac{s^2}{2 } \
\sum_{\langle \langle i j \rangle \rangle} \left( e^{\dagger}_i e_j + h.c. \right)
\Big]
\Bigg\}
\; .
\end{split}
\ee
Note the singularity in Eq.~\eqref{H_A_2} for $s=1/2$, which reflects
the incompleteness of manifold A in this case.

In a hopping Hamiltonian of the form of  Eq.~\eqref{H_A_2}, the magnon
eigenstates are delocalized Bloch states.  The spectrum of these states
in general depends in detail upon the structure of the lattice of
minority sites.  However, if we are interested in the minimum energy
state only, there is an amusing simplification.  Since the hopping above
is everywhere negative, the lowest energy state is expected to be
nodeless,  Thus the minimum energy state is simply a {\sl constant
  amplitude} (i.e. zero quasimomentum) Bloch state.   For any
configuration of the minority spins 
$\ket{\Psi_0}$ in the ground state manifold, the corresponding state in
the A manifold is
\be
\ket{\Psi} = \sum_j e^{\dagger}_j \ket{\Psi_0}
\; .
\ee
Direct calculation yields
\be
\sum_{\langle \langle i j \rangle \rangle} \left( e^{\dagger}_i e_j + h.c. \right) \ket{\Psi}
 = 6 \ket{\Psi}
\; ,
\ee
where $6$ comes about as the number of next nearest neighbors with spin
$S_j^z = -s$ the excitation can hop over to.  Because  in \emph{any} 3:1
configuration there are always precisely $6$ minority spins two links
away, we are therefore able to obtain the lowest-energy $\Delta S^z=+1$
magnon energy \emph{irrespective of the particular arrangement of the
  minority sites}!  

The minimum energy of \eqref{H_A_2} is therefore
\be
E_{A,min}^2 - E_{3:1}^2= 
- J_z \alpha^2 \frac{3 s \left(8 s^2-4 s+1\right)}{(2 s-1) (4 s-1)}
\; .
\ee

The combined gap to manifold A is now
\be
\Delta E_{A} = 
J_z 2 (3 s - h) - J_z \alpha^2 \frac{3 s \left(8 s^2-4 s+1\right)}{(2 s-1) (4 s-1)}
\; .
\ee
The gap vanishes at a magnetic field of
\be
h_A = 3 s -\alpha^2 \frac{3 s \left(8 s^2-4 s+1\right)}{2 (2 s-1) (4
  s-1)} 
\; .
\ee
Note that for $s \gg 1$, the $O(\alpha^2)$ correction is well-behaved,
i.e. it scales in the same way as the zeroth order threshold field.

Excitations with $\Delta S = -1$ (manifold B) can be realized by
replacing a spin $S_j^z = +s$  
with a spin $S_j^z = s-1$. The 0th order energy difference from the
ground state manifold is 
calculated in an identical manner as in the $\Delta S = +1$ case, and
yields 
\be
E_B^0 - E_{3:1}^0 = 2 J_z (h - s)
\; .
\ee
At this order, the $\Delta S = \pm 1$ excitation spectra are symmetric
about $h = 2 s$. 

Oppositely from the behavior at the high field edge of the plateau, the
gap shrinks as $h$ {\sl decreases}, and vanishes at a 
magnetic field $h_B = s$. Once again, due to a huge degeneracy ( the defect can reside on any one of
$\frac{3 N}{4}$ majority sites) we must
resort to DPT in order to break the massive degeneracy of manifold B,
expecting to lower the excitation energy. 

In contrast to the $\Delta S^z=+1$ excitations above, it is already
possible at
first order to hop the spin $S^z = s-1$ onto other sites and stay
within the 
B manifold.  This is because the majority sites are not isolated ($4$ of
the $6$ neighboring sites of a majority site are also majority sites).
Therefore, for any given 3:1 state, one can immediately obtain the
effective Hamiltonian to $O(\alpha)$:
\be
{\mathcal H}_B^1 = J_z s \alpha
\sum_{i j} W_{ij} h^{\dagger}_i h_j 
\; .\label{eq:hopminus}
\ee
Here $W_{ij}$ is the connectivity matrix of the lattice of majority
sites for the particular 3:1 state in consideration, and  the operator
$h^{\dagger}_j$ creates a local 
excitation -- it replaces a spin $S_j^z = s$ with spin $S_j^z = s-1$ on
site $j$:
\be
h^{\dagger}_j = \ket{s-1}_j \bra{s}_j
\; .
\ee

Since $\alpha > 0$, the
hopping amplitudes in this case are positive, rather than negative as
above.  Thus, unfortunately, the lowest-energy eigenstate is a
non-trivial Bloch state, whose energy depends upon the precise form of
$W_{ij}$, i.e. it is different for each of the 3:1 states.  Therefore it
is difficult to say anything specific about the $O(\alpha)$ correction
to the minimum $\Delta S^z=-1$ magnon energy.  Because one can easily
construct variational states with negative hopping energy (e.g. an
antibonding state with support only on two sites), the $O(\alpha)$
correction must be negative, and clearly from Eq.~\eqref{eq:hopminus} is
of the form $E_B^1 = - J_z s \alpha c$, with $c>0$ a dimensionless
constant (the largest eigenvalue of  $W_{ij}$).  This gives
\be
h_B  = s(1 + \alpha \, c)  > s.
\ee

Thus the upper edge of the plateau decreases and the lower edge of the
plateau increases as $\alpha$ is increased, narrowing the plateau with
increasing quantum fluctuations.  It is plausible that for sufficiently
large $\alpha$ the plateau is obliterated entirely, the upper and lower
edge meeting.  Unfortunately, this is beyond the realm of this
perturbative analysis.  Even a na\"ive extrapolation of the above
lowest-order results is non-trivial, since the lower plateau edge
depends upon the non-trivial constant $c$.  It can in principle be
computed for the various states obtained in Sec.\ref{diagonal_gs}, but
we do not do so here.


\section{Alternative calculation of the degenerate perturbation theory}
\label{app:Other_DPT}

In this appendix we present an alternative way of calculating the DPT 6th order diagonal term,
providing a check on the calculation described in Section~\ref{sec:DPT_Leon}.


The perturbation \eqref{H_1} can be viewed as a sum over spin transfer operators
\be
{\mathcal H}_1 = \frac{\alpha}{2} J_z \sum_{\langle i,j \rangle} \left(
{\hat S}_i^+ {\hat S}_j^- + h.c.
\right) =
\frac{\alpha}{2} J_z \sum_{\ell = \langle i,j \rangle} \left( {\hat h}_{\ell} + h.c. \right)
\; ,
\ee
where ${\hat h}_{\ell}$ transfers one quantum of spin angular momentum from site $i$ to site $j$ if these are on adjacent 
sites $i,j$.
As a result, the $n$-th order virtual processes can be classified by choosing $n$ links to act on with
spin transfer operators, in a particular order. In order to return to the initial 3:1 configuration,
the links must form closed loops of spin
transfer. These include self retracting loops, defined as loops where after following some path in the lattice,
we turn back and travel the same path in the reverse order back to the origin.
The chosen links can be represented by a graph on the lattice,
by coloring these links. Apart from the graph, we must also specify the order by which the links operate.

The virtual states in a DPT process will have a number of sites in a different spin state 
relative to the initial 3:1 state. We denote $S_j^z = \sigma_j (s - m_j)$,  
where the variables $m_j$ are non-negative,
and take on any integer value between $0$ and $2s$. 
Let us denote the set of modified 
sites by $M$. Thus $m_j \neq 0$ only if $j \in M$. 
Using the fact that $\sum_j \sigma_i = \frac{1}{2} \sum_j 1$
and $\sum_{\langle i j \rangle} \sigma_i \sigma_j = 0$, both due to the 3:1 constraint,
we find the inverse resolvent can be rewritten as a sum involving \emph{only} the modified sites
\be
\begin{split}
{\mathcal H}_0 - E_0 & = J_z \sum_{\langle i j \rangle} S_i^z S_j^z - E_0'
\\ & =
J_z \sum_{\langle i j \rangle} \sigma_i \sigma_j
(s - m_i)(s - m_j) - E_0'
\\ & =
J_z \left( \sum_{\langle i j \rangle \in M} \sigma_i \sigma_j m_i m_j + 2 s \sum_{j \in M} m_j \right)
- E_0''
\; ,
\end{split}
\ee
where we have absorbed all constant energies into $E_0''$. With all $m_j = 0$, this resulting energy should
vanish (energy difference to a state in the 3:1 manifold) and therefore conclude that the energy $E_0'' = 0$
in the final expression above. Finally, our virtual state energy is
\be\label{virtualE}
{\mathcal H}_0 - E_0 =
J_z \left( \sum_{\langle i j \rangle \in M} \sigma_i \sigma_j m_i m_j + 2 s \sum_{j \in M} m_j \right)
\; .
\ee

At 1st order, the perturbation \eqref{H_1} takes any initial state out of the 3:1 manifold, and therefore
gives no contribution (equivalently no loop can be closed with only one link).

In 2nd order, we must act on the same link twice to undo the spin transfer, and return to the 3:1 manifold.
Therefore all non--vanishing processes are confined to one tetrahedron. The same process can act on any one
of the links of this tetrahedron, \emph{with the same resolvents for a given order of the link operators}. 
Therefore, by summing up these diagrams, we will get a function of the spins
on this tetrahedron which is invariant under any permutation of the $4$ sites
(the sum is represented in a diagrammatic way drawn in Fig.~\ref{fig:dpt1}). 
Therefore the 3:1 configuration on any single tetrahedron 
has the same energy at this order in DPT, no matter which corner is chosen to be the minority site. 
For this reason, the 2nd order contribution gives a constant shift in energy to all the 3:1 states.

\begin{figure}
\centering
\includegraphics[width=2.0in]{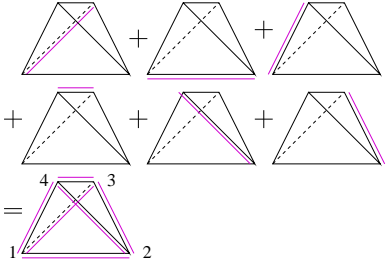}
\caption{(Color online) Second order process.}
\label{fig:dpt1}
\end{figure}

In 3rd order, the only way one can return to the 3:1 manifold is by forming a single closed loop of spin transfer.
around  a triangular side of a single tetrahedron. As in the 2nd order processes, we can sum
over all such processes occurring on the single tetrahedron.
One again we end up with a function that is invariant under any permutation of the $4$ sites
of the single tetrahedron (the sum is represented in a diagrammatic way drawn in Fig.~\ref{fig:dpt2}).
Once again, we cannot distinguish energetically between the different 3:1 states 
defined on this single tetrahedron. 
As a result, the 3rd order term must produce a constant shift in energy. 

\begin{figure}
\centering
\includegraphics[width=2.0in]{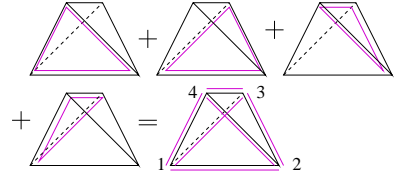}
\caption{(Color online) Third order process.}
\label{fig:dpt2}
\end{figure}

As exemplified in the 2nd and 3rd order terms, all DPT processes (of any order) that are confined to one tetrahedron
(meaning all the spin transfer operators act on links in the same tetrahedron) all add up to 
constant shifts in energy.
In every such case the argument is the same as above.
Starting with a given process defined on a single tetrahedron,
sum over all the processes of the \emph{same structure} on the same tetrahedron.
Then one always arrives at a function that is invariant under any permutation of the $4$ corners. 
Within the 3:1 manifold, these processes cannot favor one configuration over the other, and must produce a mere  
shift in the overall energy. More explicitly, every process results in a function of the four
Ising variables on a tetrahedron, as defined in Section~\ref{easy_axis}. 
The sum of all diagrams with same structure  
then takes on the generic form 
\be 
\begin{split}
  \label{B4}
& f(\sigma_1, \sigma_2, \sigma_3, \sigma_4) + \left( {\textrm{all permutations of 1...4}} \right) \\
& = a_0 + a_1 \left( \sigma_1 + \sigma_2 + \sigma_3 + \sigma_4 \right) \\
& + a_2 \left(
\sigma_1 \sigma_2 + \sigma_1 \sigma_3 + \sigma_1 \sigma_4 +
\sigma_2 \sigma_3 + \sigma_2 \sigma_4 + \sigma_3 \sigma_4
\right) \\
& + a_3 \sigma_1 \sigma_2 \sigma_3 \sigma_4 \left( \sigma_1 + \sigma_2 + \sigma_3 + \sigma_4 \right)
+a_4 \sigma_1 \sigma_2 \sigma_3 \sigma_4
\; .
\end{split}
\ee
The 3:1 constraint gives $\sigma_1 \sigma_2 \sigma_3 \sigma_4 = -1$
as well as $\left( \sigma_1 + \sigma_2 + \sigma_3 + \sigma_4 \right) = 2$.
The generic 4 spin function then reduces to
\be
\begin{split}
  \label{B5}
  f(\sigma_1, \sigma_2, \sigma_3, \sigma_4) &
= const + a_2 \sum_{i < j = 1}^4 \sigma_i \sigma_j \\
& = const + \frac{a_2}{2} \left( \sum_{j = 1}^4 \sigma_j \right)^2 = const
\; .
\end{split}
\ee
Given this result, we can always ignore virtual processes 
confined to one tetrahedron for the remainder of our discussion.

Similar arguments can be applied to processes confined to $2$ adjacent tetrahedra. 
Summing over all possible processes confined to pairs of 
adjacent tetrahedra, one always ends up with a constant energy shift in the 3:1 manifold. 
A pair of adjacent tetrahedra can only be in one of 3 configurations in the 3:1 manifold 
(see Fig.~\ref{fig:configs}). The adjacent tetrahedron pairs can be enumerated by the site 
connecting them. Therefore the number of distinct pairs is equal to the number of pyrochlore sites. 
In the 3:1 manifold, the number of configurations 
where the shared site is a minority site (see Fig.~\ref{fig:conf1})
is \emph{fixed} to be $\frac{1}{4}$ of the sites. In the remaining $\frac{3}{4}$
of the sites, the shared site is a majority site.
The two configurations in Fig.~\ref{fig:conf1} with a majority site connecting the pair of 
tetrahedra are distinguishable physically
because of the different directions of the links, but  
in terms of the diagrams defining DPT processes Fig.~\ref{fig:conf2} and Fig.~\ref{fig:conf3}
are \emph{indistinguishable} (they are identical in the graph sense - the link structure is the same). 
Therefore, the summation over all possible processes confined to these two adjacent tetrahedra always results  
in a function of the Ising variable defined on the shared site
\begin{eqnarray}
f(\sigma_{i})=a_{0} + a_{1}\sigma_{i}. \label{B6}
\end{eqnarray}

For each pair of adjacent tetrahedra, we have the same function \eqref{B6} of  
the shared site Ising variable. Summing over all pairs of adjacent tetrahedra, is equivalent
to summing over the sites of the pyrochlore lattice. Therefore, the sum of all the DPT processes
of this sort results in
\be
\sum_{i} f(\sigma_{i})=N a_{0} + a_{1}\sum_{i}\sigma_{i} = Na_{0} + \frac{N}{2}a_{1}, \label{B7}
\ee
where $N$ denotes the total number of pyrochlore lattice points.
Once again, we are led to the conclusion that all such processes can only give a constant shift in energy,
and we shall ignore all such instances in the remainder of our discussion.

\begin{figure}
\centering
\includegraphics[width=2.5in]{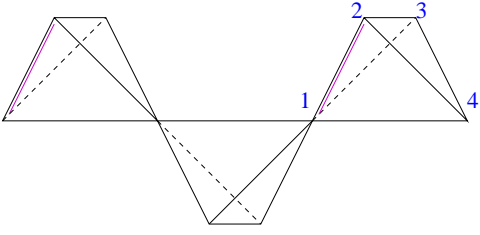}
\caption{(Color online) 4th order process.}
\label{fig:dpt4}
\end{figure}

At 4th order we have a number of possible processes. The only spin transfer processes
comprising a single loop will be confined to 2 adjacent tetrahedra.
According to the arguments above, this single loop process will 
give a constant shift in energy, within the 3:1 manifold.
Apart from single closed loop, we can only also have processes with 2 separate closed loops,
comprising two non-overlapping links which get acted on twice.
Each chosen link occupies its own tetrahedron (if the two links are on the same tetrahedron, one \emph{must}
act on a bond with two majority spins, and automatically produces zero). 
The case in which these 2 tetrahedra are adjacent, will result in a constant, by our arguments above.
Otherwise, there are only two possibility. In one case
these two tetrahedra share a neighboring tetrahedron. In the 2nd case, 
they are separated by more than one tetrahedra. This  
classification is necessary because, in the former case, different resolvents may
show up through the interaction term of the virtual state energy~\eqref{virtualE}.  
If the two links are separated by more than one tetrahedron, this does not occur.

In the case where the two tetrahedra are sufficiently well separated, with no mutual interaction in the virtual state energy,
we can once again invoke the trick of summing over all such processes on each single tetrahedron containing a link,
and the DPT term cannot distinguish between the four 3:1 configurations on each tetrahedron. Once again we end up
with a constant shift in the overall energy for all the 3:1 configurations.

Now we consider the processes where the two tetrahedra share a neighboring tetrahedron. 
One such diagram is pictured in
Fig.~\ref{fig:dpt4}. The structure spans a chain of three tetrahedra. In the present case,
when summing over diagrams one must be cautious to sum over diagrams with the same resolvents.
Summing over equivalent
diagrams on the chain of three tetrahedra, such as those found by permuting site $2$ with $3$ and $4$,
yields a function that treats the 3 sites at the outer side of each
edge tetrahedron on equal footing. We graphically indicate the sum of diagrams, by coloring all the links
that appeared in one of the diagrams we summed over (see Fig.~\ref{fig:dpt5}).
\begin{figure}
\centering
\includegraphics[width=2.5in]{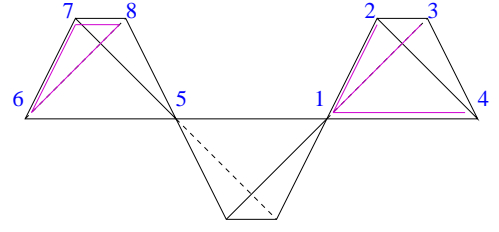}
\caption{(Color online) Sum of 4th order processes.}
\label{fig:dpt5}
\end{figure}

Formally, the sum of processes gives rise to 
a function of the 8 Ising variables on the two edge tetrahedra
$
{\tilde f}(\sigma_1, ... ,\sigma_8)
$.
Because of the above arguments, this
function has to
be invariant under permutations of the 3 outer edge spins. Considering all possible expressions 
we can form out of the 3 edge spins in a ``symmetric'' way
\be
\begin{split}
f(\sigma_1,\ldots,\sigma_8) & = a_0 + a_1 (\sigma_2 + \sigma_3 + \sigma_4) \\
& + a_2 (\sigma_2 \sigma_3 + \sigma_2 \sigma_4 + \sigma_3 \sigma_4)
+ a_3 \sigma_2 \sigma_3 \sigma_4,
\end{split}
\ee
where the dependences of $a_{0,1,2,3}$ on $\sigma_{1,5,6,7,8}$ are implicit. 
Using the 3:1 constraint on this 
tetrahedron $(\sigma_1 + \sigma_2 + \sigma_3 + \sigma_4) = 2$, as well as the 
identities it implies $\sigma_1 \sigma_2 \sigma_3 \sigma_4 = -1$ (or $\sigma_2 \sigma_3 \sigma_4 = -\sigma_1 $) 
and $(\sigma_2 \sigma_3 + \sigma_2 \sigma_4 + \sigma_3 \sigma_4) = 1-2 \sigma_1$,  
we find that $ f(\sigma_1,\ldots,\sigma_8)$ can be rewritten as a function of $\sigma_{1,5,6,7,8} $ alone; 
\be
\begin{split}
f(\sigma_1,\ldots,\sigma_8) & = a_0 + a_1 (2 - \sigma_1) + a_2 (1-2 \sigma_1)
- a_3 \sigma_1 \\ &
= {\tilde a}_{0} + {\tilde a}_1 \, \sigma_1. 
\end{split}
\ee
Repeating the same simplification for the three Ising variables of the other edge, $\sigma_{6,7,8}$,
one finds that the sum of diagrams above can only produce a function depending on the 2 spins connecting the edge 
tetrahedra to the third tetrahedron in the middle of the chain (sites $1$ and $5$ in Fig.~\ref{fig:dpt5}). 
This function is therefore a function of 2 neighboring spins on a single tetrahedron
${\tilde f}(\sigma_1,\sigma_5)$. Each three tetrahedron chain process can be represented by the pyrochlore link
between the two sites shared between the tetrahedra ( sites $1$ and $5$ in Fig.~\ref{fig:dpt5}).
Summing over all the processes represented by the 6 links in the central tetrahedron which contains sites $1$ and $5$, 
one ends up with a function which is symmetric under an permutation of the 4 sites
on this tetrahedron. Repeating the steps outlined in Eqs.~\eqref{B4} and ~\eqref{B5}, we find the
resulting function is constant.


\begin{figure}
\centering
\includegraphics[width=3.0in]{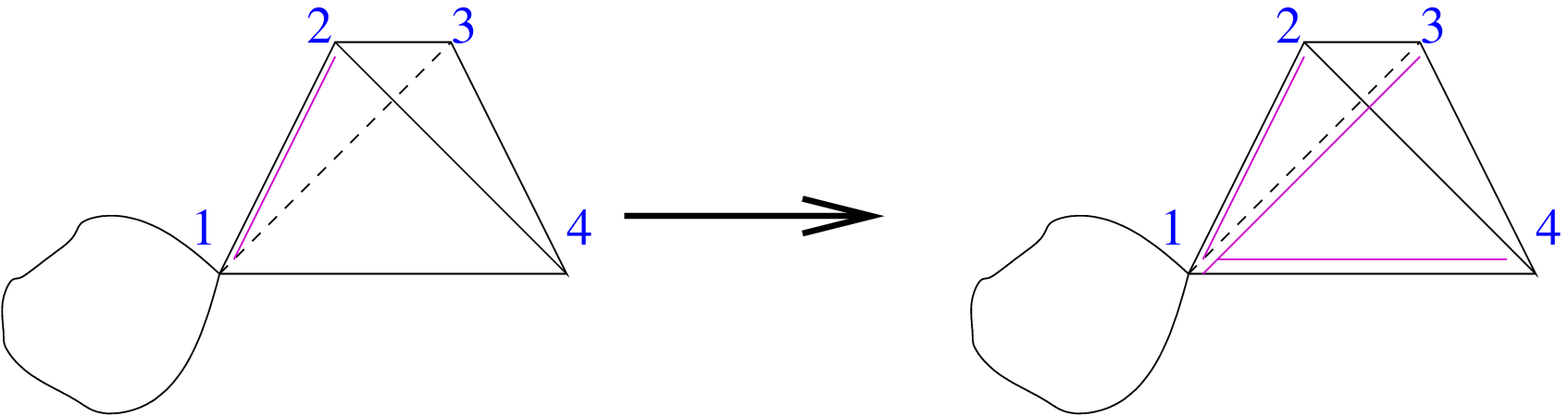}
\caption{(Color online) Summing over equivalent edges of a diagram
to contract the dependence on sites in the cluster.}
\label{fig:dpt10}
\end{figure}

The particular case analyzed above is only an example of a vastly more general case, which we now describe.
Given a set of diagrams that reside on chains of tetrahedra, with the two edge tetrahedra \emph{not} sharing a
neighboring tetrahedron between them in any one of the diagrams (except through the chain itself), 
one can sum over sites at the edge that we are free to permute
(see examples in Fig.~\ref{fig:dpt10} and Fig.~\ref{fig:dpt11}). The summation in both examples will result
in a function of the spins at site $1$. This procedure of {\it contracting} the diagram can be continued from both edges, and
we will always end up with a function of 2 sites residing on one tetrahedron. We then sum over all the different choices
of the last 2 remaining sites on this tetrahedron. We then
end up with a symmetric function of the 4 sites on a single tetrahedron, which is
always a constant, from Eqs.~\eqref{B4} and \eqref{B5}. 
Therefore, we conclude that all such diagrams can be summed over to give constant shifts in energy.
We shall refer to these sets of diagrams as ``retractable chains''.

\begin{figure}
\centering
\includegraphics[width=3.0in]{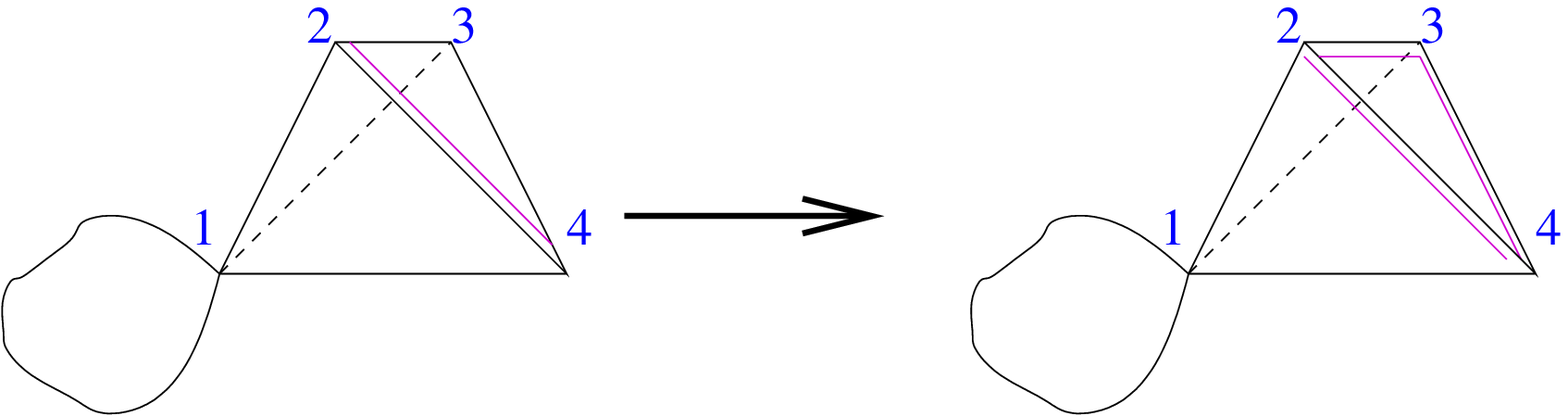}
\caption{(Color online) Summing over equivalent edges of a diagram to contract the dependence on sites in the cluster.}
\label{fig:dpt11}
\end{figure}


Bearing the above arguments in mind, we
now turn to 5th order in DPT. Any diagram  
which consists of only one closed loop of spin 
transfer, is confined to two adjacent tetrahedra. As elaborated above, the different 3:1 
configurations on the pair of adjacent tetrahedra are 
indistinguishable in the graph sense. These diagrams can therefore contribute only a constant shift in energy.
Other than these diagrams, the only possibility is to have two separate closed loops of spin transfer. 
One must involve a single link being acted on twice, and the other must be a 3--link loop. 
Both loops reside on two different tetrahedra. If the two tetrahedra are adjacent, or share
a neighboring tetrahedron (as in in Fig.~\ref{fig:dpt6}), the tetrahedra are part of a 
retractable chain. Our general analysis for retractable chains then applies here as well. Therefore,
these diagrams must also sum up to a constant in the 3:1 manifold.
Otherwise the two tetrahedra are sufficiently well separated, and we can sum over the various diagrams 
on each tetrahedron separately. The summation gives
an expression in which each of the 4 tetrahedron corners are on equal footing, and cannot distinguish between
different 3:1 states, resulting in a constant energy shift.

\begin{figure}
\centering
\includegraphics[width=2.5in]{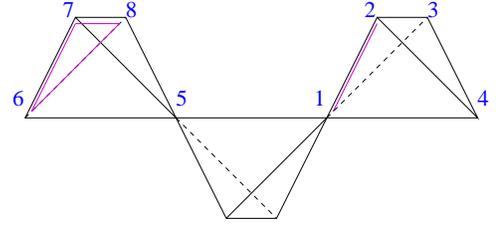}
\caption{(Color online) 5th order process.}
\label{fig:dpt6}
\end{figure}

In agreement with the results of Section~\ref{easy_axis}, DPT terms of order less than $6$ will not split the energy of the 3:1 states. Finally, we turn to 6th order in DPT
\be
{\mathcal H}_6 = -
{\mathcal P}
\left( {\mathcal H}_1 {\mathcal R} \right)^5 {\mathcal H}_1
{\mathcal P}
\ee
where ${\mathcal R} = \frac{\left( 1 - {\mathcal P}\right)}{{\mathcal H}_0 - E}$ is the resolvent.
At this order we have a new class of single closed loops - loops around hexagonal plaquettes of the pyrochlore lattice
(see Fig.~\ref{fig:dpt7}), which will split the energy of different 3:1 configurations. These processes can be enumerated
by the hexagonal plaquettes they act on. So we can write the contribution from these processes
$ \sum_{\mathcal P} {\tilde g}(\sigma_1 \ldots \sigma_6 )$ where the spins around the hexagonal plaquette
are denoted $1 \ldots 6$.

\begin{figure}
\centering
\includegraphics[width=2.5in]{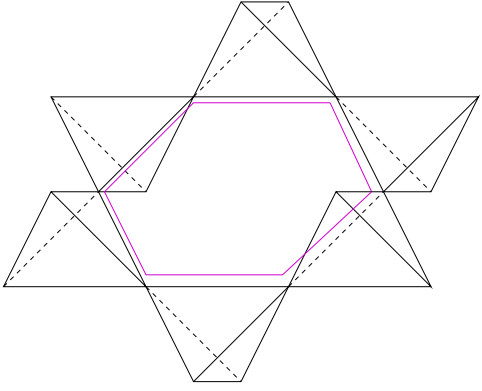}
\caption{(Color online) Hexagonal plaquette loop process.}
\label{fig:dpt7}
\end{figure}
All other single closed loop diagrams are confined to $1$, $2$ or $3$ adjacent tetrahedra, the first two we already know
will sum up to constant shifts in energy. The self-retracting loop residing on a chain of $3$ tetrahedra is depicted
in Fig.~\ref{fig:dpt8}, and is also summed over to produce a constant, since this is a retractable chain.

\begin{figure}
\centering
\includegraphics[width=2.5in]{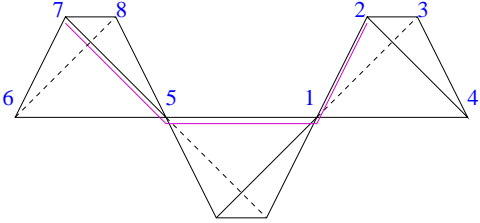}
\caption{(Color online) Single self retracting loop on a chain of three tetrahedra.}
\label{fig:dpt8}
\end{figure}

Next we consider the diagrams which comprise 2 closed loops. There must be two of length 3, or one of length 4 and one
of length 2. In the case of 2 loops of length 3, each loop must reside on a single tetrahedron.
If the two tetrahedra are adjacent, or even share a neighboring tetrahedron (as in Fig.~\ref{fig:dpt9}), 
they form a retractable chain, and result in a constant.
Otherwise the tetrahedra are sufficiently well separated so that 
we can sum over diagrams on the two tetrahedra
separately, and end up with a constant.
\begin{figure}
\centering
\includegraphics[width=2.5in]{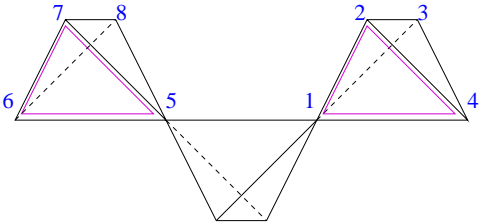}
\caption{(Color online) 6th order diagram with 2 loops of length 3.}
\label{fig:dpt9}
\end{figure}

We now turn our attention to the case of one loop of length 4 and one of length 2. 
The loop of length 2 must be one link appearing twice, and is therefore confined to 
a single tetrahedron. The loop of length 4 must be confined to 2 adjacent tetrahedra (or even just 1). 
If the two loops are sufficiently well 
separated (by more than one tetrahedra), we can sum over all 2-loop diagrams in the tetrahedron
where the loop of length 2 resides, resulting in a constant. 
We can then sum over diagrams on the tetrahedron cluster containing the 4-loop to produce another constant.   
If the two clusters are adjacent, or separated by just one tetrahedron, we have
a retractable chain, which we have already shown must result in a constant. 

The last set of 6th order diagrams are those of 3 closed loops, each comprising a single link, acted on twice.
Each loop is therefore confined to a single tetrahedron. Using the same arguments as before, if all three tetrahedra are
separated by more than one tetrahedra from one another, 
we can sum over all such diagrams contained in the same tetrahedra. In particular, sum over diagrams
related by permuting over sites on each single 
tetrahedron separately results in a constant. Even if only one tetrahedron 
is well separated from the other two, we can first sum over diagrams 
on the isolated tetrahedron to produce a constant, and then deal with 
the other two tetrahedra. The remaining two will reside on a 
retractable chain of tetrahedra, again resulting in a constant.

The only cases which we must deal with more carefully are those diagrams where each pair of tetrahedra either shares
a neighboring tetrahedron, or the two are adjacent. When at least one pair of loops resides on adjacent tetrahedra, the cluster of tetrahedra will always be a retractable chain, with either 4 or 3 tetrahedra.
We are left only with diagrams where all three tetrahedra containing the loops are \emph{not} adjacent, but rather have
shared neighboring tetrahedra. There are 3 tetrahedron clusters possible, where this occurs.
Fig.~\ref{fig:dpt12} shows a diagram residing on a closed chain of 6 tetrahedra, enclosing a hexagonal plaquette.
This is \emph{not} a retractable chain.
In Fig.~\ref{fig:dpt13} the three tetrahedra containing the loops all share one single neighboring tetrahedron.
Finally, in Fig.~\ref{fig:dpt14} the three loops reside on a linear chain of 5 tetrahedra. However, it is important
to note at this point that the tetrahedra at the two edges of this chain (1 and 7) can be identified to give the cluster
shown in Fig.~\ref{fig:dpt12}.

\begin{figure}
\centering
\includegraphics[width=2.0in]{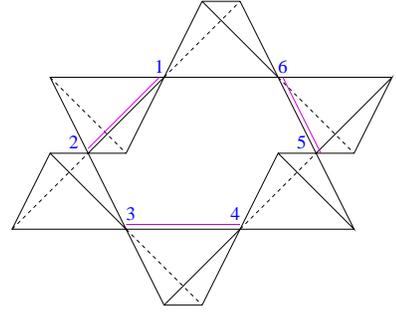}
\caption{(Color online) 3 loop diagram residing on a closed chain of tetrahedra, enclosing a hexagonal plaquette. This is the ``f'' process.}
\label{fig:dpt12}
\end{figure}

\begin{figure}
\centering
\includegraphics[width=2.0in]{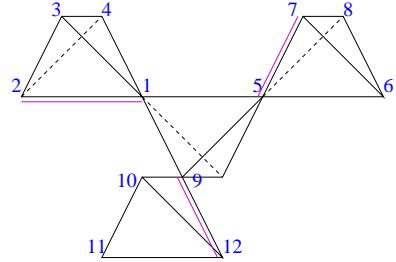}
\caption{(Color online) Diagram contributing to the function ``g''.}
\label{fig:dpt13}
\end{figure}

\begin{figure}
\centering
\includegraphics[width=3.5in]{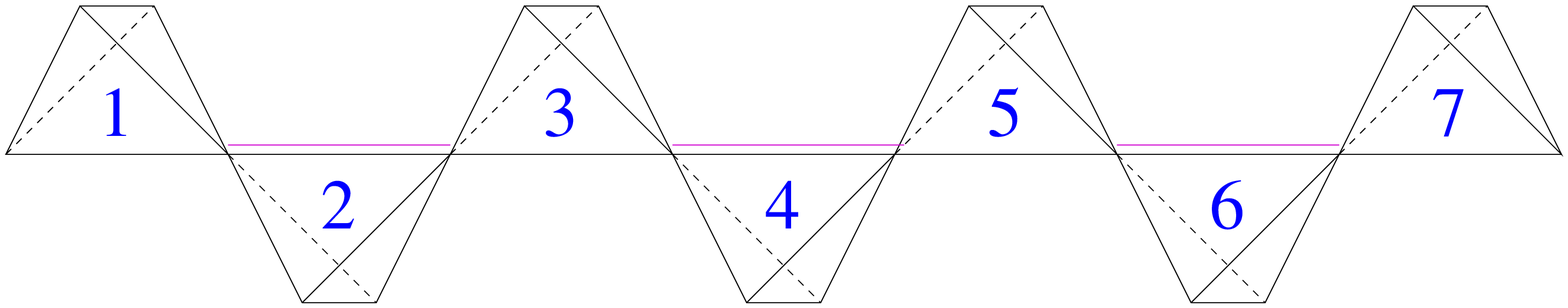}
\caption{(Color online) Diagram contributing to the function ``r''. The loops are embedded in the tetrahedra marked 2,4, and 6.}
\label{fig:dpt14}
\end{figure}

The processes on the cluster shown in Fig.~\ref{fig:dpt12} cannot be contracted, and will split the energy of the
different 3:1 states. We denote the sum of processes with the 3 loops on the same tetrahedra as in Fig.~\ref{fig:dpt12}
$f(\sigma_1 \ldots \sigma_6 )$. There is also a contribution from all the processes where the 3 loops reside
on the complementary set of tetrahedra, which must result in $f(\sigma_6, \sigma_1 \ldots \sigma_5 )$ (the same function
with the spins cyclically permuted once). Cyclically permuting the spins again, takes us back to the set of processes we
summed over in $f(\sigma_1 \ldots \sigma_6 )$, and therefore $f$ must be invariant under two permutations.
This is a good check for the correctness of the result we find.
The sum $f(\sigma_1 \ldots \sigma_6 ) + f(\sigma_6, \sigma_1 \ldots \sigma_5 ) $ accounts for all the diagrams 
on a given cluster enclosing a plaquette, and the total contribution to the effective Hamiltonian can be written
\begin{eqnarray}
\sum_{\mathcal P} \left[ f(\sigma_1 \ldots \sigma_6 ) + f(\sigma_6, \sigma_1 \ldots \sigma_5 ) \right]. \label{B11}
\end{eqnarray}

The processes on the cluster shown in Fig.~\ref{fig:dpt13} can be contracted from the ``loose ends''
of the diagram, similar to the contraction we have implemented for the retractable chains. For the particular example in Fig.~\ref{fig:dpt13}, we sum over the corresponding diagrams with $2$ permuted with $3$ and $4$.
The resulting sum
can only be a function of site $1$, $5$, $7$, $9$, and $12$. 
We then sum over the corresponding diagrams with $7$ permuted with $6$ and $8$, and then sum over $12$ permuted 
with $10$ and $11$. After these summations, the resulting function 
can only be a function of the sites $1$,$5$ and $9$.
Finally, we sum over all choices of three such sites on the central 
tetrahedron, which must result in a constant (a symmetric function of the 4 
sites of a single tetrahedron).

Next, we turn to the processes on the chain in Fig.~\ref{fig:dpt14}. 
At first glance, it would seem that these are retractable diagrams, that will amount to a 
constant. Such a summation would include 81 different diagrams (we retract the tetrahedra
2,3,5, and 6, each time summing over three permutations of sites). However, two of these diagrams
are of the form of Fig.~\ref{fig:dpt12}, in which case the tetrahedra 1 and 7 at the edges of the 
chain are \emph{identified}. As a result, these two diagrams may have \emph{different} resolvents
from the other 79 diagrams in this sum, since the sites on tetrahedra 1 and 6 may now interact.
These 2 terms have in fact already been taken into account in $f(\sigma_1 \ldots \sigma_6 )$,
and therefore should not be added again to the effective Hamiltonian.
The summation that makes contractible diagrams give constant energy shifts 
requires that the resolvents be \emph{identical} for the entire set
of diagrams. 

Rather than explicitly calculating the remaining 79 diagrams, we employ the following trick. 
In order to produce a constant, we add to the 79 diagrams
two terms having the \emph{same} resolvent, but replacing the spin variables
$\sigma_j$ with those at the positions $1$ through $6$ in Fig.~\ref{fig:dpt12}
All these 81 terms have the same resolvent as that of the process in Fig.~\ref{fig:dpt14}
and therefore will result in a constant. The two additional terms
\emph{do not} represent actual DPT processes, and therefore must be subtracted from the 
effective Hamiltonian. The terms we need to subtract
are enumerated by the choice of central tetrahedron in the chain in Fig.~\ref{fig:dpt14} (tetrahedron number 4), which
we can identify with the tetrahedron containing sites $3$ and $4$ in Fig.~\ref{fig:dpt12}. Given 
a choice of the central tetrahedron, each term of this sort
is a function of the 6 spins $ r(\sigma_1 \ldots \sigma_6 ) $ as defined in Fig.~\ref{fig:dpt12}. 
Given a hexagonal plaquette there are 6 choices of the central tetrahedron,
and correspondingly 6 terms 
and we must take into account all of these. Choosing a different central tetrahedron is equivalent to cyclically
permuting the spins $1 \ldots 6$, and so the total contribution of all these process is
\be
\begin{split}
- \sum_{\mathcal P} \Bigg( &
r(\sigma_1 \ldots \sigma_6 )
+ r(\sigma_6 \ldots \sigma_5 )
\\ &
+ r(\sigma_5 \ldots \sigma_4 )
+ r(\sigma_4 \ldots \sigma_3 )
\\ &
+ r(\sigma_3 \ldots \sigma_2 )
+ r(\sigma_2 \ldots \sigma_1 )
\Bigg).
\end{split}
\ee

The 6th order term in the effective Hamiltonian
finally reads
\be\label{H6}
\begin{split}
{\mathcal H}_6 = &
\sum_{\mathcal P} g(\sigma_1 \ldots \sigma_6 )
\\
+ & \sum_{\mathcal P} \Big( f(\sigma_1 \ldots \sigma_6 ) + f(\sigma_6, \sigma_1 \ldots \sigma_5 ) \Big)
\\
- & \sum_{\mathcal P} \Big(
r(\sigma_1 \ldots \sigma_6 )
+ r(\sigma_6 \ldots \sigma_5 )
\\ &
+ r(\sigma_5 \ldots \sigma_4 )
+ r(\sigma_4 \ldots \sigma_3 )
\\ &
+ r(\sigma_3 \ldots \sigma_2 )
+ r(\sigma_2 \ldots \sigma_1 )
\Big).
\end{split}
\ee
Calculating these functions explicitly, we find the \emph{exact} same results as in Section~\ref{easy_axis}.

\section{Spin $s=\frac{3}{2}$ diagonal term ground state degeneracy. }
\label{app:root3_degeneracy}

In this appendix we analyze the lowest energy states found for the diagonal term in the effective Hamiltonian
for spin $s = \frac{3}{2}$ discussed briefly in Section~\ref{sec:spin-s=32}. 
The lowest energy states turn out to be massively degenerate. We have found what is at the very least a 
subset of this manifold, which already exhibits a degeneracy that grows with the system size, diverging in the 
thermodynamic limit.
  
First we consider all the possible states we can construct with all the Kagome layers in the pyrochlore taking on the
$\sqrt{3}\times\sqrt{3}$ configuration in Fig.~\ref{fig:root3_state3}.
Given a plane in this
${\sqrt 3} \times {\sqrt 3}$ configuration, there are 3 different ways
to place the following layer (also in the ${\sqrt 3} \times {\sqrt
  3}$ configuration) above it, as described in Fig.~\ref{fig:stacking}.
This freedom in the way the planar configurations are stacked is the
source of a massive degeneracy -- there are $3^L$ possible stacking
choices, where $L$ is the linear dimension of the system. Also, there
are 4 plaquette directions in which to choose the direction of the
stacking, resulting in an overall number of such states $4 \times 3^L$.
For now, we shall work with this subset of the entire degenerate manifold, 
since these states are rather easy to handle. We shall refer to his set of states as the 
${\sqrt 3} \times {\sqrt 3}$ states.

Now we turn to calculate the energy of these states, proving these are degenerate states.
Each plaquette in the ${\sqrt 3} \times {\sqrt 3}$ planes shares links with three
plaquettes above it. These 4 plaquettes enclose an up-headed cell, as defined in Section~\ref{sec:strict-easy-axis}. 
For any one of the 3 ${\sqrt 3} \times {\sqrt 3}$ spin configurations of the next Kagome plane,
we always get the same up-headed cell types above the type 1 and type 0 plaquettes in the ${\sqrt 3} \times {\sqrt 3}$ plane.
The clusters enclosing these up-headed cells above the type 1 and type 0 plaquettes are depicted in Fig.~\ref{fig:stack1} and
Fig.~\ref{fig:stack2}, respectively. Since we are only considering up-headed cells, every one of the plaquettes appears
in only one cell. 

Inspection of Fig.~\ref{fig:stack1} shows that it includes one plaquette of type 1, one
of type 2, one of type 3, and one of type 4 (This is just the type 1 cell in Table~\ref{cell_table}).
For the cluster in Fig.~\ref{fig:stack2}, we find it includes two type 0
plaquettes, and two type 2 (type 8 cell in  Table~\ref{cell_table}). 
Of the plaquettes in the planes, which
comprise $\frac{1}{4}$ of the plaquettes in the lattice, $\frac{2}{3}$
are in a type 1 configuration, and $\frac{1}{3}$ are in a type 0
configuration. Since each planar plaquette determines the configuration of
the unique up-headed cell it is a part of, we find therefore, that 
$\frac{2}{3}$ of the up-headed cells are type 1 cells ($y_1 = \frac{2}{3}$)
and the remaining $\frac{1}{3}$ are type 8 cells ($y_8 = \frac{1}{3}$).
The corresponding plaquette type fractions are 
$x_0 = \frac{1}{6}, x_1 = \frac{1}{6}, x_2 = \frac{1}{3}, x_3 = \frac{1}{6}$ 
and $x_4 = \frac{1}{6}$. Finally, the energy
per plaquette of the ${\sqrt 3} \times {\sqrt 3}$ states is
\be\label{root3_E} 
\frac{1}{N} E_{{\sqrt 3} \times {\sqrt 3}} =
\frac{1}{6} \left( V_1 + 2 V_2 + V_4 \right).  
\ee
We expect the fractions of plaquettes in all the states in this degenerate manifold
to be the same as in this subset, since otherwise, it is rather unlikely (though not impossible)
that a different combination of fractions will yield the same energy.


Having analyzed this set of states, we turn to an additional set of states, with the same energy.
Consider a particular subset of ${\sqrt 3} \times {\sqrt 3}$ states, with every two Kagome planes stacked 
in the \emph{same} manner. This subset of the ${\sqrt 3} \times {\sqrt 3}$ states
has a total of $4 \times 3^2 = 36$ states - factor 4 for choosing the direction of the layering, a factor 3
for the choice of how to position the planar configuration on one plane, and another factor of 3 from the 
freedom to choose how to stack the next planar layer. We shall refer to these as the uniformly stacked
states.

Starting from any one of these uniformly stacked states we note that the type 1 cell can swap 
its type 1 and type 3 plaquettes by changing the position 
of only one minority site. In Fig.~\ref{fig:stack1} this can be accomplished by moving the minority 
site at site 4 to site 5. In addition, site 5 is part of a type 8 cell, and it is denoted also in 
Fig.~\ref{fig:stack2}, where it is evident that in order to maintain the 3:1 constraint, we must also shift
the minority site at site 1 in Fig.~\ref{fig:stack2} to site 6. Because the stacking of the next layer is 
\emph{exactly} the same, the same shifting of minority sites must occur along the entire straight line 
passing through site 4 in the direction from site 4 to site 5. Closing this chain at infinity makes 
this chain into an infinite length loop of alternating minority and majority sites, that are \emph{flipped}, 
and therefore this maintains the 3:1 constraint. One can convince one self from Fig.~\ref{fig:stack2} that the type 8 cell 
will remain a type 8 cell under these minority site shifts. Since all the cells remain in the same configuration type, 
the plaquette type fractions remain the same as in the ${\sqrt 3} \times {\sqrt 3}$ states, have the same energy,
and are therefore degenerate. 

We now turn to calculate the degeneracy of this new set of states.
Each type 8 cell has exactly one site it can shift in this manner,
but since the chains are not shared between different type 0 plaquettes in the plane, one
can convince oneself that these straight line 
chains of sites can be flipped \emph{independently}. Therefore, starting from a particular 
uniformly stacked state, since there are $\frac{N}{4L}$ plaquettes in every plane 
(where N is the number of pyrochlore sites, and the number of hexagonal plaquettes), 
there are $\frac{N}{12L}$ chains we can independently flip between two 
configurations, resulting in a degeneracy of $36 \times 2^{\frac{N}{12L}}$.
However, an additional subtlety must be addressed.

Every ${\sqrt 3} \times {\sqrt 3}$ state has all the up pointing tetrahedra in one of three 3:1 
configurations. Flipping any chain in a uniformly stacked state will introduce some number of up 
pointing tetrahedra in the 4th possible 3:1 configuration of a single tetrahedron. Starting 
from one uniformly stacked state, flipping all 
possible chains, we will change \emph{all} the tetrahedra of one of the three 3:1 configurations into 
the 4th tetrahedron configuration type, which was absent in the initial uniformly stacked state.
Once again we will find ourselves with all tetrahedra in only three possible different 3:1 configurations. 
This suggests, that perhaps this final state we have reached is also a uniformly stacked state.
$ $From inspection we find this indeed is the case, so that from one uniformly stacked state
we can reach one other such state by flipping all possible chains in the manner described above.
We must therefore correct the degeneracy to $18 \times 2^{\frac{N}{12L}}$ to account for this double counting.

In total we have found the degeneracy of these two sets of states to be
\be
18 \times 2^{\frac{N}{12L}} + 4 \times 3^L - 36
\; ,
\ee
where we have subtracted 36 since this is the number of states that appear in both sets of states we 
have analyzed (these are simply the uniformly stacked states). We shall refer to the combined set of states as
the $1-8$ manifold of states, since it involves only type 1 and type 8 cells.
The degeneracy we have calculated matches \emph{precisely}
the degeneracies in Table~\ref{table2} for all the clusters where we have found the lowest energy state for $s = \frac{3}{2}$,
namely the $3 \times 3 \times 1$, $6 \times 3 \times 1$, $3 \times 3 \times 2$ and $3 \times 3 \times 3$ clusters.
Our analysis clearly shows that this set of states is massively degenerate, and despite exhausting all the 
states we have found numerically with this energy, we cannot be certain that these exhaust all possible states with this
energy.

\begin{figure}
	\centering
	\subfigure[Stacking choices above a type 1 plaquette.]{
	\label{fig:stack1}
		\includegraphics[width=1.5in]{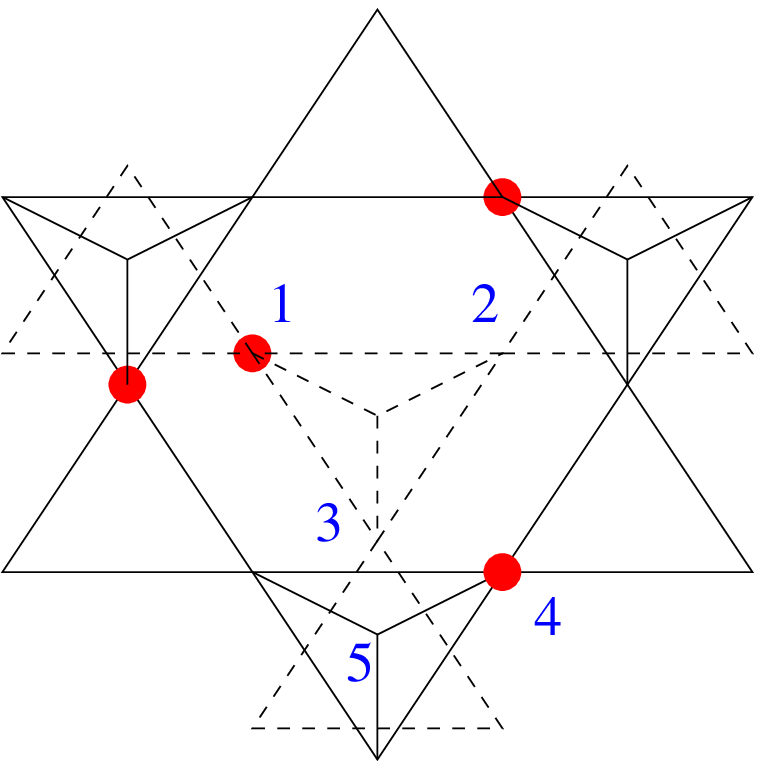}}
	\subfigure[Stacking choices above a type 0 plaquette.]{
	\label{fig:stack2}
		\includegraphics[width=1.5in]{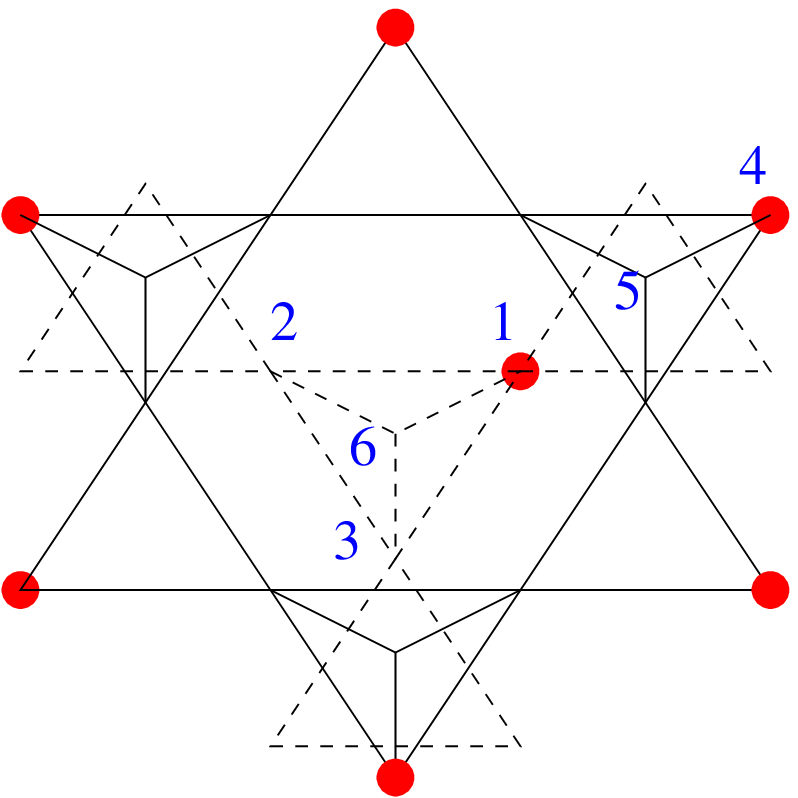}}	
              \caption{(Color online) Stacking of two planar ${\sqrt 3}
                \times {\sqrt 3}$ states. The solid lines represent the
                tetrahedra of the lower plane. The dashed lines
                represent the tetrahedra of the upper layer. The same
                convention regarding up pointing and down pointing
                tetrahedra applies here as in Fig.~\ref{fig:trig7},
                for both the solid and dashed line triangles.  In both
                figures, the dashed tetrahedron with the corners $1,2,3$
                must have one of these three corners a minority site,
                and any one of these three can be chosen. Once the
                minority site has been chosen from $1,2,3$, the ${\sqrt
                  3} \times {\sqrt 3}$ state of the upper layer is then
                uniquely determined. The choice of site 1 is explicitly
                shown in both cases.}
	\label{fig:stacking}
\end{figure}

In a previous publication,\cite{Bergman:prb06} analyzing the same 3:1
degenerate manifold of states, the authors ascertained the maximum
fraction of type 1 plaquettes that can be placed on a pyrochlore lattice
is $\frac{1}{4}$.
The $1-8$ states do not realize this limit, but come fairly close to it with $\frac{1}{6}$ of
the plaquettes in the type 1 configuration.
Therefore, the ${\sqrt 3} \times {\sqrt 3}$ states are a ``compromise''
between the energy gain of $V_1$ (which favors the type 1 plaquettes)
and the energy loss of $V_2$ (which disfavors the type 2 plaquettes),
taking into account the constraint $x_1<x_2$.
This becomes evident when calculating the energy of the various cell types.
The state realizing the maximum fraction of type 1 plaquettes found in 
Ref.~\cite{Bergman:prb06} is comprised of only type 2 cells.
For $s = \frac{3}{2}$, type 1 cells are far more favorable in energy than type 2 cells,
and type 1 cells are abundant in the $1-8$ states.


\end{document}